\documentclass{aa}  

\usepackage{amsmath, amssymb, amsfonts, amscd}
\usepackage{natbib}
\usepackage{graphicx}
\usepackage{txfonts}
\usepackage[final]{hyperref}
\usepackage{mathrsfs}
\usepackage{color}
\usepackage{mathabx}

\hypersetup{
   colorlinks=true,
   citecolor=blue,
   linkcolor=blue,
   urlcolor=black
   }

\newcommand\ed{{\rm e}}
\newcommand\dd{{\rm d}}
\newcommand{\vect}[1]{\textbf{#1}}
\newcommand{\bs}[1]{\boldsymbol{#1}}

\definecolor{emerald}{rgb}{0.3,0.85,0.2}
\definecolor{smcolor}{rgb}{0.7,0.3,0.0}

\newcommand{\smc}[1]{#1}
\newcommand{\jlc}[1]{#1}
\newcommand{\padc}[1]{#1}
\newcommand{\rec}[1]{#1}
\newcommand{\smcc}[1]{#1}

\begin{document} 
  \title{Oceanic tides from Earth-like to ocean planets}
 
  \author{P. Auclair-Desrotour\inst{1,2,4}
     \and
           S. Mathis\inst{2,3}\fnmsep
     \and 
          J. Laskar\inst{1}
     \and
          J. Leconte\inst{4}
          }

  \institute{
  IMCCE, Observatoire de Paris, CNRS UMR 8028, PSL, 77 Avenue Denfert-Rochereau, 75014 Paris, France\\
\email{pierre.auclair-desrotour@u-bordeaux.fr, jacques.laskar@obspm.fr}
\and Laboratoire AIM Paris-Saclay, CEA/\smc{DRF} - CNRS - Universit\'e Paris Diderot, IRFU/\smc{DAp} Centre de Saclay, F-91191 Gif-sur-Yvette, France
\and LESIA, Observatoire de Paris, PSL Research University, CNRS, Sorbonne Universit\'es, UPMC Univ. Paris 06, Univ. Paris Diderot, Sorbonne Paris Cit\'e, 5 place Jules Janssen, F-92195 Meudon, France\\
\email{stephane.mathis@cea.fr}
\and Laboratoire d'Astrophysique de Bordeaux, Univ. Bordeaux, CNRS, B18N, allée Geoffroy Saint-Hilaire, 33615 Pessac, France \\
\email{jeremy.leconte@u-bordeaux.fr}
    }

  \date{Received ...; accepted ...}

  \abstract
   {Oceanic tides are a major source of tidal dissipation. They drive the evolution of planetary systems and the rotational dynamics of planets. However, 2D models commonly used for the Earth cannot be applied to extrasolar telluric planets hosting \smc{potentially} deep oceans because they ignore the three-dimensional effects related to the ocean vertical structure.}
   {Our goal is to investigate in a consistant way the importance of the contribution of internal gravity waves in the oceanic tidal response and to propose a modeling allowing to treat a wide range of cases from shallow to deep oceans. }
   {A 3D ab initio model is developed to study the dynamics of a global planetary ocean. This model takes into account compressibility, stratification and sphericity terms, which are usually ignored in 2D approaches. An analytic solution is computed and used to study the dependence of the tidal response on the tidal frequency and on the ocean depth and stratification.}
   {In the 2D asymptotic limit, we recover the frequency-resonant behaviour due to surface \rec{inertial-gravity} waves identified by early studies. As the ocean depth and Brunt-Väisälä frequency increase, the contribution of internal gravity waves grows in importance and the tidal response become three-dimensional. In the case of deep oceans, the stable stratification induces resonances that can increase the tidal dissipation rate by several orders of magnitude. \padc{It is thus able to affect significantly the evolution time scale of the planetary rotation.} }
   {}

  \keywords{hydrodynamics -- planet-star interations -- planets and satellites: oceans -- planets and satellites: terrestrial planets.}

\maketitle


\section{Introduction}

Even though the number of detected extrasolar planets located in the habitable zone of their host stars has kept growing continuously for the \rec{past two} decades, two major discoveries recently aroused excitement within the community of exoplanets. First, in \rec{Summer, 2016}, a telluric extrasolar planet was found orbiting the Sun's closest stellar neighbour, the red dwarf Proxima Centauri \citep[][]{AE2016,Ribas2016}. This planet, Proxima b, has a minimum mass of $ 1.3~M_\Earth $ and its equilibrium temperature is within the range where water could be liquid on its surface, which makes it appear as a \smc{potential} twin sister of the Earth. A few months later, \smc{another} star was in the spotlight, the ultra-cool dwarf star TRAPPIST-1 \citep[][]{Gillon2017}. The system hosted by TRAPPIST-1 drew attention for its remarkable architecture, composed of eight telluric planets of masses between $ 0.7~M_\Earth $ and $ 1.2~M_\Earth $ and radii between $ 0.1~R_\Earth $ and $ 1.5~R_\Earth $ \citep[][]{Gillon2017,Wang2017}. Most of these planets (i.e. b, c, d, e, f, g, h) exhibit small densities suggesting that water stands for a large fraction of their masses \citep[][]{Wang2017}. Hence, Proxima b and several of the telluric planets orbiting TRAPPIST-1 could host deep oceans of liquid water.

Oceans play a crucial role in the evolution of planetary systems. Similarly to rocky cores, they are distorted by the tidal gravitational forcings of perturbers such as stars \smc{or} satellites. The energy dissipated by the resulting oceanic tides can be of \smc{the} same order of magnitude and even greater than that associated with the solid tide, as proved by the case of the Earth. Observed as an exoplanet, the Earth appears as a very dry rocky planet with an oceanic layer of depth approximatively equal to $ 6 \times 10^{-4}~R_\Earth $ \citep{ES2010}. Yet, the Earth's ocean is today the main contributor to the total energy tidally dissipated by the Earth-Moon-Sun system. It accounts for roughly 95 \% of the planetary dissipation rate of 2.54 TW generated by the Lunar semidiurnal tide \cite[see][and references therein]{Lambeck1977,Ray2001,ER2001}. Moreover, the oceanic dissipation rate can vary significantly owing to the frequency-resonant behaviour of the fluid layer \citep[e.g.][]{Webb1980}, thus leading to the \rec{\smc{present-day high transfer}} of angular momentum from the spin of the Earth to the orbit of the Moon \citep[][]{Lambeck1980,BR1999}. \smc{More generally, oceanic tides have an impact on the history of the spin rotation and states of equilibrium of the Earth \citep[][]{NSL1997} and terrestrial exoplanets \citep[][]{Correia2008,Leconte2015,ADLM2017b}.} \smc{Therefore, the oceanic tidal dissipation rate} needs to be quantified in the case of extrasolar planets to study the history of observed planetary systems and constrain their physical properties. 

Since Laplace's pioneering works \citep[][]{Laplace1798}, the Earth's oceanic tides have been examined through different approaches. The dissipation rate they induce is now well constrained thanks to the fit of the altimetric data provided by the TOPEX/Poseidon satellite with numerical simulations made using General Circulation Models (or GCM) \citep[][]{ER2001,ER2003}. Moreover, numerous studies have characterized their dependence on the planet parameters by developing linear ab initio global models, such as the hemispherical ocean model \citep[][]{Proudman1936,Doodson1938,LH1970,Webb1980}. This approach has been used lately to estimate the tidal dissipation rate in the oceans of icy satellites orbiting Jupiter and Saturn \citep[][]{Tyler2011,Tyler2014,Chen2014,Matsuyama2014}. 

Simplified ab initio models are very convenient to explore the domain of parameters because they involve a small number of control parameters and require a computational cost relatively small compared to numerical models taking into account local features (e.g. topography, mean flows). They are commonly based on a two dimensional spherical geometry and the \rec{thin shell hypothesis}. \rec{In this approach, the tidal response of the ocean is controlled by the Laplace Tidal Equations (LTE), which describe its barotropic response, that is the component associated with the propagation of surface gravity waves. This is typically the tidal response that the ocean will have if it is unstratified (uniform density). Nevertheless, solutions to the barotropic equations can be applied to the case of stably stratified oceans to describe the contribution of internal gravity waves, restored by the Archimedean force. As it has been shown for long \citep[e.g.][in the case of the atmosphere]{CL1970}, the tidal response of a stratified fluid layer in the classical tidal theory is characterized by a frequency-dependent equivalent depth, which is the analogous of the barotropic ocean depth \citep[see e.g.][]{Tyler2011}. Results obtained in the barotropic approximation can thus be reinterpreted by replacing the ocean depth by the equivalent depth associated with the studied mode.  However, the complete solution controlling the tidal response of a deep stratified ocean, including its vertical component, has never been explicitly resolved nor analyzed to our knowledge.}

Therefore, following along the line of \cite{Webb1980} \rec{and \cite{Tyler2011}}, we develop here a linear global model of oceanic tides including three-dimensional effects related to the ocean sphericity and vertical structure. This model, by taking into account the vertical stratification of the ocean with respect to convection \smc{(i.e. its convective stability)}, allows us to quantify in a consistent way the contribution of internal gravity waves to the oceanic tidal response \rec{through explicit solutions derived in simplified cases}. The obtained results can be used as a tool to (1) characterize the tidal behaviour of deep global oceans, (2) constrain the structure and history of observed telluric planet from the architecture of their host planetary system, and (3) provide a \rec{simplified} diagnosis of the oceanic tidal dissipation rate.

In Section~\ref{sec:dynamics_ocean}, we establish the equations governing the dynamics of the 3D oceanic tidal response, identify the possible tidal regimes, and express the tidal Love numbers and torque exerted on the planet in the general case. In Section~\ref{sec:uniform_ocean}, we simplify the dynamics by assuming a uniform stratification \rec{as a first step} and derive an analytic solution for the tidal response and the resulting dissipation rate. This solution is then applied to idealized Earth and TRAPPIST-1 f planets in Section~\ref{sec:application_cases} in order to illustrate the difference between shallow \smc{and} deep oceans. We use it in Section~\ref{sec:explo_para} to explore the dependence of the dissipation rate on the ocean depth and stratification. Finally, we discuss the simplifications assumed in the modeling in Section~\ref{sec:discussion} and give our conclusions in Section~\ref{sec:conclusion}.

\section{Dynamics of a thick ocean forced gravitationally}
\label{sec:dynamics_ocean}

\begin{figure}[htb]
 \centering
  \includegraphics[width=0.45\textwidth,clip]{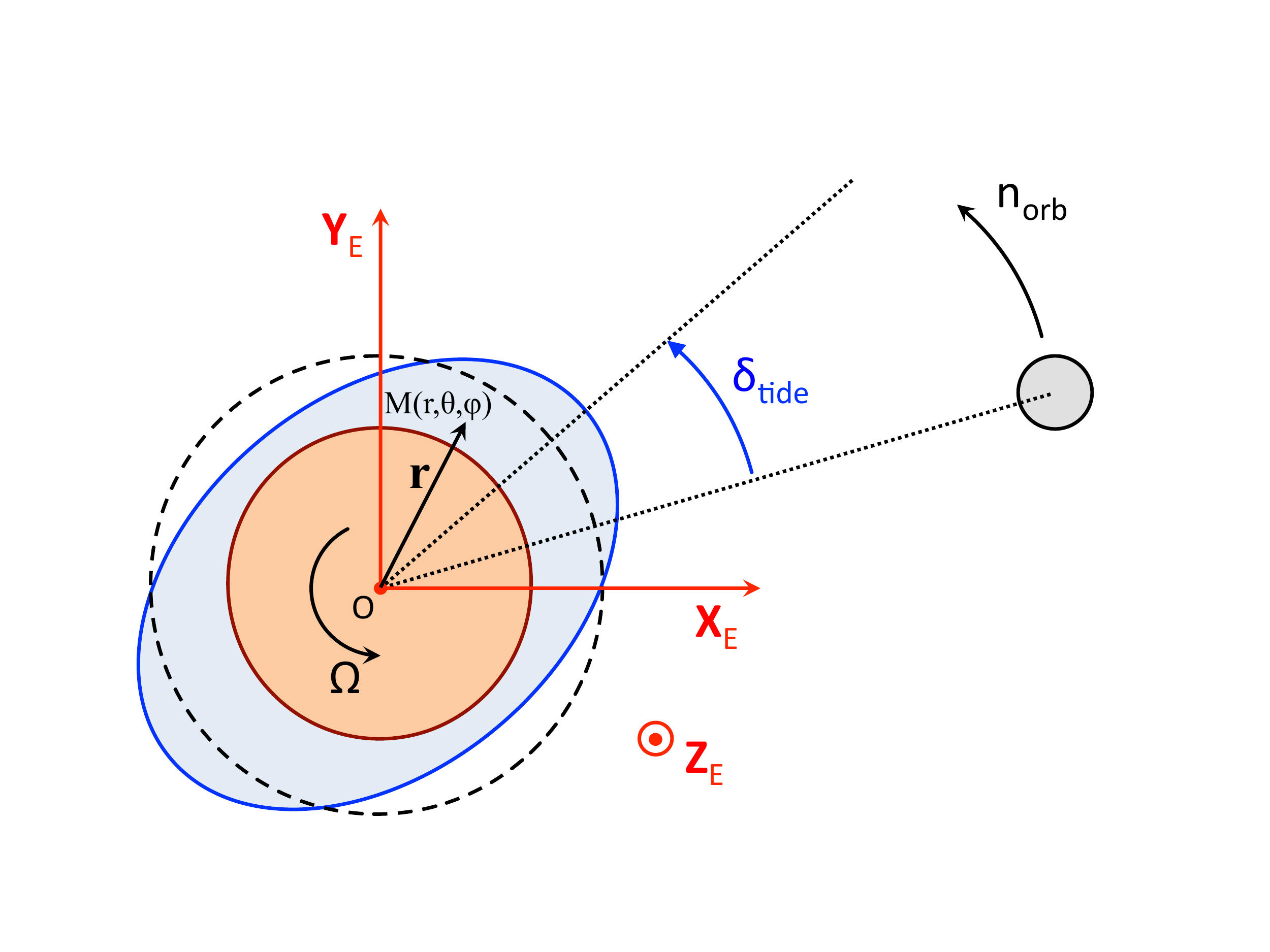} 
\caption{\label{fig:schema_setup} Quadrupolar semidiurnal oceanic tide in a rotating terrestrial planet gravitationally forced by a perturber (star of satellite). Reference frame and system of coordinates. The spin frequency of the planet and the orbital frequency of the perturber are denoted $ \Omega $ and $ n_{\rm orb} $ respectively. The notation $ \delta_{\rm tide} $ designates the angular lag of the tidal bulge.  }  
\end{figure}

We establish in this section the equations that govern the dynamics of tides in a thick oceanic shell submitted to the gravitational tidal potential of a perturber. The considered system is a spherical telluric planet of external radius $ R $ uniformly covered by an oceanic layer of depth $ H $ (Fig.~\ref{fig:schema_setup}). The radius of the solid part is designated by $ R_{\rm c} = R - H $. We assume that the ocean rotates uniformly with the rocky part at the angular velocity $ \Omega $, the corresponding spin vector being denoted $ \boldsymbol{\Omega} $. This allows us to write the fluid dynamics in the natural equatorial reference frame rotating with the planet, $ \mathscr{R}_{\rm E;T} : \left\{ O, \textbf{X}_{\rm E}, \textbf{Y}_{\rm E}, \textbf{Z}_{\rm E}  \right\} $, where $ O $ is the centre of the planet, $  \textbf{X}_{\rm E} $ and $ \textbf{Y}_{\rm E} $ define the equatorial plane and $ \textbf{Z}_{\rm E} = \boldsymbol{\Omega} / \left| \boldsymbol{\Omega} \right|  $. We use the spherical basis $ \left( \textbf{e}_r , \textbf{e}_\theta, \textbf{e}_\varphi \right) $ and coordinates $ \left( r, \theta , \varphi \right) $ where $ r $ is the radial coordinate, $\theta $ the colatitude and $ \varphi $ the longitude (the position vector is $ \textbf{r} = r \, \textbf{e}_r $). Finally, the time is denoted $ t $.  \\

\subsection{Forced dynamics equations}
\label{subsec:dynamics_equations}

The planet generates a gravity field, denoted $ g $, oriented radially. The centrifugal acceleration due to rotation tends to make the effective acceleration depend on latitude. We ignore this dependence by considering that $ \Omega \ll \Omega_{\rm c} $, where $ \Omega_{\rm c} = \sqrt{g / r} $ represents the Keplerian critical rotation velocity (the rotation velocity of the planet is far slower than the critical velocity for which the distortion due to the centrifugal acceleration destroys the oceanic layer). We assume then that the ocean is stratified radially in pressure $ p $ and density $ \rho $. These quantities are written 

\begin{equation}
\begin{array}{ll}
  p \left( \textbf{r} , t \right) = p_0 \left( r \right) + \delta p \left( \textbf{r} , t \right),  & \! \!  \rho \left( \textbf{r} ,t \right) = \rho_0 \left( r \right) + \delta \rho \left( \textbf{r} , t \right),
\end{array}
\end{equation} 

\noindent where the superscript $ _0 $ refers to the background distribution and $ \delta $ to a small fluctuation at the vicinity of the equilibrium. Similarly, the velocity of the flows generated by the perturbation is denoted $ \textbf{V} \left( \textbf{r} , t \right) = \left( V_r , V_\theta , V_\varphi \right) $ and the associated displacement $ \bs{\xi} $ \smc{\citep[such that $ \textbf{V} = \partial_t \, \bs{\xi} $,][]{Unno1989}}. Note that, because of the solid rotation assumption, there is no \smc{meridional or zonal} background circulation. We will consider in the following that the tidal perturbation can be approximated by a linear model and, therefore, ignore terms of order greater than 1 \rec{with respect to} $ \bs{\xi} $, $ \textbf{V} $, $ \delta p $ and $ \delta \rho $. Moreover, we assume the \emph{Cowling approximation} \citep[][]{Cowling1941}, i.e. the self gravitational effect of the variation of mass distribution is not taken into account. \rec{Finally, following early works \citep[e.g.][]{Webb1980,ER2001,ER2003,Tyler2011},} we introduce dissipation by using a Rayleigh friction, \smc{$ \sigma_{\rm R} \textbf{V} $}, where $ \sigma_{\rm R} $ is a constant effective \rec{frequency associated with the damping} \rec{\citep[note that][consider two different models to describe friction, the Rayleigh friction model and a quadratic one depending on the total velocity of the flow and including all tidal components]{ER2001,ER2003}}. As demonstrated by \cite{Ogilvie2009}, the results obtained with such a simplified approach are a reasonably good approximation of those obtained by taking into account the complete viscous force. \smc{It also seems to be a reasonable approach for situations such as exoplanets, where the solid-fluid coupling is naturally unknown.}

Under these assumptions, the Navier-Stokes equation describing the distortion of the ocean forced by the tidal potential $ U $ can be expressed in the frame rotating with the planet as \citep[][]{GZ2008}

\begin{equation}
\partial_t \textbf{V} + 2 \boldsymbol{\Omega} \wedge \textbf{V} = - \frac{1}{\rho_0} \bs{\nabla} \delta p - \frac{g}{\rho_0} \delta \rho \textbf{e}_r + \bs{\nabla} U - \sigma_{\rm R} \textbf{V}. 
\label{NS}
\end{equation}

\noindent It is completed by the equation of mass conservation

\begin{equation}
\partial_t \delta \rho + \frac{d \rho_0}{dr} V_r + \rho_0 \bs{\nabla} \cdot \textbf{V} = 0,
\label{mass_cons}
\end{equation}

\noindent and the equation of buoyancy

\begin{equation}
\partial_t \delta \rho + \frac{d \rho_0}{dr} V_r = \frac{1}{c_s^2} \left( \partial_t \delta p  + \frac{d p_0}{dr} V_r \right),
\label{buoyancy}
\end{equation}

\noindent where $ c_s \left( r \right) = \Gamma \left( p_0 , \rho_0 , S_0 \right) $ designates the sound velocity, $ \Gamma $ being any state function and $ S_0 $ the salinity of the fluid. It shall be noted here that we only consider isohaline and isentropic processes. Diabatic processes such as the thermal expansion of the ocean are ignored. The radial stratification of the oceanic layer is characterized by the Brunt-Väisälä frequency \citep[e.g.][]{GZ2008},

\begin{equation}
N^2 = - \frac{g}{H} \left[ \frac{d \ln \rho_0}{dx} + \frac{g H}{c_s^2}  \right],
\label{N2}
\end{equation} 

\noindent where we have introduced the \rec{non-dimensional} reduced altitude $ x$, defined by $ r = R_{\rm c} + H x $. The three components of the momentum equation given by Eq.~\ref{NS} are strongly coupled by the Coriolis acceleration. To make the problem tractable analytically, we assume the commonly used \emph{traditional approximation} \citep[e.g.][]{Eckart1960,Mathis2008,ADLM2017a}. This simplification consists in ignoring the latitudinal component of the rotation vector in the Coriolis acceleration (i.e. $2 \Omega \sin \theta V_r$ and $ 2 \Omega \sin \theta V_\varphi $ \smc{along $\textbf{e}_\varphi $ and $\textbf{e}_r $ respectively}), giving

\begin{equation}
\begin{array}{rcl}
    \displaystyle \partial_t V_\theta - 2 \Omega \cos \theta V_\varphi & = &  \displaystyle - \frac{1}{r} \partial_\theta \left( \frac{\delta p}{\rho_0} - U \right)  - \sigma_{\rm R} V_\theta, \\[0.3cm]
   \displaystyle  \partial_t V_\varphi + 2 \Omega \cos \theta V_\theta & = &  \displaystyle - \frac{1}{r \sin \theta} \partial_\varphi \left( \frac{\delta p}{\rho_0} - U \right) - \sigma_{\rm R} V_\varphi, \\[0.3cm]
    \displaystyle \partial_t V_r & = &  \displaystyle - \frac{1}{\rho_0} \partial_r \delta p - \frac{g}{\rho_0} \delta \rho + \partial_r U - \sigma_{\rm R} V_r.
\end{array}
\label{eq1}
\end{equation}

\noindent It thus makes possible the separation of the $ \theta $ and $ r $ coordinates in solutions, as showed thereafter.  As discussed in Section~\ref{sec:discussion}, the traditional approximation is usually considered to be satisfactory in stably stratified layers ($ 2 \Omega \ll N^2 $) and for tidal frequencies satisfying the condition $ 2 \Omega < \sigma \ll N^2 $, which corresponds to the regime of super-inertial waves \smc{\citep[e.g.][]{Mathis2009,ADML2017c}}. 

\rec{Finally,} the equations of mass conservation (\ref{mass_cons}) and of the buoyancy (\ref{buoyancy}) can be written respectively

\begin{equation}
\partial_t \delta \rho + \frac{1}{r^2} \partial_r \left( r^2 \rho_0 V_r \right) = - \frac{\rho_0}{r \sin \theta} \left[ \partial_\theta \left( \sin \theta V_\theta \right) + \partial_\varphi V_\varphi \right],
\end{equation}

\noindent and

\begin{equation}
\partial_t \delta \rho = \frac{1}{c_s^2} \partial_t \delta p + \frac{\rho_0 N^2}{g} V_r. 
\label{eq3}
\end{equation}

The tidal perturbation is \smc{periodic} in time and longitude. Therefore, introducing the longitudinal wavenumber $ m $ and the tidal frequency $ \sigma $, we expand the perturbed quantities in Fourier series of $ t$ and $ \varphi $. Any quantity $ f $ is written

\begin{equation}
f = \sum_{m,\sigma} f^{m,\sigma} \left( r , \theta \right) \ed^{i \left( \sigma t + m \varphi \right)},
\label{Fourier}
\end{equation}

\noindent where the $ f^{m,\sigma} $ are the spatial distributions of $ f $ associated with the doublet $ \left( m , \sigma \right) $. In the case of a perfect fluid ($ \sigma_{\rm R} = 0 $), the latitudinal structure of the perturbation is fully characterized by $ m $ and the real parameter $ \nu = 2 \Omega / \sigma $, which is the so-called spin parameter \citep[e.g.][]{LS1997}. Viscous friction modifies slightly the latitudinal structure by introducing a dependence on $ \sigma_{\rm R} $. Thus, we define the complex tidal frequency and spin parameter as

\begin{equation}
\begin{array}{rcl}
\displaystyle \tilde{\sigma} = \sigma - i \sigma_{\rm R}  & \mbox{and} & \displaystyle \tilde{\nu} = \frac{2 \Omega}{\tilde{\sigma} },
\end{array}
\label{nu_tilde}
\end{equation}

\noindent and the latitudinal operators 

\begin{equation}
\begin{array}{lr}
    \displaystyle \mathcal{L}_\theta^{m,\tilde{\nu}} = \frac{1}{1 - \tilde{\nu}^2 \cos^2 \theta} \left[ \partial_\theta + m \tilde{\nu} \cot \theta \right]  & \mbox{and} \\[0.3cm]
   \displaystyle \mathcal{L}_\varphi^{m,\tilde{\nu}} =  \frac{1}{1 - \tilde{\nu}^2 \cos^2 \theta} \left[ \tilde{\nu} \cos \theta \partial_\theta + \frac{m}{\sin \theta} \right] , &
\end{array}
\end{equation}

\noindent such that

\begin{equation}
\begin{array}{lr}
  \displaystyle V_\theta^{m,\sigma} = \frac{i}{\tilde{\sigma} r } \mathcal{L}_\theta^{m,\tilde{\nu}} \left( \frac{\delta p^{m,\sigma}}{\rho_0} - U^{m,\sigma} \right) & \mbox{and} \\[0.3cm]
  \displaystyle V_\varphi^{m,\sigma} = - \frac{1}{\tilde{\sigma}r} \mathcal{L}_\varphi^{m,\tilde{\nu}} \left( \frac{\delta p^{m,\sigma}}{\rho_0} - U^{m,\sigma} \right). 
\end{array}
\end{equation}

\begin{figure*}[htb]
 \centering
 \includegraphics[height=0.3cm]{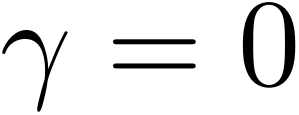} \hspace{2.8cm}
  \includegraphics[height=0.3cm]{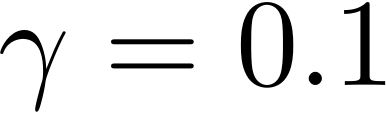} \hspace{2.8cm}
  \includegraphics[height=0.3cm]{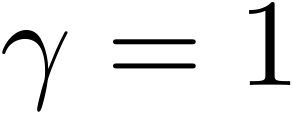} \hspace{2.8cm}
  \includegraphics[height=0.3cm]{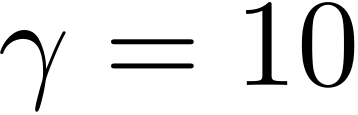} \hspace{2.8cm} \\
 \raisebox{1.5\height}{\includegraphics[width=0.018\textwidth]{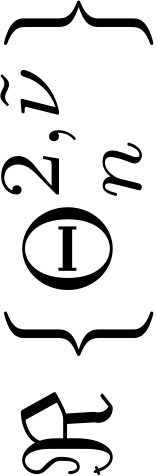}}
  \includegraphics[width=0.21\textwidth,trim = 3.2cm 3cm 6.5cm 2cm,clip]{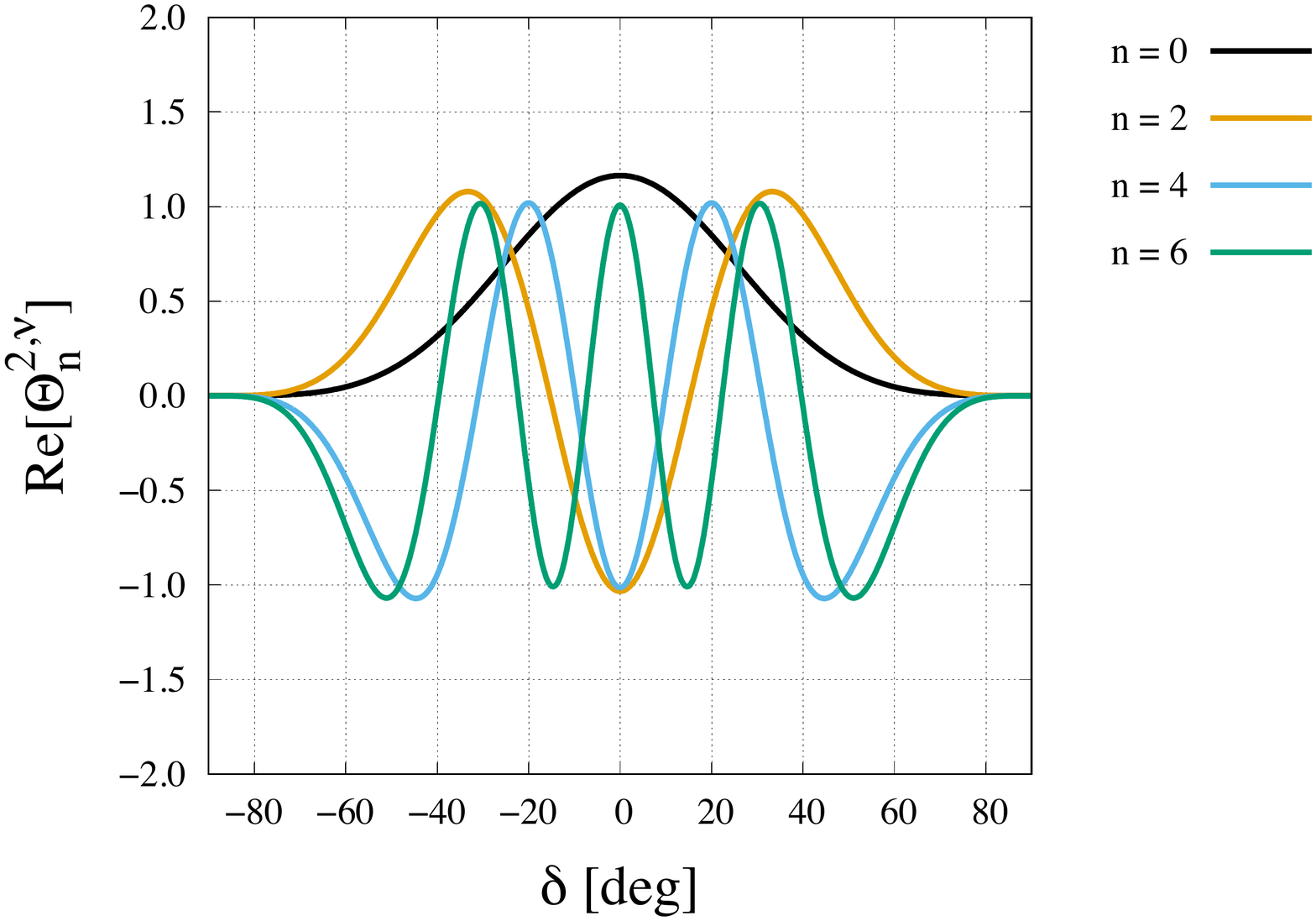} 
  \includegraphics[width=0.21\textwidth,trim = 3.2cm 3cm 6.5cm 2cm,clip]{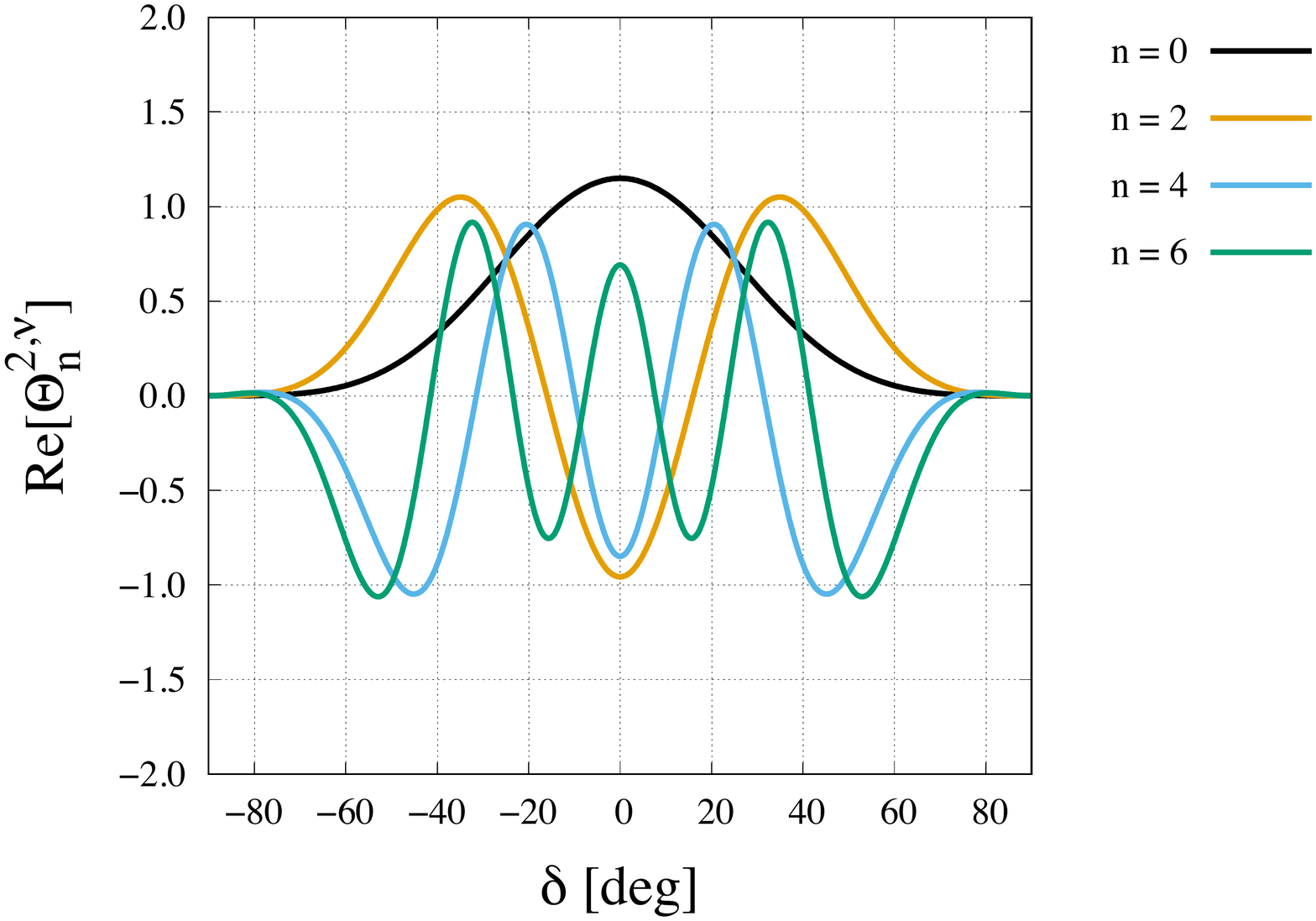} 
  \includegraphics[width=0.21\textwidth,trim = 3.2cm 3cm 6.5cm 2cm,clip]{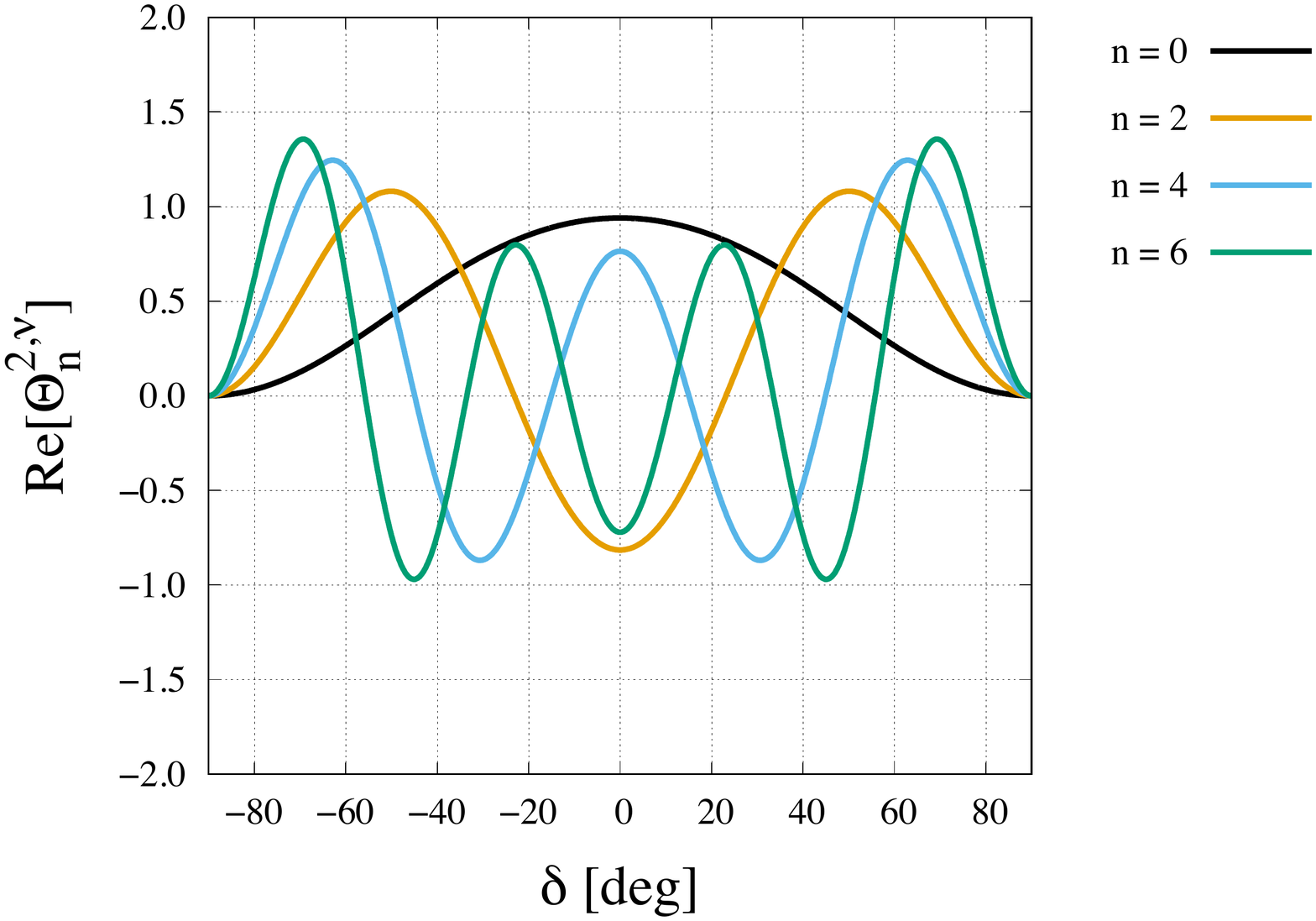} 
  \includegraphics[width=0.21\textwidth,trim = 3.2cm 3cm 6.5cm 2cm,clip]{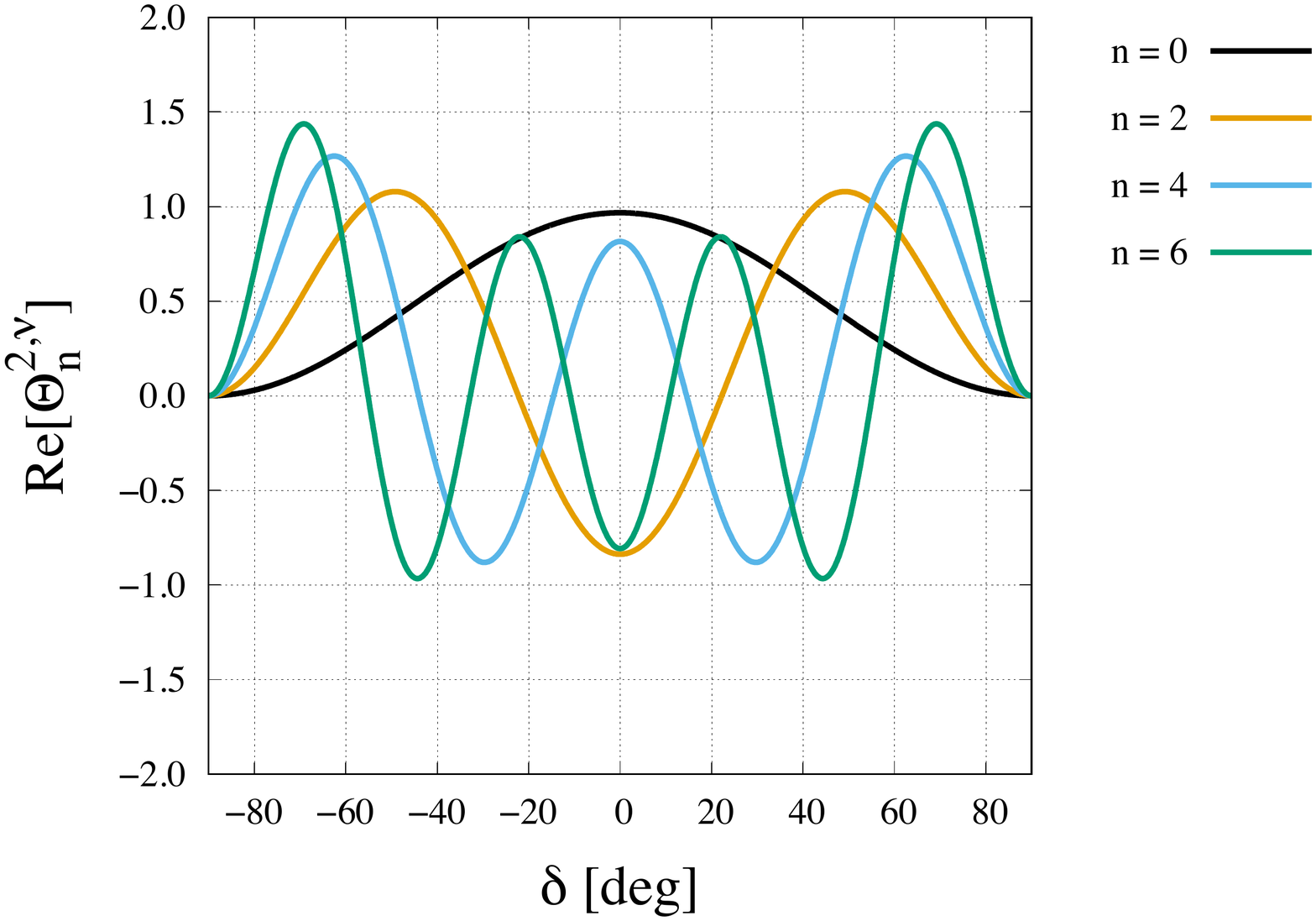}
  \includegraphics[width=0.08\textwidth,trim = 22cm 9cm 1.6cm 2cm,clip]{auclair-desrotour_fig2i.pdf} \\
  \raisebox{1.5\height}{\includegraphics[width=0.018\textwidth]{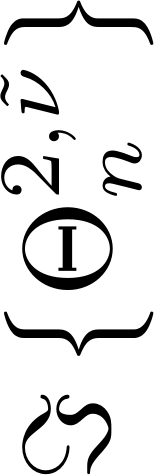}}
  \includegraphics[width=0.21\textwidth,trim = 3.2cm 3cm 6.5cm 2cm,clip]{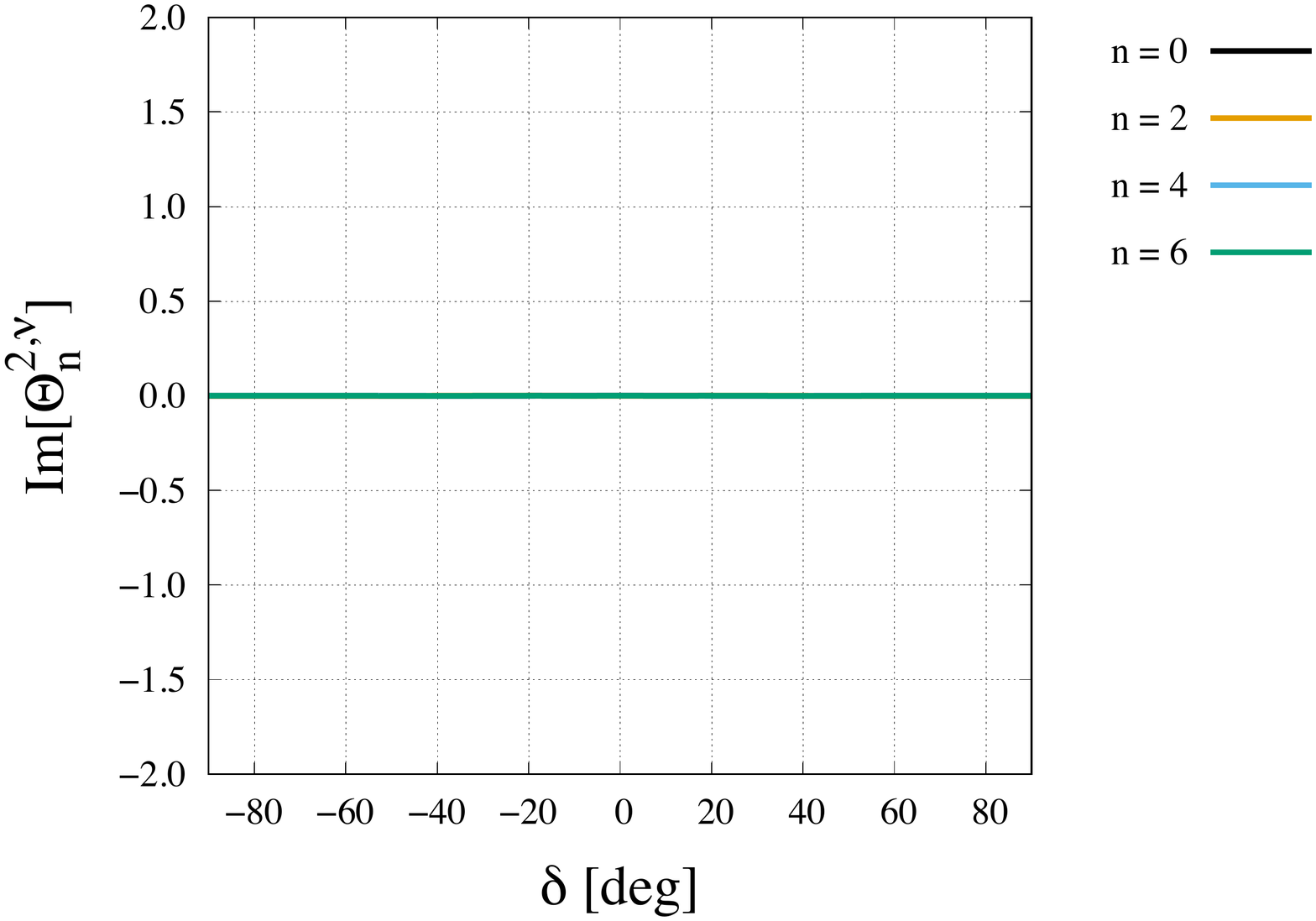}  
  \includegraphics[width=0.21\textwidth,trim = 3.2cm 3cm 6.5cm 2cm,clip]{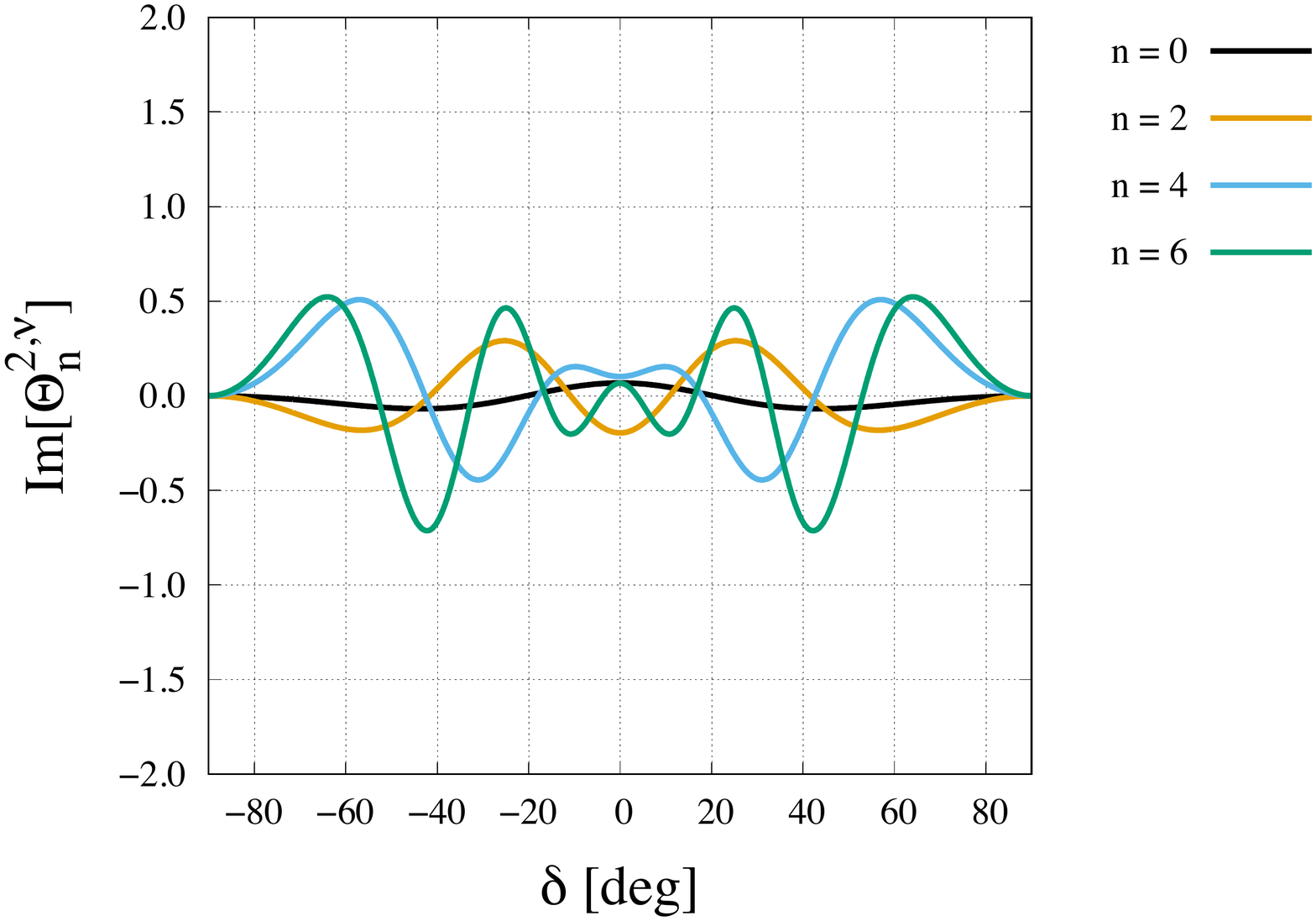} 
  \includegraphics[width=0.21\textwidth,trim = 3.2cm 3cm 6.5cm 2cm,clip]{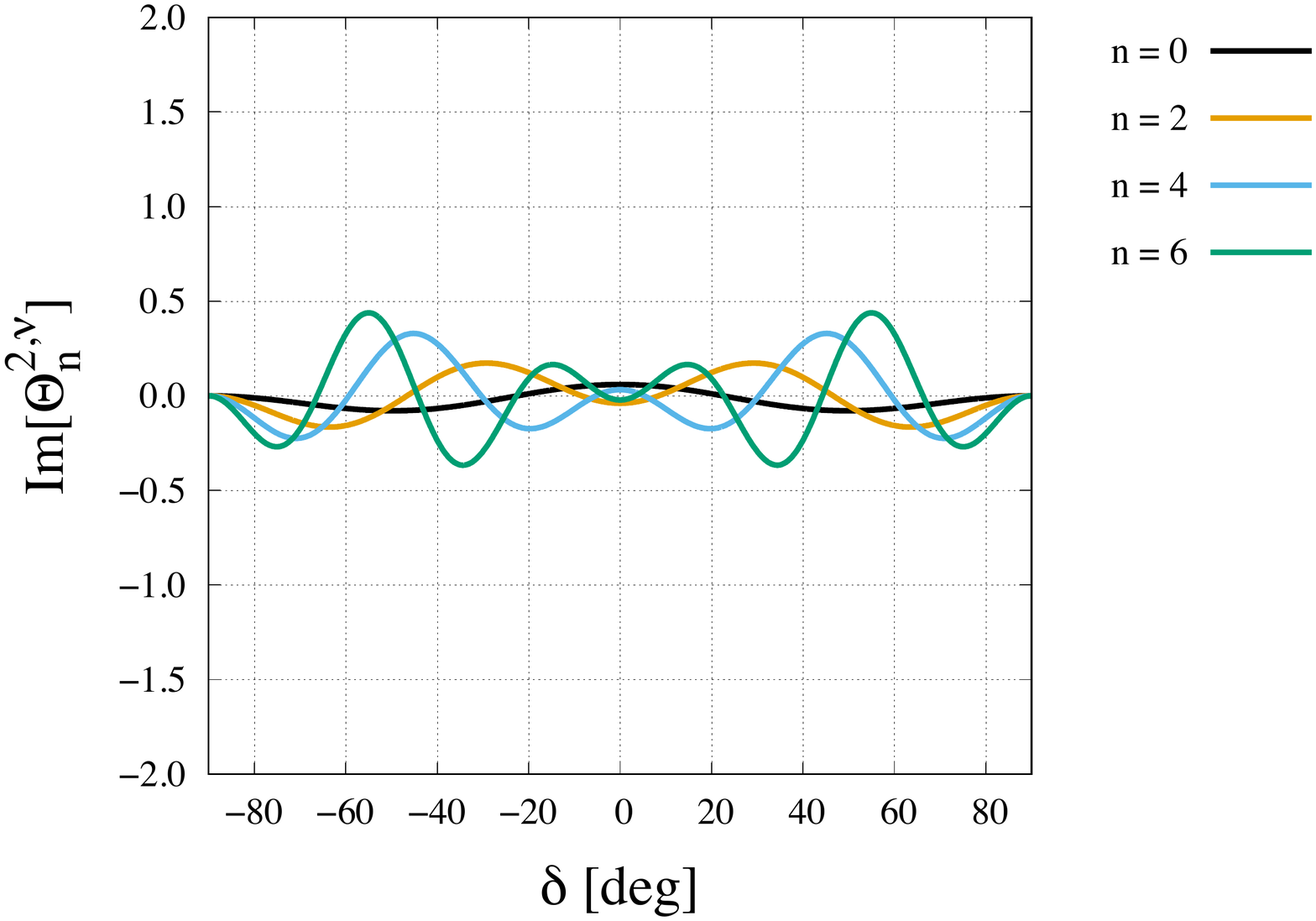} 
  \includegraphics[width=0.21\textwidth,trim = 3.2cm 3cm 6.5cm 2cm,clip]{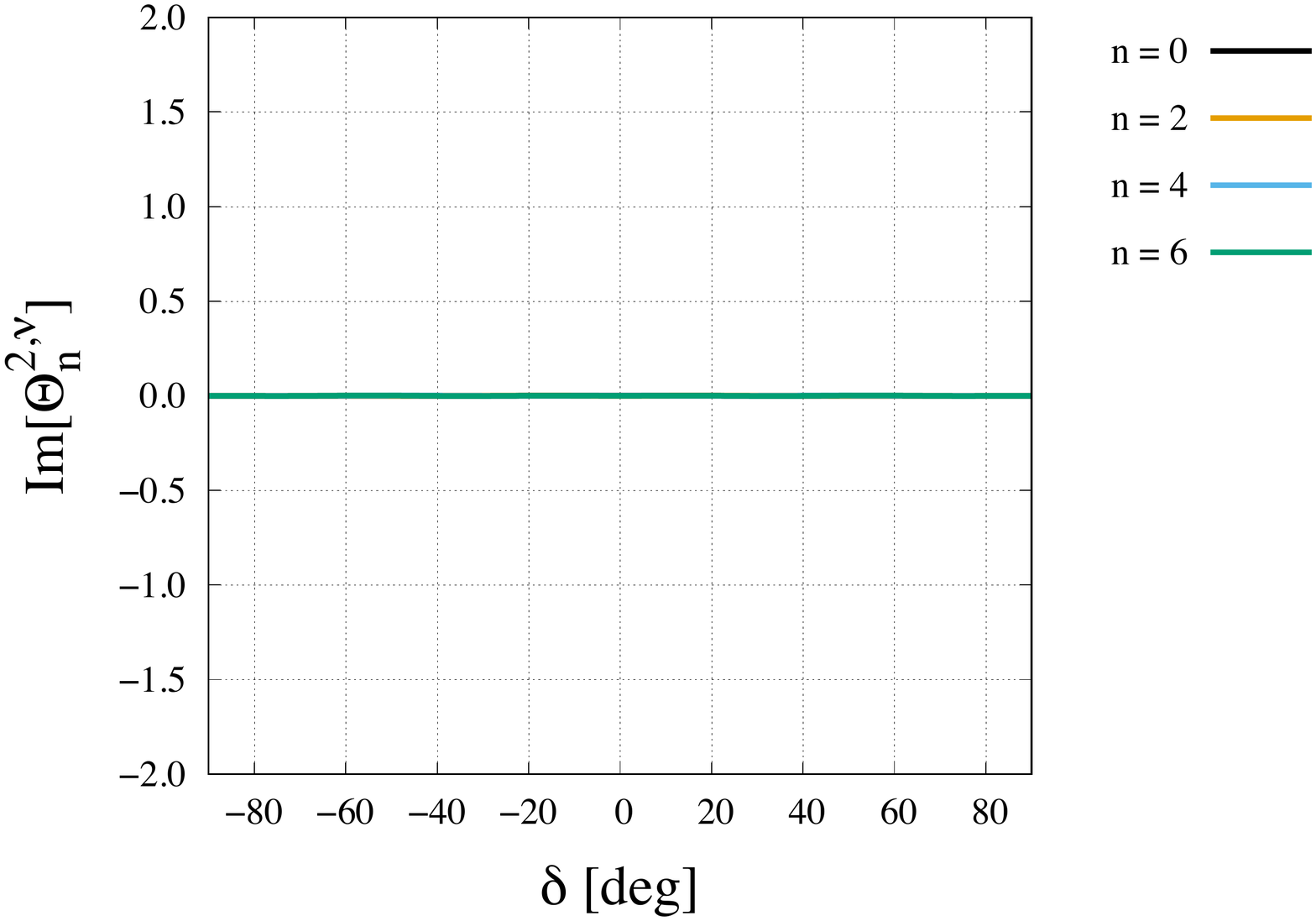} 
     \hspace{0.08\textwidth} \mbox{} \\
     \hspace{0.10\textwidth}
 \includegraphics[height=0.3cm]{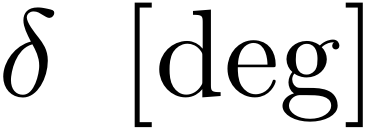} \hspace{3.0cm}
  \includegraphics[height=0.3cm]{auclair-desrotour_fig2o.pdf} \hspace{3.0cm}
  \includegraphics[height=0.3cm]{auclair-desrotour_fig2o.pdf} \hspace{3.0cm}
  \includegraphics[height=0.3cm]{auclair-desrotour_fig2o.pdf} \hspace{3.0cm}
\caption{\label{fig:Hough_functions} Real (top) and imaginary (bottom) parts of even Hough functions ($ n $ even) associated with a quadrupolar tidal perturbation ($ m = 2 $) and defined by $ \tilde{\nu} = \nu / \left( 1 - i \gamma \right) $ with $ \nu = 1 $ (ordinary/gravity modes) and various positive values of $ \gamma = \sigma_{\rm R} / \sigma$. From left to right, $\gamma = 0$ (non-frictional regime), $ \gamma = 0.1 $ (weakly frictional regime), $ \gamma = 1 $ and $ \gamma = 10 $ (strongly frictional regime). Hough functions are plotted as functions of the latitude $ \delta $ (degrees).}
\end{figure*} 

As mentioned above, the traditional approximation allows us to separate the coordinates $ r $ and $ \theta $ in the Fourier coefficients of \smc{Eq.~(\ref{Fourier})}. Therefore, we write these functions as

\begin{equation}
\begin{array}{ll}
  \displaystyle \xi_r^{m,\sigma } = \sum_n \xi_{r;n}^{m,\sigma} \left( r \right) \Theta_n^{m,\tilde{\nu}} \left( \theta \right), &
  \displaystyle V_r^{m,\sigma } = \sum_n V_{r;n}^{m,\sigma} \left( r \right) \Theta_n^{m,\tilde{\nu}} \left( \theta \right), \\[0.3cm] 
  \displaystyle \xi_\theta^{m,\sigma } = \sum_n \xi_{\theta;n}^{m,\sigma} \left( r \right) \Theta_{\theta ; n}^{m,\tilde{\nu}} \left( \theta \right), &
  \displaystyle V_\theta^{m,\sigma } = \sum_n V_{\theta;n}^{m,\sigma} \left( r \right) \Theta_{\theta ; n}^{m,\tilde{\nu}} \left( \theta \right), \\[0.3cm] 
  \displaystyle \xi_\varphi^{m,\sigma } = \sum_n \xi_{\varphi;n}^{m,\sigma} \left( r \right) \Theta_{\varphi ; n}^{m,\tilde{\nu}} \left( \theta \right), &
  \displaystyle V_\varphi^{m,\sigma } = \sum_n V_{\varphi;n}^{m,\sigma} \left( r \right) \Theta_{\varphi ; n}^{m,\tilde{\nu}} \left( \theta \right), \\[0.3cm]
  \displaystyle \delta p^{m,\sigma } = \sum_n \delta p_n^{m,\sigma} \left( r \right) \Theta_n^{m,\tilde{\nu}} \left( \theta \right), &
  \displaystyle \delta \rho^{m,\sigma } = \sum_n \delta \rho_n^{m,\sigma} \left( r \right) \Theta_n^{m,\tilde{\nu}} \left( \theta \right), \\[0.3cm]
  \displaystyle U^{m,\sigma } = \sum_n U_n^{m,\sigma} \left( r \right) \Theta_n^{m,\tilde{\nu}} \left( \theta \right), &  
\end{array}
\label{decompo_Hough}
\end{equation}

\noindent where $ n $ designates the latitudinal wavenumber of a component ($ n \in \mathbb{N} $ if $ \left| \nu \right| \leq 1 $ ; $ n \in \mathbb{Z} $ if $ \left| \nu \right| > 1 $), the $ \Theta_n^{m,\tilde{\nu}} $ are the so-called Hough functions \citep[][]{Hough1898}, defined on the interval $\theta \in \left[ 0 , \pi \right] $, and the $ \Theta_{\theta ; n}^{m,\tilde{\nu}} \left( \theta \right) = \mathcal{L}_\theta^{m,\tilde{\nu}} \left[ \Theta_n^{m,\tilde{\nu}} \left( \theta \right) \right] $ and $ \Theta_{\varphi ; n}^{m,\tilde{\nu}} \left( \theta \right) = \mathcal{L}_\varphi^{m,\tilde{\nu}} \left[ \Theta_n^{m,\tilde{\nu}} \left( \theta \right) \right] $ stand for the latitudinal functions associated with the latitudinal and longitudinal \smc{velocities and} displacements respectively. Hough functions are the solutions of the eigenfunctions-eigenvalues problem defined by the Laplace's tidal equation \citep[][]{Laplace1798},

\begin{equation}
\mathcal{L}^{m,\tilde{\nu}} \Theta = - \Lambda \Theta,
\end{equation}

\noindent where $ \Theta $ is a function, $ \Lambda \in \mathbb{C} $ and $ \mathcal{L}^{m,\tilde{\nu}} $ the operator

\begin{equation}
\begin{array}{rl}
\displaystyle \mathcal{L}^{m,\tilde{\nu}} = & \displaystyle \frac{1}{\sin \theta} \frac{d}{d \theta} \left( \frac{\sin \theta}{1 - \tilde{\nu}^2 \cos^2 \theta} \frac{d}{d \theta} \right) \\[0.3cm]
 & \displaystyle - \frac{1}{1 - \tilde{\nu}^2 \cos^2 \theta} \left( m \tilde{\nu} \frac{1 + \tilde{\nu}^2 \cos^2 \theta}{1- \tilde{\nu}^2 \cos^2 \theta} + \frac{m^2}{\sin^2 \theta} \right).
\end{array}
\end{equation}

Hence, for a given $ n $, $ \Theta_n^{m,\tilde{\nu}} $ is such that $ \mathcal{L}^{m,\tilde{\nu}} \Theta_n^{m,\tilde{\nu}} = -  \Lambda_n^{m,\tilde{\nu}} \Theta_n^{m,\tilde{\nu}} $, the parameter $ \Lambda_n^{m,\tilde{\nu}} $ being the associated eigenvalue. We can note here that the only difference with the usual case where friction is not taken into account \citep[e.g.][]{LS1997} is the nature of $ \tilde{\nu} $ which is a complex number and not a real one (see Eq.~(\ref{nu_tilde})). It follows that the $ \Theta_n^{m,\tilde{\nu}} $ and $ \Lambda_n^{m,\tilde{\nu}} $ are complex functions and parameters in the general case \citep[e.g.][]{Volland1974a,Volland1974b}. Looking at the expression of $ \tilde{\nu} $, one can identify two dissipation regimes:
\begin{itemize}
  \item[$\bullet$] a \emph{weakly frictional regime} ($ \left| \sigma \right| \gg  \sigma_{\rm R} $), where Hough functions and eigenvalues are real (they are denoted $ \Theta_n^{m,\nu} $ and $ \Lambda_n^{m,\nu} $). This regime corresponds to the case treated usually in the theory of tides \smc{\citep[][]{CL1970,LS1997,ADLPM2014}}. Tidal modes can be divided into two families: the so-called gravity modes ($ n \geq 0 $), which are defined both in the regime of super-inertial waves ($ \left| \nu \right| \leq 1 $) and in the regime of sub-inertial waves ($\left| \nu \right| > 1$), and the inertial modes ($n < 0$), defined only in the regime of sub-inertial waves \citep[see][]{LS1997}. The Hough functions associated with gravity modes degenerate in the associated Legendre polynomials\footnote{The $P_l^m$ represent here the normalized associated Legendre polynomials, expressed as \smc{$ P_l^m   = \left[ \left( - 1 \right)^m / \left( 2^l l ! \right) \right] \left( 1 - x^2 \right)^{m/2} \left( d^{l+m} / dx^{l+m} \right) \left( x^2 - 1 \right)^l $} \citep[][]{AS1972}.} $ P_l^m $ (with $ l = m + n $) when $ \nu \rightarrow 0 $.
   \item[$\bullet$] a \emph{strongly frictional regime} ($ \left| \sigma \right| \lesssim \sigma_{\rm R}  $), where friction affects significantly the horizontal structure of tidal waves. In this regime, the hierarchy of eigenvalues can change, as demonstrated by \cite{Volland1974a} who notes that the corresponding critical points appears for $ \left| \sigma \right| \sim \sigma_{\rm R} $. For $ \sigma_{\rm R} \gtrsim \left| \Omega \right| $, inertial modes converge towards gravity modes. Gravity and Rossby Hough functions merge asymptotically \smc{when} $ \sigma_{\rm R} \rightarrow + \infty $, and converge to the associated Legendre polynomials. Thus, a strong friction \smc{makes} the Coriolis effects negligible, as if the planet were not rotating. \smc{In this asymptotic regime, the angular lag between the tidal bulge and the direction of the perturber is given by the imaginary part of the vertical profiles of perturbed quantities.} The behaviour of tidal modes in the dissipative regime is discussed thoroughly by \cite{VM1972b} and \cite{Volland1974a}.
 \end{itemize}
 
 Hough functions associated with symmetric modes of degree $m = 2 $ are plotted on Fig.~\ref{fig:Hough_functions} for various values of the ratio $ \sigma_{\rm R} / \sigma $. The equations describing the vertical structure are obtained by substituting the expansions given by Eqs.~(\ref{Fourier}) and (\ref{decompo_Hough}) in Eqs.~(\ref{eq1}-\ref{eq3}). In order to lighten expressions, the superscripts $ \left( m , \sigma \right) $ will be omitted up to the end of Section~\ref{subsec:dynamics_equations}, where no confusion will arise. Let us introduce the notation $ y_n = \delta p_n / \rho_0 $ and assume the ocean to be at the hydrostatic equilibrium ($dp_0 / dx = - H g \rho_0$), \jlc{where $x$ is the reduced altitude introduced in Eq.~(\ref{N2})}. After some manipulations, one gets
 
\begin{equation}
\begin{array}{rl}
\displaystyle
  \frac{d \vect{Y}}{dx} =
\begin{bmatrix}
  A_1 \left( x \right) & B_1 \left( x \right) \\
  A_2 \left( x \right) & B_2  \left( x \right)
\end{bmatrix}
\vect{Y} +
\begin{bmatrix}
  C_1\left( x \right)  \\
  C_2 \left( x \right) 
\end{bmatrix},
& \mbox{with} \ 
\displaystyle \vect{Y} = 
\begin{bmatrix}
y_n   \\
 r^2 \xi_{r;n}
 \displaystyle  
\end{bmatrix}
\end{array}
\! \! 
\label{sv_1}
\end{equation}

\noindent and the coefficients

\begin{equation}
\begin{array}{ll}
    \displaystyle A_1 = \frac{N^2 H}{g}, &
    \displaystyle A_2 = H \left( \frac{\Lambda_n^{m,\tilde{\nu}}}{\sigma \tilde{\sigma}} - \frac{r^2}{c_s^2} \right), \\[0.3cm]
    \displaystyle B_1 = \frac{H}{r^2} \left( \sigma \tilde{\sigma} - N^2 \right), &
    \displaystyle B_2 = \frac{g H}{c_s^2}, \\[0.3cm]
    \displaystyle C_1 =  \frac{d U_n}{dx}, &
    \displaystyle C_2 = - \frac{H \Lambda_n^{m,\tilde{\nu}}}{\sigma \tilde{\sigma}} U_n.
\end{array}
\end{equation}

\noindent This system is \smc{then} written as a single \smc{second-order ordinary differential} equation,

\begin{equation}
\frac{d^2 y_n}{dx^2} + A \left( x \right) \frac{d y_n}{dx}  + B \left( x \right) y_n = C \left( x \right),
\label{struc_vert1}
\end{equation}

\noindent with the coefficients


\begin{align}
 \displaystyle A \left( x \right) = & \, \displaystyle \frac{d \ln \rho_0}{dx} - K_\circ , \\[0.3cm]
\displaystyle B \left( x \right)  = & \,  \displaystyle \frac{H^2}{r^2} \Lambda_n^{m,\tilde{\nu}} \left( \frac{N^2}{\sigma \tilde{\sigma}} - 1 \right) \left( 1 - \varepsilon_{s ; n} \right)   \\
  & \displaystyle  - \left( \frac{d}{dx} - \frac{gH}{c_s^2} - K_\circ \right) \left( \frac{N^2 H}{g} \right), \nonumber \\[0.3cm]
\displaystyle C \left( x \right)  = & \,  \displaystyle  \frac{H^2}{r^2} \Lambda_n^{m,\tilde{\nu}} \left( \frac{N^2}{\sigma \tilde{\sigma}} - 1 \right) U_n - \left( \frac{g H}{c_s^2} + K_\circ \right) \frac{d U_n}{dx} + \frac{d^2 U_n}{dx^2} .
\label{ABC}
\end{align}

\noindent Here, we have introduced the frequency-dependent\rec{,} dimensionless acoustic and sphericity parameters

\begin{equation}
\begin{array}{rcl}
  \displaystyle \varepsilon_{s;n} = \frac{r^2 \sigma \tilde{\sigma}}{\Lambda_n^{m,\tilde{\nu}} c_s^2} & \mbox{and} & \displaystyle K_\circ = \frac{r^2}{N^2 - \sigma \tilde{\sigma}} \frac{d}{dx} \left(\frac{N^2 - \sigma \tilde{\sigma} }{r^2} \right). 
\end{array}
\label{eps_Ks}
\end{equation}

\begin{figure}[b]
 \centering
  \includegraphics[width=0.48\textwidth,clip]{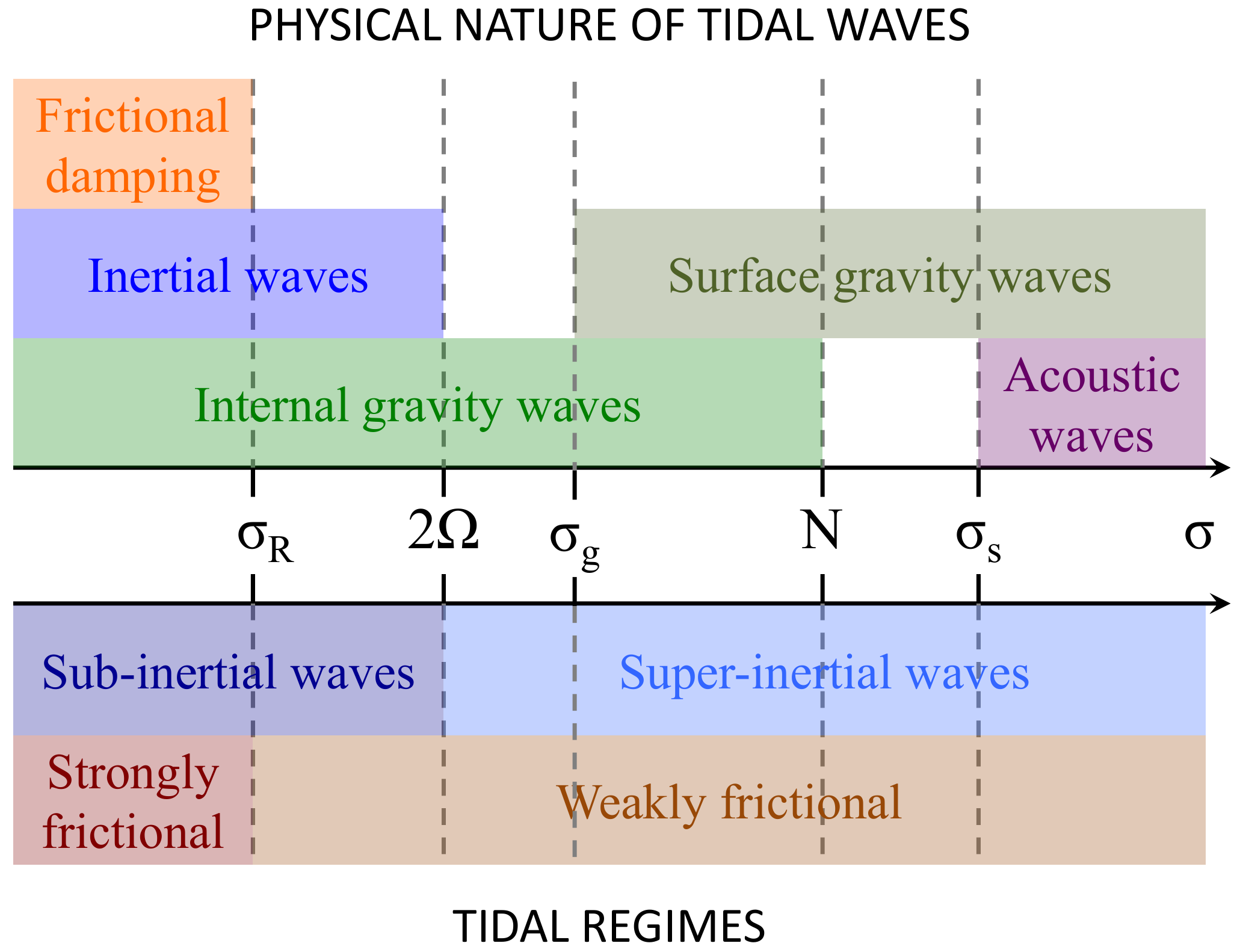} 
\caption{\label{fig:spectre_regimes} Frequency spectrum of waves likely to be excited by the tidal perturbation and of the corresponding tidal regimes. The characteristic frequencies of the system are, from left to right, the drag frequency $ \sigma_{\rm R} $, the inertia frequency $ 2 \Omega $, the surface gravity waves cutoff frequency $ \sigma_{\rm g} = \sqrt{gH} / R $, the Brunt-Väisälä frequency $ N $, and the acoustic cutoff frequency $ \sigma_{\rm s} = c_s / r $. Their hierarchy can vary as a function of the physical and dynamical properties of the planet. In the case of the Earth's ocean, $ \sigma_{\rm R} \approx 10^{-5} \ {\rm s^{-1}} $ \citep{Webb1980}, $ 2 \Omega = 1.46 \times 10^{-4} \ {\rm s^{-1}} $, $ \sigma_{\rm g} = 3.1 \times 10^{-5} \ {\rm s^{-1}} $, $ N \approx 2.2 \times 10^{-3} \ {\rm s^{-1}} $ \citep{GZ2008}, and $ \sigma_{\rm s} = 6.3 \times 10^{-3} \ {\rm s^{-1}} $. }  
\end{figure} 

The acoustic parameter can be written $ \varepsilon_{s;n} = \sigma \tilde{\sigma} / \sigma_{s;n}^2 $, where $ \sigma_{s;n} = \left( \Lambda_n^{m,\tilde{\nu}} \right)^{1/2} c_s /r $ is the cutoff frequency of acoustic waves (also called the \emph{Lamb frequency}) for the \smc{degree-$n$ mode} when $ \Lambda_n^{m,\tilde{\nu}} \in \mathbb{R}^+ $. The regime of acoustic waves thus corresponds to $ \left| \varepsilon_{s;n} \right| \gtrsim 1 $. Typically, for an Earth-like planet of radius $ R_\Earth = 6371 $ km and angular velocity $ \Omega_\Earth = 7.25 \times 10^{-5} {\rm s^{-1}} $ (\href{https://nssdc.gsfc.nasa.gov/planetary/factsheet/earthfact.html}{NASA fact sheets}), with a water oceanic layer characterized by $ c_s \approx 1.545 \ {\rm km.s^{-1}} $ \citep[][]{GZ2008}, the \smc{gravest acoustic modes} of the quadrupolar semidiurnal tide ($ m =2 $, $ n = 0 $, $ \Lambda_0^{2,\tilde{\nu}} \sim 11 $) appear for $ \Omega \gtrsim 5.5 \ \Omega_\Earth $. The other parameter, $ K_\circ $, corresponds to the effect of the spherical geometry of the layer on the structure of tidal waves. It is usually ignored in \rec{two-}dimensional modelings where $ r = R $. We will see in the next section that $ K_\circ \, \propto \, H/R $ when $ H/R \ll 1 $. Eventually, using the change of variables $ y_n = \Phi_n^{m,\sigma} \Psi_n^{m,\sigma} $, where $ \Phi_n^{m,\sigma} $ is the radial function 


\begin{equation}
\Phi_n^{m,\sigma} \left( x \right) = {\rm exp} \left[ - \frac{1}{2} \int_{0}^x A \left( x' \right) \dd x' \right],
\label{Phin}
\end{equation}

\noindent we write the vertical structure equation (Eq.~\ref{struc_vert1}) as the Schrödinger-like equation

\begin{equation}
\frac{d^2 \Psi_n^{m,\sigma}}{dr^2} + \hat{k}_n^2 \Psi_n^{m,\sigma} =  \left( \Phi_n^{m,\sigma} \right)^{-1} C ,
\label{Vstruct_general}
\end{equation}

\noindent in which $ \hat{k}_n $ represents the vertical wavenumber of the mode, defined as

\begin{equation}
\hat{k}_n^2 =  B - \frac{1}{2} \left( \frac{d A}{dx} + \frac{A^2}{2} \right). 
\label{kn2_general}
\end{equation}

The vertical profiles of \smc{the} other quantities are deduced from $ \Psi_n^{m,\sigma} $ straightforwardly \smc{thanks to polarization relations}. We get, for the velocity field

\begin{equation}
V_{r ; n}^{m,\sigma} = - \frac{i \sigma}{H \left( N^2 - \sigma \tilde{\sigma} \right)} \left[ \Phi_n \left( \frac{d \Psi_n}{dx} + \mathcal{A}_n \Psi_n \right) - \frac{d U_n}{dx} \right],
\label{eqpolar_Vr}
\end{equation}

\begin{equation}
V_{\theta ; n}^{m,\sigma} = \frac{i}{\tilde{\sigma} r} \left( \Phi_n \Psi_n - U_n \right),
\end{equation}

\begin{equation}
V_{\varphi ; n}^{m,\sigma} = - \frac{1}{\tilde{\sigma} r} \left( \Phi_n \Psi_n - U_n \right) ,
\end{equation}

\noindent for the displacement

\begin{equation}
\xi_{r ; n}^{m,\sigma} = - \frac{1}{H \left( N^2 - \sigma \tilde{\sigma} \right)} \left[ \Phi_n \left( \frac{d \Psi_n}{dx} + \mathcal{A}_n \Psi_n \right) - \frac{d U_n}{dx} \right],
\label{eqpolar_xir}
\end{equation}

\begin{equation}
\xi_{\theta ; n}^{m,\sigma} = \frac{1}{\sigma \tilde{\sigma} r} \left( \Phi_n \Psi_n - U_n \right),
\end{equation}

\begin{equation}
\xi_{\varphi ; n}^{m,\sigma} = \frac{i}{\sigma \tilde{\sigma} r} \left( \Phi_n \Psi_n - U_n \right),
\end{equation}

\noindent and for scalar quantities

\begin{equation}
\delta p_n^{m,\sigma} =  \rho_0 \Phi_n \Psi_n,
\end{equation}

\begin{equation}
\delta \rho_n^{m,\sigma} = - \frac{\rho_0}{gH} \frac{N^2}{N^2 - \sigma \tilde{\sigma}} \left[ \Phi_n \left( \frac{d \Psi_n}{dx} + \mathcal{B}_n \Psi_n \right) - \frac{d U_n}{dx} \right],
\label{eqpolar_rho}
\end{equation}

\noindent with the frequency-dependent factors

\begin{align}
\label{An}
 & \displaystyle \mathcal{A}_n \left( x \right)  =  \frac{1}{2} \left[  \frac{gH}{c_s^2} - \frac{N^2 H}{g} + K_\circ  \right] , \\[0.3cm]
 \label{Bn}
 & \displaystyle \mathcal{B}_n \left( x \right)  = \frac{1}{2} \left[ \frac{Hg}{c_s^2} \left( 2 \frac{\sigma \tilde{\sigma}}{N^2} - 1 \right) - \frac{H N^2}{g} + K_\circ \right]. 
\end{align}

Through equations describing the tidal dynamics, we identify the families of waves involved in the tidal response. First, because of rotation, inertial waves are generated in the frequency range $ \left| \sigma \right| < 2 \Omega $. Then, we identify surface gravity waves, which are restored by gravity and are characterized by the cutoff frequency $ \sigma_{\rm g} = \sqrt{gH} / r $. If the ocean is stably-stratified ($ N > 0 $), internal gravity waves can propagate in the frequency range $ \left| \sigma \right| < N $. These waves are restored by the Archimedean force. Finally, in the high-frequency range delimited by the acoustic cutoff frequency $ \sigma_{\rm s} = c_{\rm s} / r $, the tidal response is partly composed of horizontally propagating \smc{acoustic} Lamb modes, restored by compressibility. The possible regimes of oceanic tides are determined by the hierarchy of the characteristic frequencies of the system, namely $ \sigma $, \smc{$ \sigma_{\rm R} $}, $ 2 \Omega $, $ N $, $ \sigma_{\rm g} $ and $ \sigma_{\rm s} $. The frequency spectrum of waves composing the oceanic tidal response is \smc{summarized} by Fig.~\ref{fig:spectre_regimes}.

\subsection{Tidal potential, Love numbers and tidal torque}

The variation of mass distribution due to the tidal distortion modifies the self-gravitational potential of the planet. The tidal potential of the excitation is usually expanded in Fourier series and spherical harmonics \citep[see for instance \smc{the} Kaula's multipolar expansion;][]{Kaula1962}. It is thus convenient to write the fluctuations of the self-gravitational potential, denoted $ \mathcal{U} $, in the same form,

\begin{equation}
\mathcal{U} = \sum_{\sigma,l,m} \mathcal{U}_l^{m,\sigma} \left( x \right) P_l^m \left( \cos \theta \right) \ed^{i \left( \sigma t + m \varphi \right)}.
\end{equation}

\noindent The radial profiles of the $ \left( m , \sigma \right) $-components are themselves written
 
\begin{equation}
\mathcal{U}_l^{m,\sigma} = \mathcal{U}_{\xi ; l}^{m,\sigma} + \mathcal{U}_{\rho ; l}^{m,\sigma},
\end{equation} 

\noindent where we have introduced the contribution of the surface displacement $ \mathcal{U}_{\xi ; l}^{m,\sigma} $ and of the internal density variations $ \mathcal{U}_{\rho ; l}^{m,\sigma} $. In the following, we will use the subscripts $ \xi $ and $ \rho $ to make the distinction between the two contributions in a systematic way. \smc{At the planet surface} ($ x = 1 $), the two components of the potential are expressed as 

\begin{align}
& \mathcal{U}_{\xi ; l}^{m,\sigma} \left( 1 \right) =  \frac{4 \pi \mathscr{G} R}{2 l + 1}  \rho_0 \left( 1 \right)  \xi_{r ; l}^{m,\sigma} \left( 1 \right), \\
& \mathcal{U}_{\rho ; l}^{m,\sigma} \left( 1 \right)  =  \frac{4 \pi \mathscr{G} R}{2 l + 1} H \left[ \int_{0}^1 \left( 1 + \frac{H}{R} x' \right)^{l+2} \delta \rho_l^{m,\sigma}  \dd x' \right],
\end{align}

\noindent with $ \mathscr{G} $ representing the universal gravity constant. Expanded in Hough functions, they write

\begin{equation}
\mathcal{U}_{\xi ; l}^{m,\sigma} \left( 1  \right) = \frac{4 \pi \mathscr{G} R}{2 l + 1}  \rho_0 \left( 1 \right) \sum_{n } \sum_{k \geq  m } C_{l,n,k}^{m,\tilde{\nu}}  \xi_{r ; n,k}^{m,\sigma} \left( 1 \right) ,
\label{Uxi}
\end{equation}

\begin{equation}
\mathcal{U}_{\rho ; n}^{m,\sigma} \left( 1 \right) = \frac{4 \pi \mathscr{G} R}{2 l + 1} H \sum_{n } \sum_{k \geq m} C_{l,n,k}^{m,\tilde{\nu}} \int_{0}^1 \left( 1 + \frac{H}{R} x' \right)^{l+2} \delta \rho_{n,k}^{m,\sigma} \dd x'  .
\label{Urho}
\end{equation}

%

In these expressions, the complex weighting coefficients $ C_{l,n,k}^{m,\tilde{\nu}} $ are expressed as 

\begin{equation}
C_{l,n,k}^{m,\tilde{\nu}} = B_{k,n}^{m,\tilde{\nu}} A_{n,l}^{m,\tilde{\nu}},
\end{equation}


\noindent where the $ A_{n,l}^{m,\tilde{\nu}} $ and $ B_{k,n}^{m,\tilde{\nu}} $ designate the mutual projection coefficients of the Hough functions on normalized associated Legendre polynomials, defined respectively by the expansions

\begin{equation}
 \Theta_n^{m,\tilde{\nu}} \left( \theta \right) = \sum_{l \geq m}  A_{n,l}^{m,\tilde{\nu}} P_l^m \left( \cos \theta \right),
\end{equation}

\begin{equation}
P_k^m \left( \cos \theta \right) = \sum_n B_{k,n}^{m,\tilde{\nu}} \Theta_n^{m,\tilde{\nu}} \left( \theta \right).
\end{equation}

The $ C_{l,n,k}^{m,\tilde{\nu}} $ coefficients quantify the coupling induced by the Coriolis effects in the dynamics of the tidal response. They are real in the non-frictional case ($ \sigma_{\rm R} = 0 $ and $ \tilde{\nu} = \nu $), where $ A_{n,l}^{m,\nu} = \langle P_l^m , \Theta_n^{m,\nu} \rangle $ and $ B_{k,n}^{m,\nu} = \langle \Theta_n^{m,\nu} , P_k^m \rangle = A_{n,k}^{m,\nu} $, the notation $ \langle \cdot , \cdot \rangle $ standing for the scalar product defined for any $ P_l^m $ and $ \Theta_n^{m,\nu} $ as

\begin{equation}
 \langle P_l^m , \Theta_n^{m,\nu} \rangle = \int_{0}^{\pi} P_l^m \left( \cos \theta \right) \Theta_n^{m,\nu} \left( \theta \right) \sin \theta \, \dd \theta.
\end{equation}

\noindent  In the case of a non-rotating planet ($ \Omega = 0 $ and $\nu = 0$), there is no coupling. Therefore, $ C_{l,n,k}^{m,0} =  1 $ if $ l = n + m = k $ and $ C_{l,n,k}^{m,0} =  0 $ else. The notations $ \xi_{r ; n,k}^{m,\sigma} = \xi_{r ; n}^{m,\sigma} \left( U_k^{m,\sigma}  \right) $ and $ \delta \rho_{n,k}^{m,\sigma} = \delta \rho_n^{m,\sigma} \left( U_k^{m,\sigma} \right) $ in Eqs.~(\ref{Uxi}) and (\ref{Urho}) stand for the vertical displacement and density variations due to the $ P_k^m $-component of the forcing projected on the $ \left( n , m, \tilde{\nu} \right) $-Hough function. The expressions of the components of the tidal potential (Eqs.~\ref{Uxi} and \ref{Urho}) allow us to compute \smc{complex} Love numbers, which are commonly used to quantify tidal dissipation in celestial bodies \smc{\citep[][]{Tobie2005,RMZL2012,Ogilvie2014}}. Love numbers are defined as the ratio between the gravitational potential due to the distortion and the excitating tidal gravitational potential taken at the external surface of the body ($r = R$). Thus, oceanic Love numbers associated with the triplet $ \left( l , m , \sigma \right) $, denoted $ k_l^{m,\sigma} $, are defined as  

\begin{equation}
k_l^{m,\sigma} = \left. \frac{\mathcal{U}_l^{m,\sigma}}{U_l^{m,\sigma}} \right|_{x=1} . 
\label{klm}
\end{equation} 

\noindent \smc{Here, $\Re \left\{ k_l^{m,\sigma} \right\}$ and  $\Im \left\{ k_l^{m,\sigma} \right\} $ stand for the tidal response of the ocean to the $\left( l, m \right)$-component of the excitating tidal potential.} The $ U_l^m $ are provided by the Kaula's multipolar expansion of the tidal gravitational potential and are expressed as functions of the Keplerian elements of the planet-perturber system \citep[][]{Kaula1966,MLP2009}. 

Because of internal dissipation, the variation of mass distribution due to the tidal perturbation generates a tidal torque. The torque exerted by the perturber on the oceanic layer with respect to the spin axis of the planet, denoted $ \mathcal{T} $, is given by

\begin{equation}
\mathcal{T} = \sum_{m,\sigma} \left( \mathcal{T}_\xi^{m,\sigma} + \mathcal{T}_\rho^{m,\sigma} \right).
\end{equation}

\noindent The components $ \mathcal{T}_\xi^{m,\sigma} $ and $ \mathcal{T}_\rho^{m,\sigma} $ of the $ \left( m , \sigma \right) $-mode are defined as 

\begin{align}
& \mathcal{T}_\xi^{m,\sigma} = \Re \left\{ \frac{1}{2} \rho_{\rm s} \int_{\partial \mathscr{V}_0} \left. \frac{\partial U^{m,\sigma}}{\partial \varphi} \right|_{r=R} \left[  \xi_{r}^{m,\sigma} \left( 1 \right) \right]^* \dd S  \right\}, \\[0.3cm]
& \mathcal{T}_\rho^{m,\sigma} = \Re \left\{ \frac{1}{2} \int_{\mathscr{V}_0} \frac{\partial U^{m,\sigma}}{\partial \varphi} \left( \delta \rho^{m,\sigma} \right)^* \dd \mathscr{V}  \right\},
\end{align}

\noindent where $ ^* $ stands for the conjugate of a complex number and $ \Re $ its real part ($ \Im $ will be used for the imaginary part), $ \rho_{\rm s} = \rho_0 \left( 1 \right) $ denotes the density at the surface ocean,  $ \mathscr{V}_0 $ the spatial domain filled by the oceanic shell at rest, $ \partial \mathscr{V}_0 $ its upper boundary, and $ \dd S $ and $ \dd \mathscr{V} $ infinitesimal surface and volume parcels respectively. Similarly as the tidal potential due to the distortion, these two components can be expanded on Hough functions and associated Legendre polynomials. One thus obtains, for $ l \geq m $ and $ k \geq m $,

\begin{equation}
\mathcal{T}_\xi^{m,\sigma} = - m \pi R^2 \rho_{\rm s} \sum_{l,n,k} \Im \left\{   \left( C_{l,n,k}^{m,\tilde{\nu}} \right)^* U_l^{\sigma,m} \left( 1 \right) \left[ \xi_{r ; n,k}^{m,\sigma} \left( 1 \right)  \right]^*  \right\},
\label{torque_xi}
\end{equation}

\begin{equation}
\mathcal{T}_\rho^{m,\sigma} = - m \pi R^2 H\sum_{l,n,k} \Im \left\{ \left( C_{l,n,k}^{m,\tilde{\nu}} \right)^* \int_{0}^1 U_l^{m,\sigma} \left[ \delta \rho_{n,k}^{m,\sigma}  \right]^* \dd x  \right\}.
\label{torque_rho}
\end{equation}

\section{Tides in a thin stably-stratified ocean}
\label{sec:uniform_ocean}

In \rec{some of} observed cases, the oceanic layer of terrestrial planets and satellites is thin compared to the radius of the body \citep[for instance, the Earth's ocean has an aspect ratio of $ 6 \times 10^{-4} $\smc{;}][]{ES2010}. Therefore, in this section, we apply \smc{our general} modeling to the simplified case of an oceanic layer of \smc{small} depth $ H \ll R $, and compute an analytic solution of the oceanic tidal response, Love numbers and tidal torque.

\subsection{Tidal waves dynamics}
\label{ssec:tidal_waves}

The \rec{thin shell} hypothesis allows us to simplify the physical setup. Given that the relative difference of gravity between the lower and the upper boundary is proportional to the ratio $ H /R $, $ g $ is supposed to be constant in the oceanic layer. Similarly, the radial variations of the tidal gravitational potential with $ x$ can be neglected ($ d U_n /dx \approx 0 $ and $ d^2U_n /dx^2 \approx 0 $). Furthermore, we assume \rec{for simplicity} uniform stratification and compressibility, i.e. that the Brunt-Väisälä frequency ($ N $) and the sound velocity ($c_s$) are constants. \rec{We note that the case of the Earth's ocean does not well follow this assumption since $ N $ decays over several orders of magnitude from the pycnocline (the upper region of the ocean where the density is changing most rapidly) where its typical values are about $ 0.01 \ {\rm s^{-1}} $, to the abyssal ocean, where they fall to $ 0.001 \ {\rm s^{-1}} $ \citep[e.g.][]{GZ2008}. Although a more realistic model would be more appropriate in this case, the uniform stratification appears as a useful first step to solve the vertical structure equation and derive explicit solutions controlling the tidal response of a stratified ocean.} As shown by Eq.~(\ref{N2}), \rec{setting $ N^2 $ to a constant} implies a density decreasing exponentially with altitude,

\begin{equation}
\rho_0 \left( x \right) = \rho_{\rm s} \ed^{\tau \left( 1 - x \right)},
\end{equation}

\noindent where $ \tau $ designates the decreasing rate given by 

\begin{equation}
\tau = H \left( \frac{N^2}{g} + \frac{g}{c_s^2} \right). 
\label{tau}
\end{equation}

\noindent The acoustic and sphericity terms (Eq.~\ref{eps_Ks}) are simplified into 

\begin{equation}
\begin{array}{lll}
  \displaystyle \varepsilon_{s ; n} \approx \frac{R^2 \sigma^2}{\Lambda_n^{m,\tilde{\nu}} c_s^2} & \mbox{and} & \displaystyle K_\circ \approx - 2 \frac{H}{R}.
\end{array}
\end{equation}

\noindent It follows that $ \Phi_n $ (see Eq.~\ref{eps_Ks}) writes 

\begin{equation}
\begin{array}{lll}
 \displaystyle \Phi_n \left( x \right) = \ed^{\delta x} & \mbox{with} & \displaystyle \delta = \frac{1}{2} \left( \tau + K_\circ \right). 
\end{array}
\end{equation}

Hence, we express the vertical structure equation (Eq.~\ref{Vstruct_general}) 

\begin{equation}
\frac{d^2 \Psi_n^{m,\sigma}}{dx^2} + \hat{k}_n^2 \Psi_n^{m,\sigma} =  \frac{H^2}{R^2} \Lambda_n^{m,\tilde{\nu}} \left( \frac{N^2}{\sigma \tilde{\sigma}} - 1 \right) U_n \ed^{- \delta x} .
\label{kn2_uni1}
\end{equation}

\noindent The vertical wavenumber to square $ \hat{k}_n^2 $, given by Eq.~(\ref{kn2_general}) in the general case, is now the constant 

\begin{equation}
\hat{k}_n^2 =  \frac{H^2}{R^2} \Lambda_n^{m,\tilde{\nu}} \left( \frac{N^2}{\sigma \tilde{\sigma}} - 1 \right) \left( 1 - \varepsilon_{s ; n} \right) + \left( \frac{g H}{c_s^2} + K_\circ \right) \frac{N^2 H}{g} - \delta^2.
\end{equation}

\noindent In this expression, the first term is the vertical wavenumber of internal gravity waves. It dominates when $ \left| \sigma \tilde{\sigma} \right| \ll N^2 $, i.e. for a stable stratification. Other terms correspond to acoustic and sphericity terms usually ignored in anelastic ($ c_{\rm s} = + \infty $) and \rec{thin-shell} ($ H/R \rightarrow 0 $) approximations. Here, we will only ignore sphericity terms, i.e. assume that $ K_\circ = 0 $. We will keep acoustic terms because they do not bring supplementary mathematical complexities in the analytic treatment and can, furthermore, be comparable to those associated to stratification. For instance, in the case of the Earth's ocean, density increases from $ 1022 \ {\rm kg.m^{-3}} $ at the surface to $ 1070 \ {\rm kg.m^{-3}} $ at $ 10 $ km depth \citep[][]{GZ2008}, which gives the mean gradient $ d \rho_0 / dz = - 0.0048 \ {\rm kg.m^{-2}} $. As $ \rho_0 g / c_s^2 \approx 0.0043 \ {\rm kg.m^{-2}} $, the two terms of \smc{Eq.~(\ref{N2})} are comparable and must be \rec{retained}. To solve the vertical structure equation, two boundary conditions are required. At $ x = 0$, we set the impenetrable rigid-wall condition $ \xi_r = 0 $. At the upper boundary, we apply the usual stress-free condition \citep[e.g.][]{Unno1989}, $ \delta p = g \rho_0 \xi_r $. It follows that

\begin{equation}
\begin{array}{ll}
\displaystyle \Psi_n \left( x \right) = & \! \! \displaystyle \frac{\Psi_n^{(0)}}{\mathcal{D}_n} \left\{ \mathcal{D}_n \ed^{- \delta x} \right. \\[0.3cm]
 & \! \! \displaystyle \left. +   \left( \mathcal{C}_n - \delta \right) \ed^{-\delta} \left[ \mathcal{A}_n \sin \left( \hat{k}_n x \right) - \hat{k}_n \cos \left( \hat{k}_n x \right) \right]  \right. \\[0.3cm]
 & \! \! \displaystyle \left. +  \left( \mathcal{A}_n - \delta \right)  \hat{k}_n  \cos \left( \hat{k}_n \left( 1 - x \right) \right)  \right. \\[0.3cm]
& \! \! \displaystyle \left. +  \left( \mathcal{A}_n - \delta \right) \mathcal{C}_n \sin \left( \hat{k}_n \left( 1 - x \right) \right)  \right\} \!  ,
\end{array}
\label{Psin}
\end{equation}


\noindent where $ \Psi_n^{(0)} $ is the constant given by 

\begin{equation}
\Psi_n^{(0)} =  \frac{H^2 \Lambda_n^{m,\tilde{\nu}}}{R^2 \left( \hat{k}_n^2 + \delta^2 \right)}  \left( \frac{N^2}{\sigma \tilde{\sigma}} - 1  \right) U_n,
\end{equation}

\noindent and $ \mathcal{C}_n $ and $ \mathcal{D}_n $ the dimensionless coefficients expressed as

\begin{eqnarray}
\mathcal{C}_n &=& \mathcal{A}_n + \frac{H}{g} \left( N^2 - \sigma \tilde{\sigma} \right), \\
\label{Dn}
\mathcal{D}_n &=& \hat{k}_n \left( \mathcal{C}_n - \mathcal{A}_n \right) \cos \left( \hat{k}_n \right) - \left( \mathcal{A}_n \mathcal{C}_n + \hat{k}_n^2 \right) \sin \left( \hat{k}_n \right).
\end{eqnarray}

\begin{figure}[t]
 \centering
  \includegraphics[width=0.4\textwidth,trim = 1.0cm 6.0cm 2.0cm 1.8cm,clip]{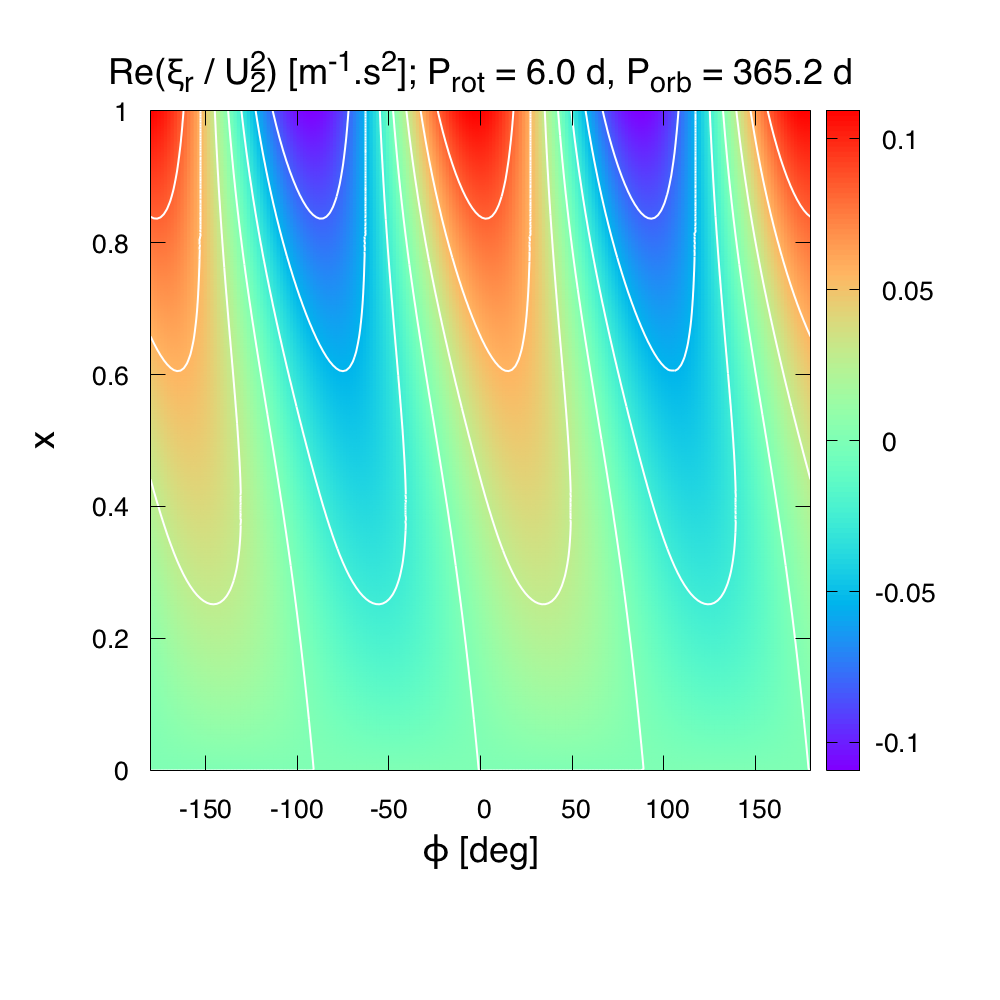} \\
  \includegraphics[width=0.05\textwidth,clip]{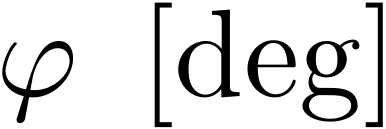} 
\caption{\label{fig:Rexir_equat_ex} Vertical displacement due to the quadrupolar semidiurnal tide ($\sigma = 2 \left( \Omega - n_{\rm orb} \right) $, $ n_{\rm orb} $ being the dynamical frequency) obtained using the analytic solution established \smc{for} the thin layer approximation (Eq.~\ref{Psin}). The displacement, normalized by the quadrupolar tidal potential, is plotted in the equatorial plane of the planet as a function of longitude (horizontal axis) and normalized altitude (vertical axis). The position $ \varphi = 0 $ corresponds to the sub-perturber point; the oceanic floor and surface are located at $ x = 0 $ and $ x = 1 $ respectively. For this computation, the rotation period of the planet $ P_{\rm spin} = 2 \pi / \Omega $ and orbital period of the perturber $ P_{\rm orb} = 2 \pi / n_{\rm orb} $ are set to $ P_{\rm spin} = 6.0 $~d and \padc{$ P_{\rm orb} = 365.25 $~d}. The used values of parameters are $ R = R_\Earth $, $ H = 100 $~km, $ g = 9.81 \ {\rm m.s^{-2}} $, $ N = 10^{-3} \ {\rm s^{-1}} $, $ c_{\rm s} = 1545 \ {\rm m.s^{-1}} $ and \padc{$ \sigma_{\rm R} = 10^{-5} \ {\rm s^{-1}} $}.  }  
\end{figure}

\subsection{Second order tidal Love number and tidal torque}

By substituting Eq.~(\ref{Psin}) in Eqs.~(\ref{eqpolar_xir}) and (\ref{eqpolar_rho}) we obtain the components of the variation of mass distribution intervening in the Love numbers and tidal torque (see Eqs.~\ref{torque_xi} and \ref{torque_rho}) as explicit functions of the internal structure parameters,

\begin{equation}
\begin{array}{rcl}
\displaystyle \xi_{r ;n,k}^{m,\sigma} \left( 1 \right) = H \mathcal{Q}_{\xi ; n}^{m,\sigma} U_k^{m,\sigma} & \mbox{and} & \displaystyle \int_0^1 \delta \rho_{n,k}^{m,\sigma} \left( x \right)  \dd x = \rho_{\rm s} \mathcal{Q}_{\rho ; n}^{m,\sigma} U_k^{m,\sigma},
\end{array}
\end{equation}

\noindent where the frequency-dependent parameters $ \mathcal{Q}_{\xi ; n}^{m,\sigma} $ and $ \mathcal{Q}_{\rho ; n}^{m,\sigma} $ are expressed as

\begin{align}
 \label{mdistrib_xi}
 \mathcal{Q}_{\xi ; n}^{m,\sigma}  = &  - \frac{\Lambda_n^{m,\tilde{\nu}}}{R^2 \sigma \tilde{\sigma} \left( \hat{k}_n^2 + \delta^2 \right)} \left\{ \mathcal{A}_n - \delta + \mathcal{D}_n^{-1} \left[ \mathcal{E}_n + \mathcal{F}_n \sin \left( \hat{k}_n \right) \right] \right\} , \\[0.3cm]
 \label{mdistrib_rho}
 \mathcal{Q}_{\rho ; n}^{m,\sigma} = &  \displaystyle - \frac{N^2 H \Lambda_n^{m,\tilde{\nu}}}{g R^2  \sigma \tilde{\sigma} \left( \hat{k}_n^2 + \delta^2 \right)} \left\{  \tau^{-1} \left( \mathcal{B}_n - \delta \right)  \left( \ed^{\tau} - 1 \right) \right. \\
  & \! \! \left. + \frac{\mathcal{G}_n + \mathcal{H}_n \cos \left( \hat{k}_n \right) + \mathcal{K}_n \sin \left( \hat{k}_n \right) }{\mathcal{D}_n \left(  \gamma^2 + \hat{k}_n^2 \right) } \right\}, \nonumber
\end{align}

\noindent with $ \gamma = \delta - \tau $ and the dimensionless coefficients

\begin{align}
  \mathcal{E}_n  = & \hat{k}_n \left( \mathcal{A}_n - \delta \right) \left( \mathcal{A}_n - \mathcal{C}_n \right) \ed^{\delta}, 
\end{align}

\begin{align}
\mathcal{F}_n = & \left( \mathcal{C}_n - \delta \right) \left( \mathcal{A}_n^2 + \hat{k}_n^2 \right) ,
\end{align}

\begin{align}
 \mathcal{G}_n  = & \hat{k}_n \left\{ \left(\mathcal{A}_n - \delta \right) e^\delta \left[ \gamma \left( \mathcal{B}_n - \mathcal{C}_n \right) + \mathcal{B}_n \mathcal{C}_n + \hat{k}_n^2 \right]  \right. \\
  & \left. + \left( \mathcal{C}_n - \delta \right) \ed^{- \gamma} \left[ \gamma \left( \mathcal{B}_n - \mathcal{A}_n \right) + \mathcal{A}_n \mathcal{B}_n + \hat{k}_n^2 \right] \right\}, \nonumber
\end{align}

\begin{align}
  \mathcal{H}_n = & - \hat{k}_n \left\{ e^\tau \left( \mathcal{A}_n - \delta \right) \left[ \gamma \left( \mathcal{B}_n - \mathcal{C}_n \right) + \mathcal{B}_n \mathcal{C}_n + \hat{k}_n^2 \right] \right. \\
   & \left. + \left( \mathcal{C}_n - \delta \right) \left[ \gamma \left( \mathcal{B}_n - \mathcal{A}_n \right) + \mathcal{A}_n \mathcal{B}_n + \hat{k}_n^2 \right]  \right\}, \nonumber
\end{align}

\begin{align}
 \mathcal{K}_n  = & \left( \mathcal{C}_n - \delta \right) \left[ \gamma \left( \mathcal{A}_n \mathcal{B}_n + \hat{k}_n^2 \right) + \hat{k}_n^2  \left( \mathcal{A}_n - \mathcal{B}_n \right) \right] \\
 & - e^\tau \left( \mathcal{A}_n - \delta \right) \left[ \gamma \left( \mathcal{B}_n \mathcal{C}_n + \hat{k}_n^2 \right) + \hat{k}_n^2 \left( \mathcal{C}_n - \mathcal{B}_n \right) \right]. \nonumber
\end{align}

Hence, $ \mathcal{Q}_{\xi ; n}^{m,\sigma} $ stands for the intrinsic response of the planet due to the variations of the oceanic surface level, while $ \mathcal{Q}_{\rho ; n}^{m,\sigma} $ corresponds to the contribution of internal \rec{inertial-gravity} waves. This later will be equal to zero in the case of an incompressible and neutrally-stratified ocean ($ \tau = 0 $). The contribution of internal gravity waves to the tidal response with respect to that of surface gravity waves is weighted by the ratio $\left| \mathcal{Q}_{\rho ; n}  / \mathcal{Q}_{\xi ; n}  \right|$. By using the expressions of the solution, we retrieve the weighting factor given in \smc{the} literature \citep[see e.g.][Eq.~10.40]{Hendershott1981}, that is $ \left| \mathcal{Q}_{\rho ; n}  / \mathcal{Q}_{\xi ; n}  \right| \sim N^2 H / g $. The oceanic second order Love number $ k_2^2 $ is then deduced from Eqs.~(\ref{Uxi}), (\ref{Urho}) and (\ref{klm}) straightforwardly. In the quadrupolar approximation, where terms of degrees $ l > 2 $ are neglected, it writes

\begin{equation}
k_2^2 = \frac{ \mathscr{G} M_{\rm oc} }{5R}  \sum_{n \in \mathbb{Z}} C_{2,n,2}^{2,\tilde{\nu}} \left( \mathcal{Q}_{\xi ; n}^{2,\sigma} + \mathcal{Q}_{\rho ; n}^{2,\sigma} \right),
\label{k22_thin}
\end{equation}

\noindent where $ M_{\rm oc} = 4 \pi R^2 H \rho_{\rm s} $ designates the total mass of the ocean in the weak stratification approximation ($ \tau \ll 1 $). Similarly, the tidal quality factor $ Q = \left|  k_2^2 / \Im \left\{ k_2^2 \right\} \right| $ \citep[][]{MLP2009} is expressed as

\begin{equation}
 Q = \left|  \frac{ \displaystyle  \sum_{n } C_{2,n,2}^{2,\tilde{\nu}} \left( \mathcal{Q}_{\xi ; n}^{2,\sigma} + \mathcal{Q}_{\rho ; n}^{2,\sigma} \right) }{ \displaystyle  \sum_{n } \Im \left\{  C_{2,n,2}^{2,\tilde{\nu}} \left( \mathcal{Q}_{\xi ; n}^{2,\sigma} + \mathcal{Q}_{\rho ; n}^{2,\sigma} \right) \right\} }   \right|,
\label{Q_thin}
\end{equation}

\noindent and the \padc{quadrupolar} tidal torque, \padc{which corresponds to the $ l = m =  2 $ component}, as

\begin{equation}
\mathcal{T}^{2,\sigma} = \frac{1}{2} M_{\rm oc}  \left| U_2^{2,\sigma} \right|^2  \sum_{n}  \Im \left\{ C_{2,n,2}^{2,\tilde{\nu}} \left(  \mathcal{Q}_{\xi ; n}^{2,\sigma} + \mathcal{Q}_{\rho ; n}^{2,\sigma}\right)  \right\}.
\label{torque_thin}
\end{equation}

\noindent where the quadrupolar component of the tidal potential $ U_2^{2,\sigma} $ is given by

\begin{equation}
U_2^{2,\sigma} = \sqrt{\frac{3}{5}} \left( \frac{R}{a} \right)^2 \frac{\mathscr{G} M_\star}{a}.
\label{U22}
\end{equation}

\noindent \padc{Expressed as a function of the degree-$2$ tidal Love number given by Eq.~(\ref{k22_thin}), the tidal torque is written }

\begin{equation}
\mathcal{T}^{2,\sigma} = \frac{3}{2} \mathscr{G} M_\star^2 \frac{R^5}{a^6} \Im \left\{ k_2^2 \right\},
\end{equation}

\noindent \padc{which is the well-known expression of the tidal torque associated with the quadrupolar semidiurnal tide \citep[e.g.][]{EW2009,Makarov2012,Correia2014}.}

\subsection{Case of the neutrally stratified ocean ($ N = 0 $)}
\label{subsec:neutral_ocean}

In the analytic modeling of oceanic tides, the stratification is usually not taken into account, the layer being \rec{two-dimensional} and the fluid assumed to be incompressible \citep[e.g.][]{Webb1980,Tyler2011,Tyler2014,Chen2014,Matsuyama2014}. \smc{This} reduces \smc{the oceanic tidal dynamics} to horizontal flows. We show here that we recover the results given by this approach with our modeling, in which the \smc{shallow water case} is a particular case.

Following the early works mentioned above, let us set $ N = 0 $ (neutral stratification) and $ c_s = + \infty $ (incompressible fluid). The vertical wavenumber thus becomes

\begin{equation}
\hat{k}_n = i \frac{H}{R} \sqrt{\Lambda_n^{m,\tilde{\nu}}}.
\label{kn_neutral}
\end{equation}

\noindent As shown by Eq.~(\ref{kn_neutral}), $ \hat{k}_n $ is now directly proportional to the  horizontal wavenumber $ \hat{k}_{\perp ; n} = \sqrt{ \Lambda_n^{m,\tilde{\nu}} / R^2 } $. In the weak-friction approximation ($ \sigma_{\rm R} \ll \left| \sigma \right| $) and the regime of super-inertial waves ($ \left| \tilde{\nu} \right| \leq 1 $), it does not depend on the tidal frequency. Indeed, in this case, the associated eigenvalues are $ \Lambda_n^{m,\tilde{\nu}} \approx \Lambda_n^{m,0} = \left( m + n  \right) \left( m + n + 1 \right) $ $ n \in \mathbb{N} $ (the relation between the degree $ l $ of associated \smc{Legendre} polynomials and the degree $ n $ has been defined as $ l = m + n $). The frequency dependence of $ \Lambda_n^{m,\tilde{\nu}} $ cannot be ignored any more in the regime of sub-inertial waves. Because of the thin-layer approximation, $ \left| \hat{k}_n \right| \ll 1 $. It follows that $ \sin \left( \hat{k}_n x \right) \approx \hat{k}_n x $ and $ \cos \left( \hat{k}_n x \right) \approx 1 $. Hence, in the case of the neutrally stratified incompressible ocean,  the analytic solution of Eq.~(\ref{Psin}) simply reduces to

\begin{equation}
\Psi_n^{m,\sigma} =  U_n \frac{ g H \hat{k}_{\perp ; n}^2 - \frac{H}{R} \sigma \tilde{\sigma} x }{\left( \sigma - \sigma_n^- \right) \left( \sigma - \sigma_n^+ \right) }, 
\label{Psin_neutre}
\end{equation}

\noindent where $ \sigma_n^{\rm -} $ and $ \sigma_n^{\rm +} $ are the frequencies of resonance of large scale ($ \left| \hat{k}_{\perp ; n} R \right| \ll 1 $) damped surface \rec{inertial-gravity} waves (see Fig.~\ref{fig:spectre_regimes}) associated with the gravity mode of degree $ n $. These frequencies characterize the tidal response of the \rec{two-dimensional} incompressible ocean, where only surface gravity waves can propagate \citep[e.g.][Eq.~2.12]{Webb1980}. They are expressed as

\begin{equation}
\sigma_n^{\rm \pm} =  i  \frac{\sigma_{\rm R}}{2} \pm  \sqrt{  g H \hat{k}_{\perp ; n}^2 - \left( \frac{\sigma_{\rm R}}{2}  \right)^2 } .
\label{sigmanpm}
\end{equation}

We recognize in this expression the dispersion relation of large scale surface gravity waves, $ \sigma^2 = g H \hat{k}_{\perp}^2 $ \citep[see e.g.][\smc{in the adiabatic limit}]{Vallis2006}. Like the vertical wavenumber, $ \sigma_n^{\rm -} $ and $ \sigma_n^{\rm +} $ formally depend on $ \sigma $ through $ \Lambda_n^{m,\tilde{\nu}} $ but this dependence can be neglected in the regime of super-inertial waves if the weak-friction approximation is assumed. The above approximations imply $ \tau = 0 $ and $ \mathcal{Q}_{\rho ; n}^{m,\sigma} = 0 $. Thus, the Love numbers and tidal torque (Eqs.~\ref{k22_thin} and \ref{torque_thin}) are determined by the contribution of the surface displacement solely, which reads

\begin{equation}
\mathcal{Q}_{\xi ; n}^{m,\sigma} = - \frac{\Lambda_n^{m,\tilde{\nu}}}{R^2 } \frac{1}{\left( \sigma - \sigma_n^- \right) \left( \sigma - \sigma_n^+ \right)}.
\label{Qxi_neutre}
\end{equation}

\noindent The expression of $ Q_{\xi,n}^{m,\sigma} $ shows that, in the absence of stable stratification, the coupling induced by the Coriolis effect will lead to a \rec{set} of resonances (one per Hough mode) enhancing the tidal dissipation and tidal torque at frequencies $ \sigma_n^{\pm} $. 

\section{Application to illustrative cases}
\label{sec:application_cases}

In this section, we apply the results established previously to representative telluric planets by using the \rec{thin-shell} approximation of Sect.~\ref{sec:uniform_ocean}. We first treat the case of the Earth, which illustrates the thin-layer asymptotic behaviour. However, we shall bear in mind that the global ocean approximation is a rough hypothesis in the case of the Earth, where the interaction of the tidal perturbation with continents plays a central role \citep[][]{Webb1980,ER2001,ER2003}. We then apply the model to Trappist-1 f, which could be covered with a deep global ocean of liquid water owing to its small density \citep[$ M_{\rm p} \approx 0.36 \ M_\Earth $ and $ R  \approx 1.045 \ R_\Earth $, see][]{Wang2017}.


\begin{figure*}[htb]
 \centering
   \includegraphics[width=0.4\textwidth,trim = 1.5cm 3cm 6.5cm 2.3cm,clip]{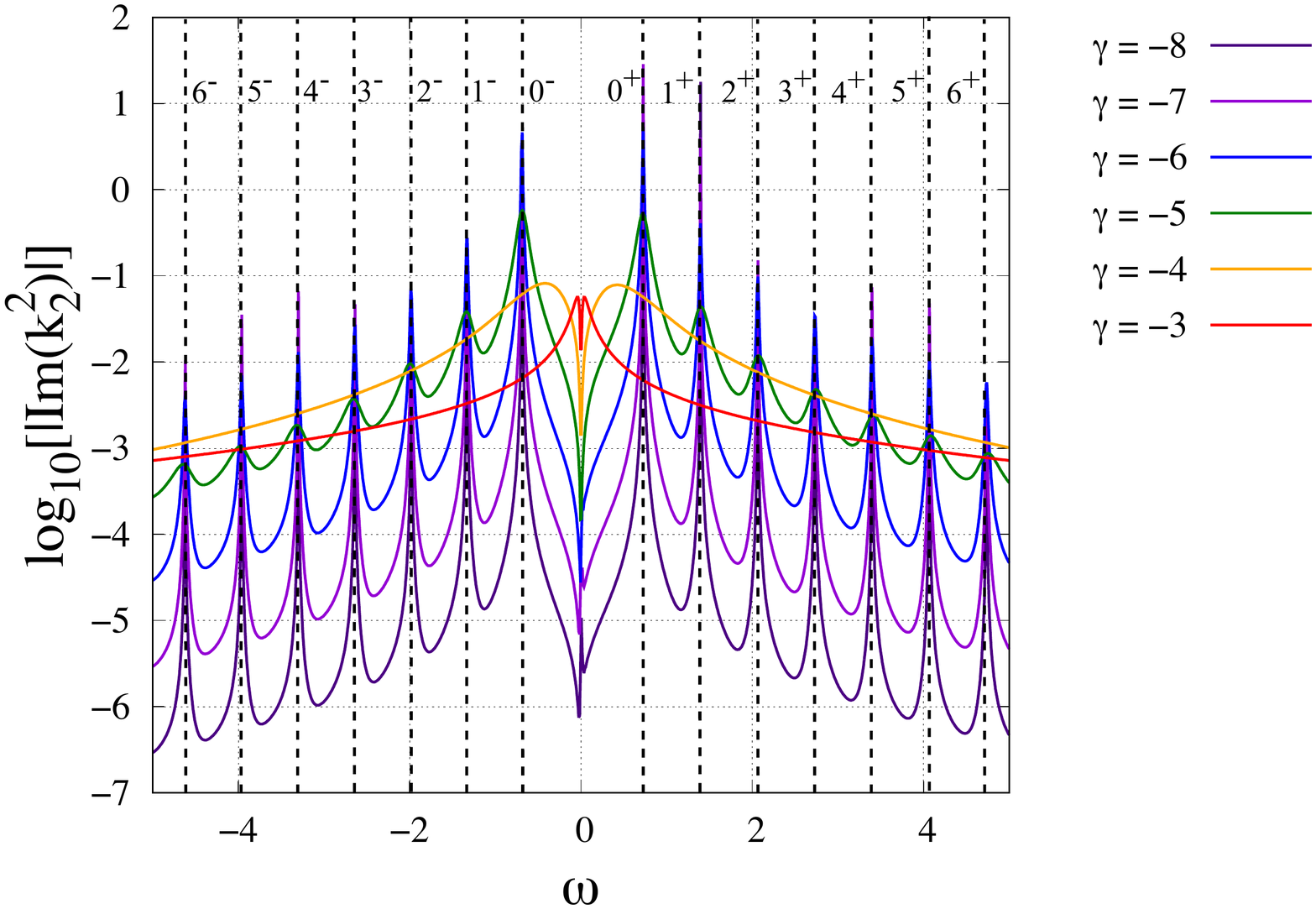} \hspace{0.5cm}
    \includegraphics[width=0.4\textwidth,trim = 1.5cm 3cm 6.5cm 2.3cm,clip]{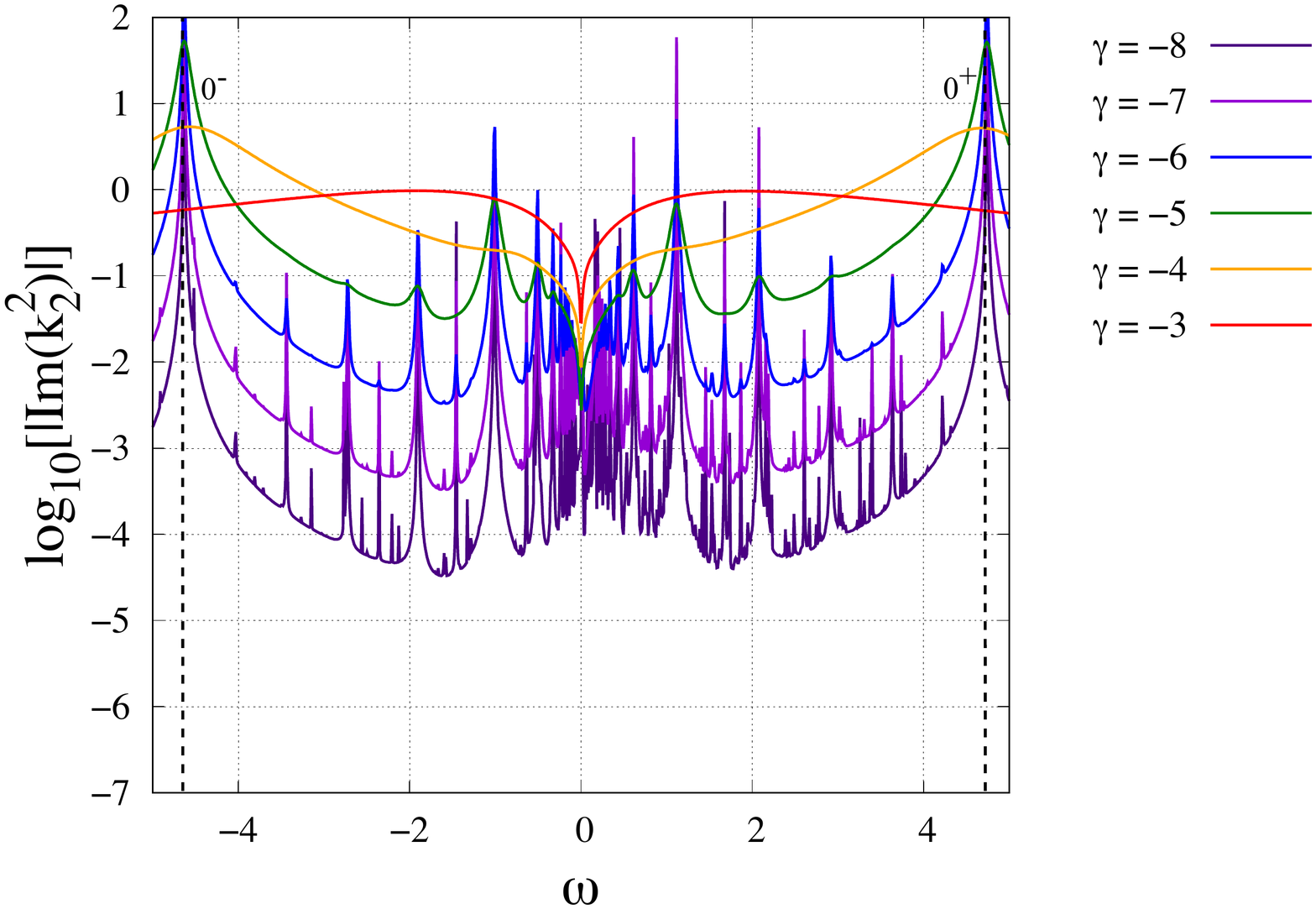}
  \includegraphics[width=0.1\textwidth,trim = 21.5cm 5cm 2.0cm 2.3cm,clip]{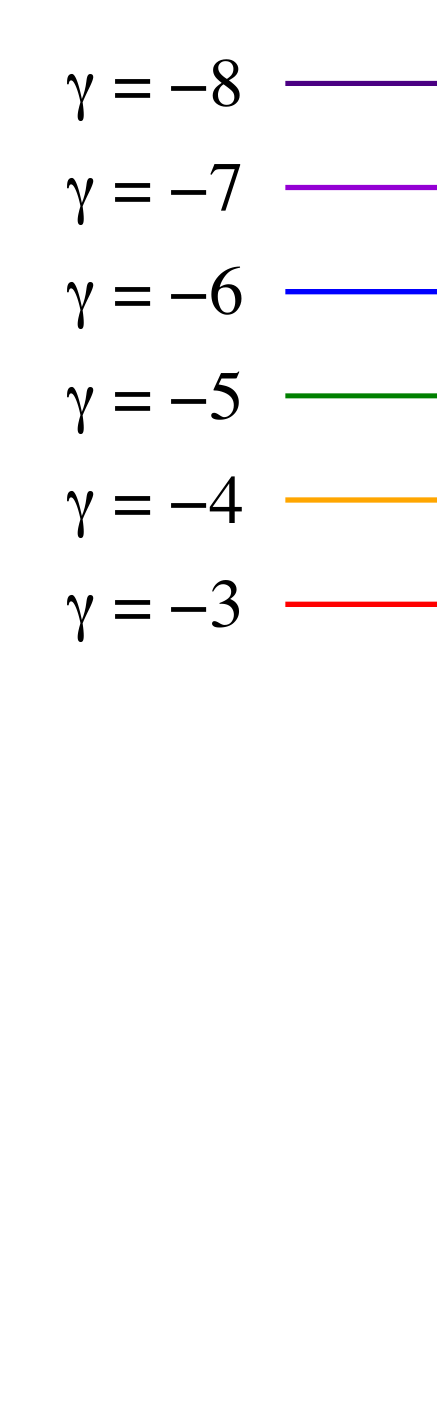} \\
    \includegraphics[width=0.4\textwidth,trim = 1.5cm 2.0cm 6.5cm 2.3cm,clip]{auclair-desrotour_fig5c.pdf}  \hspace{0.5cm}
  \includegraphics[width=0.4\textwidth,trim = 1.5cm 2.0cm 6.5cm 2.3cm,clip]{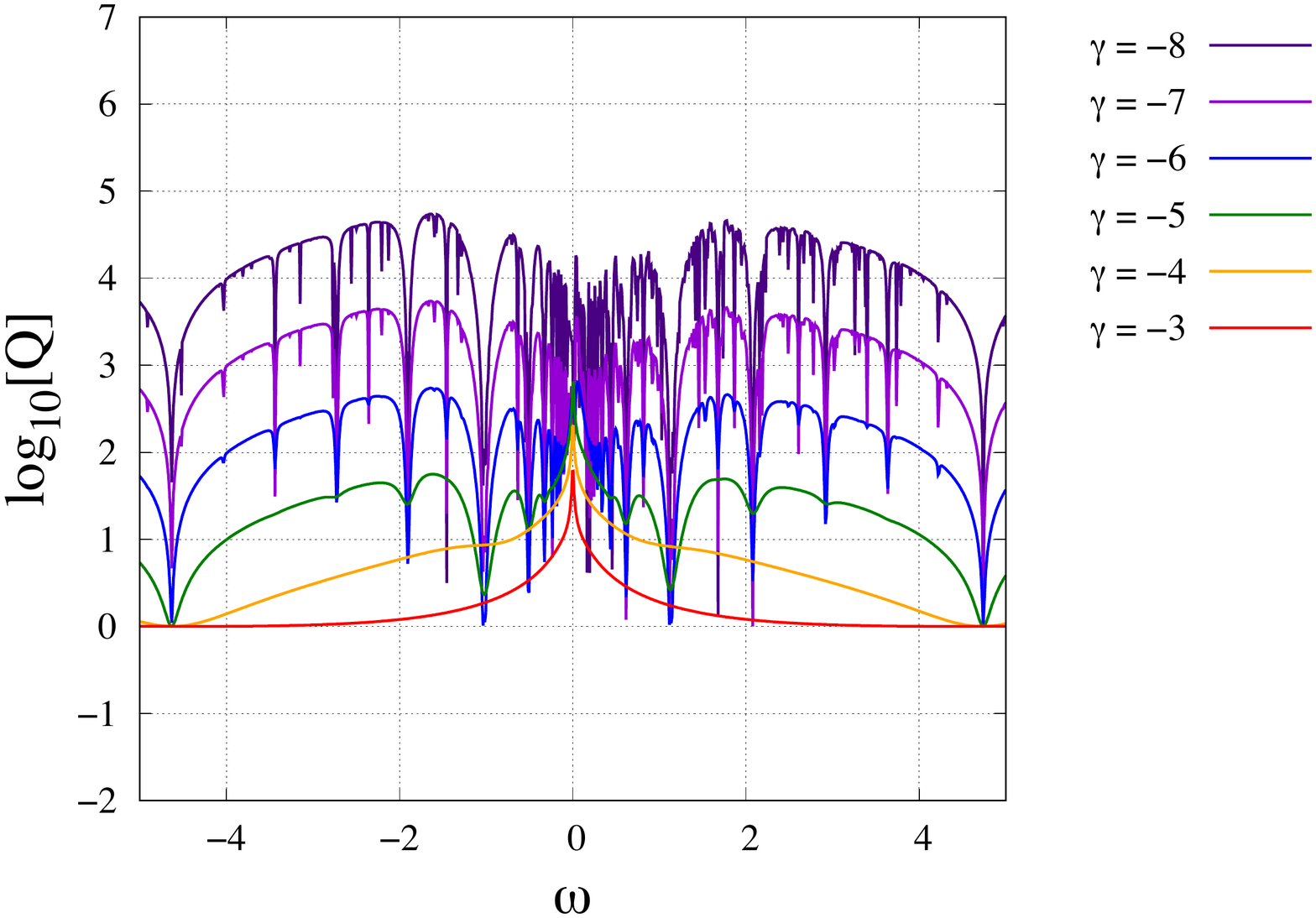} 
  \hspace{0.1\textwidth}
\caption{\label{fig:Earth_spectre} Semidiurnal oceanic tide of the Earth (left column) and Trappist-1 f (right column). The logarithms of the imaginary part of the tidal Love number (left) and tidal quality factor (right) are plotted as functions of the normalized forcing frequency $ \omega = \left( \Omega - n_{\rm orb} \right) / \Omega_\Earth $ (where $ \Omega_\Earth $ designates the today rotation rate of the Earth) for various orders of magnitude of the drag parameter $ \gamma = \log \left( \sigma_{\rm R} \right) $. These frequency spectra are \smc{computed} using the expressions given by Eqs.~(\ref{k22_thin}) and (\ref{Q_thin}). In each case, the parameter $ n_{\rm orb} $ is assumed to be constant and the rotation rate of the planet varies with the tidal frequency following the formula $ \Omega =  n_{\rm orb} + \sigma/2 $. \padc{The resonances associated with surface \rec{inertial-gravity} modes are designated by black dashed lines in the top panels and numbers indicate the degree $ n $ of the corresponding Hough modes and the sign of eigenfrequencies given by Eq.~(\ref{sigmanpm}).}The values of parameters used for this \smc{evaluation} are summarized in Table~\ref{para_cases}.}  
\end{figure*}

\subsection{The Earth}
\label{subsec:Earth}

The Sun and the Moon exert on the Earth gravitational forcings of comparable intensities. The resulting tidal perturbation is the sum of two contributions, the solid and oceanic tides, which drive the rotational evolution of the planet as well as the orbital evolution of the Earth-Moon system. We focus here on the Lunar quadrupolar tide, which corresponds to the tidal frequency $ \sigma = 2 \left( \Omega - n_{\rm orb} \right) $, where $ \Omega $ and $ n_{\rm orb} $ stand for the spin frequency of the Earth and orbital frequency of the Moon. We introduce the corresponding rotation ($ P_{\rm rot} = 2 \pi / \Omega $) and orbital ($ P_{\rm orb} = 2 \pi / n_{\rm orb} $) periods. Following \cite{Webb1980}, we use for the uniform oceanic depth the mean depth of the Earth's ocean, that is $ H = 4 $ km \citep[][]{Webb1980}. The radius of the planet is set to $ R = R_\Earth $ km (with $ R_\Earth = 6378 $ km), the surface gravity and density to $ g = 9.81 \ {\rm m.s^{-2}}  $ and $ \rho_{\rm s} = 1022 \ {\rm kg.m^{-3}} $. The Earth's ocean is characterized by $ N \sim 10^{-4} - 10^{-2} \ {\rm s^{-1}} $ and $ c_{\rm s} \approx 1545 \ {\rm m.s^{-1}} $ \citep[][]{GZ2008}, which means that $ \tau \ll 1 $.  Thus, the effects of stratification are negligible in this case. We set the Brunt-Väisälä frequency to \smc{the intermediate value} $ N = 10^{-3} \ {\rm s^{-1}} $. All these \smc{quantities} are summarized in Table~\ref{para_cases}.

\begin{table}[htb]
 \textsf{\caption{\label{para_cases} Values of parameters used in Section~\ref{sec:application_cases}. For the Earth, $ M_\Earth $, $ R_\Earth $, $ g $, $ P_{\rm orb} $ and $ M_{\rm pert} $ are given by \href{https://nssdc.gsfc.nasa.gov/planetary/factsheet/earthfact.html}{NASA fact sheets}. The oceanic parameters $ H $, $ \rho_{\rm s} $, $ c_{\rm s} $ and $ N $ come from \cite{GZ2008}. For TRAPPIST-1 f, we use the values given by \cite{Wang2017} for $ M $, $ R $, $ M_{\rm pert} $ and $ P_{\rm orb} $. The surface gravity $ g $ is estimated using the \smc{definition} $ g = \mathscr{G} M / R $ and we use the oceanic parameters of the Earth. The ocean depth is arbitrarily set to $ H = 1000 $ km. }}
\centering
    \begin{tabular}{ l  l  l  l  l l}
      \hline
      \hline
      \textsc{Parameters} & \textsc{Units} & \textsc{Earth} & \textsc{Trappist-1 f} \\
      \hline
       $ M $ & $ M_\Earth $ & $ 1 $ & $ 0.36 $\\
       $ R $ & $ R_\Earth $ & $ 1 $ & $ 1.045 $ \\ 
       $ g $ & $ {\rm m.s^{-2}} $ & $ 9.81 $ & $ 3.23 $\\
       $ H $ & km & $ 4.0 $ & $ 1000 $ \\
       $ \rho_{\rm s} $ & $ {\rm kg.m^{-3}} $ & $ 1022 $ & $ 1022 $\\
       $ c_{\rm s} $ & $ {\rm m.s^{-1}} $ & $ 1545 $ & $ 1545 $ \\ 
       $ N $ & $ {\rm s^{-1}} $ & $ 10^{-3} $ & $ 10^{-3} $ \\
       $ M_{\rm pert} $ & kg & $ 7.346 \times 10^{22} $ & $ 1.59 \times 10^{29} $ \\
       $ P_{\rm orb} $ & \smc{days} & $ 27.32 $ & $ 9.20 $
      \vspace{0.1mm}\\
       \hline
    \end{tabular}
    \end{table}

The drag frequency characterizing the effective Rayleigh friction ($ \sigma_{\rm R} $) is more difficult to specify because it models the effects of several mechanisms, such as turbulent friction, viscous friction, friction with topography and breaking of tidal waves \smc{\citep[][]{GM1979,GK2007}}. Thus, we begin by studying the dependence of the oceanic tidal response on $ \sigma_{\rm R} $. The frequency spectra of the imaginary part of the tidal Love number \smc{(Eq.~\ref{k22_thin})} and of the \smc{associated} tidal quality factor \smc{(Eq.~\ref{Q_thin})} are plotted in \smc{Fig.}~\ref{fig:Earth_spectre} (left column) as functions of the normalized frequency $ \omega = \left( \Omega - n_{\rm orb} \right) / \Omega_\Earth $ ($ \Omega_\Earth$ stands for the today Earth rotation rate) for various orders of magnitude of $ \sigma_{\rm R} $. We observe on these plots the resonances associated with surface gravity modes modified by rotation. They correspond to the eigenfrequencies $ \sigma_n^{\pm} $ given by Eq.~(\ref{sigmanpm}). As $ \sigma_{\rm R} $ decreases, the variability of $ \Im \left\{ k_2^{2} \right\} $ increases. Particularly, the level of the non-resonant background decreases proportionally to $ \sigma_{\rm R} $, while the peaks heights increase \smc{as $\sigma_{\rm R}^{-1}$}, which is in good agreement with the scaling laws derived in \cite{ADMLP2015}. In the asymptotic regime of strong friction, the behaviour of the ocean is regular. The imaginary part of the tidal Love number scales as $ \Im  \left\{ k_2^2 \right\} \, \propto \, \sigma $ in the zero-frequency limit ($ \sigma \rightarrow 0 $) and as $ \Im \left\{ k_2^2 \right\} \, \propto \, \sigma^{-3} $ at $ \left| \sigma \right| \rightarrow + \infty $.

\padc{The observed receding of the Moon is estimated to 3.82 cm per year \citep[][]{Dickey1994,BR1999}. This corresponds to $ \Im \left\{ k_2^2 \right\} \approx 2.56 \times 10^{-2} $ for the tidal Love number and $ \mathcal{T} \approx 4.50 \times 10^{16} \ {\rm N.m} $ for the tidal torque, which are the values obtained by setting the Rayleigh drag frequency to $ \sigma_{\rm R} = 10^{-5} \ {\rm s^{-1}} $ in our model (Fig.~\ref{fig:Earth_spectre}, right panel). We hence retrieve the $ \sim 30 $~hours friction time scale estimated by \cite{Webb1980} with a similar approach. This value of $ \sigma_{\rm R} $ will be used in Section~\ref{sec:explo_para} to explore the domain of parameters. }


\subsection{TRAPPIST-1 f}

To illustrate the behaviour of a deep stably-stratified ocean, we consider the case of TRAPPIST-1 f. TRAPPIST-1 f is one of the eight telluric planets recently discovered in the vicinity of the star TRAPPIST-1 \citep[][]{Gillon2017,Wang2017}. Its radius is very close to that of the Earth ($R = 1.045 \ M_\Earth $) whereas its mass is only $ M = 0.36 \ M_\Earth $. As a consequence, its mean density is equal to $ \bar{\rho} = 1740 \ {\rm kg.m^{-3}} $, that is far smaller than those of rocky planets \smc{(for instance, the mean density of the Earth is $ \bar{\rho}_\Earth = 5514 \ {\rm kg.m^{-3}} $; see \href{https://nssdc.gsfc.nasa.gov/planetary/factsheet/earthfact.html}{NASA fact sheets})}. This means that water could stand for an important fraction of the planet mass. Typically, the water fraction is estimated between 25~\% and 100~\% of the planet mass. Moreover, with a greenhouse effect, TRAPPIST-1 f is sufficiently irradiated by its host star to have a surface temperature compatible with liquid water \citep[its black body equilibrium temperature is estimated to 219~K, see][]{Wang2017}. All these features argue for the possible presence of a deep oceanic layer on TRAPPIST-1 f. In such a case, if the layer is stably stratified, the effect of stratification cannot be ignored any more because the variation rate of the density profile ($ \tau$ defined by \smc{Eq.~\ref{tau}}) is not necessary very small with respect to 1. Thus, as in the case of the Earth, the tidal response is mainly composed of surface gravity waves modified by rotation. But it is also composed of internal gravity waves, restored by the Archimedean force and inducing internal density variations (Fig.~\ref{fig:spectre_regimes}). 

We investigate this complex behaviour by considering an idealized TRAPPIST-1 f planet with a global ocean of depth $ H = 1000 $~km and a stratification similar to that of the Earth's ocean, i.e. $ N = 10^{-3} \ {\rm s^{-1}}$. Assuming the orbital period of the planet ($n_{\rm orb}$) to be constant, we study the frequency dependence of the quadrupolar component of the semidiurnal stellar tide (cf. Table~\ref{para_cases}) by letting $ \Omega $ vary with $ \sigma $. Similarly to the previous case, the corresponding spectra of $ \Im \left\{ k_2^2 \right\} $ and $ Q $ are plottted on Fig.~\ref{fig:Earth_spectre} (right column) as a function of the forcing frequency. We retrieve in the high-frequency range the resonances associated with surface gravity modes. The two peaks appearing around $ \omega \approx \pm 4.5 $ correspond to the Hough mode of degree $ n = 0 $, which is the surface gravity mode of lowest eigenfrequency. Given that $ \sigma_n^{\pm} \, \propto \, \sqrt{gH} $ in the weakly frictional regime (see \smc{Eq.~\ref{sigmanpm}}), the resonances are translated towards the high-frequency range with respect to the case of the Earth (left column). In addition to surface gravity modes, we observe resonances caused by the propagation of internal gravity waves in the low frequency range. These modes can exist for $ \left| \sigma \right| \lesssim N $ (see Fig.~\ref{fig:spectre_regimes}), which corresponds to $ \left| \omega \right| \lesssim 5 $.

\begin{figure*}[htb]
 \centering
   \includegraphics[height=0.6cm]{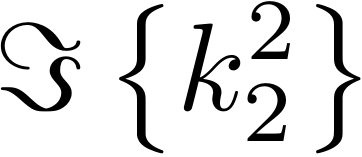} \\[0.3cm]
   \includegraphics[height=0.45cm]{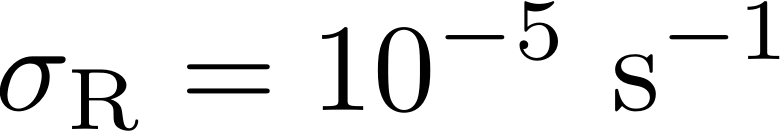} 
   \hspace{0.8cm}
   \includegraphics[height=0.3cm]{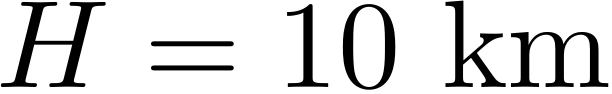} \hspace{2.5cm}
   \includegraphics[height=0.3cm]{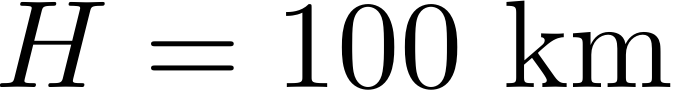} \hspace{2.5cm}
   \includegraphics[height=0.3cm]{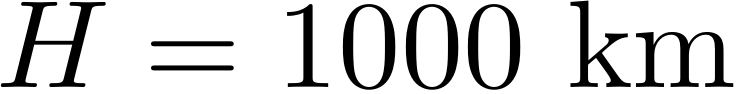} \hspace{2.0cm}~ \\[0.2cm]
   \raisebox{1.0cm}{\includegraphics[width=0.02\textwidth]{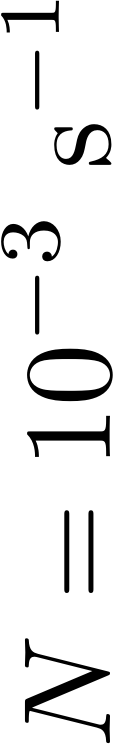}} \hspace{0.1cm}
   \raisebox{1.0\height}{\includegraphics[width=0.015\textwidth]{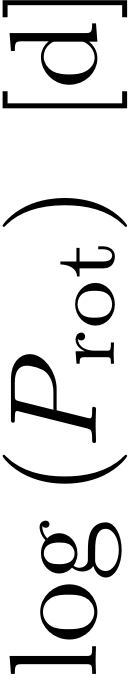}}
  \includegraphics[width=0.25\textwidth,trim = 2.5cm 6.3cm 3.0cm 3.5cm,clip]{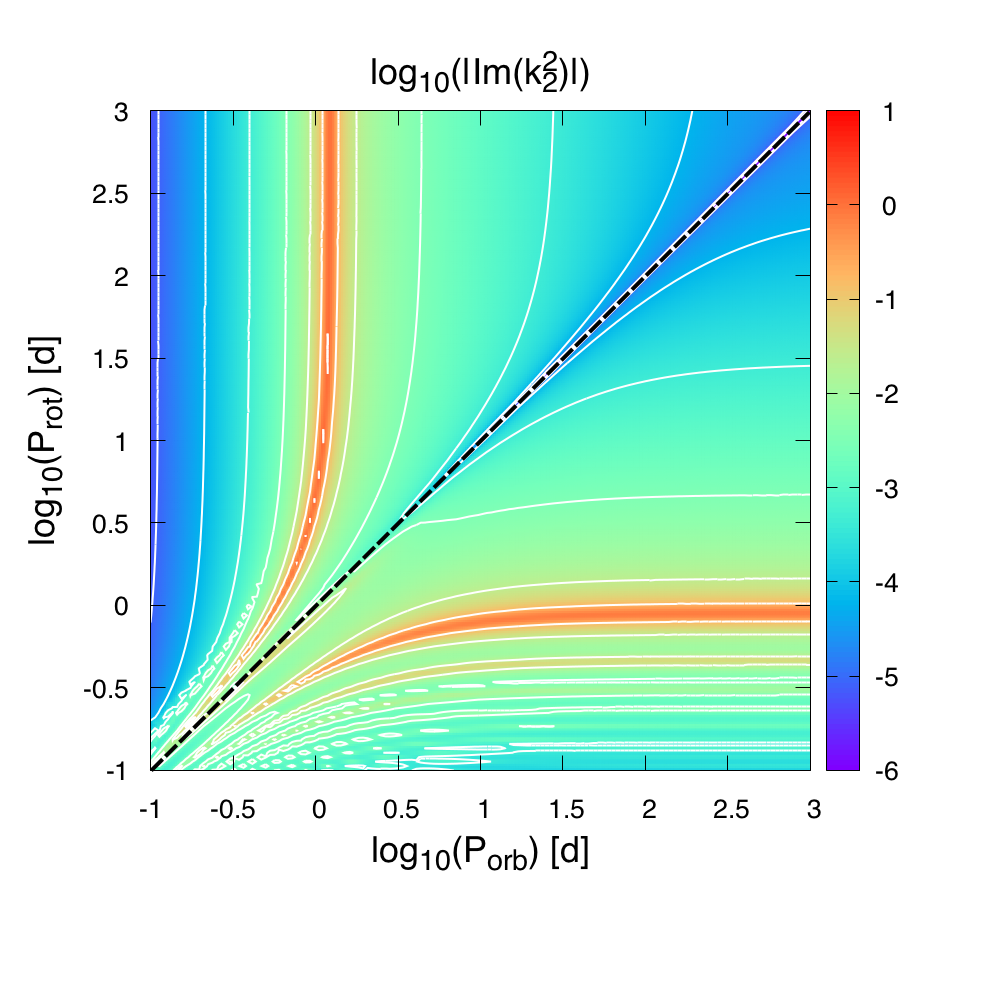} \hspace{0.1cm}
  \includegraphics[width=0.25\textwidth,trim = 2.5cm 6.3cm 3.0cm 3.5cm,clip]{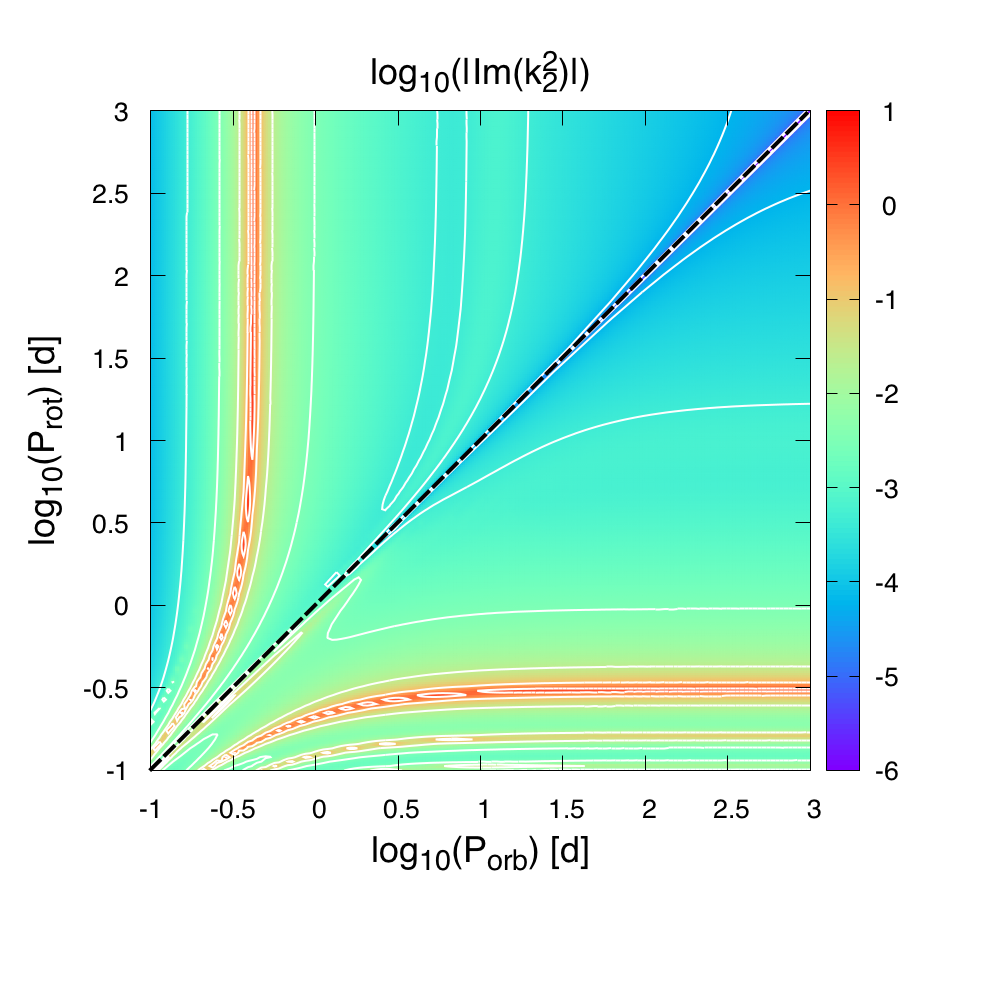} \hspace{0.1cm}
  \includegraphics[width=0.25\textwidth,trim = 2.5cm 6.3cm 3.0cm 3.5cm,clip]{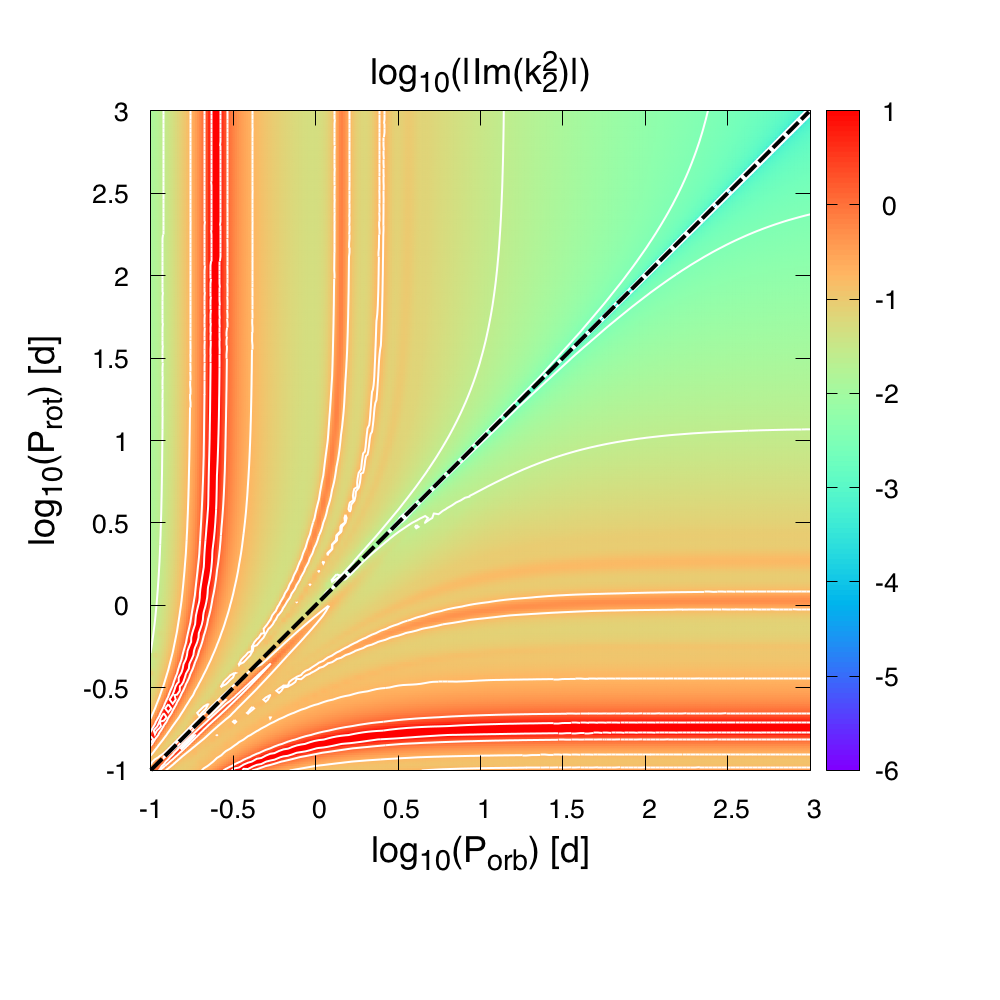} \\[0.1cm]
    \raisebox{1.0cm}{\includegraphics[width=0.02\textwidth]{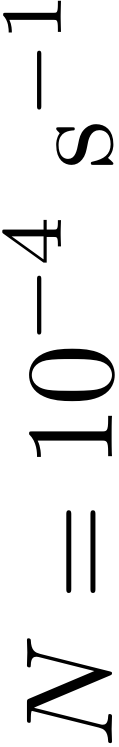}} \hspace{0.1cm}
   \raisebox{1.0\height}{\includegraphics[width=0.015\textwidth]{auclair-desrotour_fig6g.pdf}}
 \includegraphics[width=0.25\textwidth,trim = 2.5cm 6.3cm 3.0cm 3.5cm,clip]{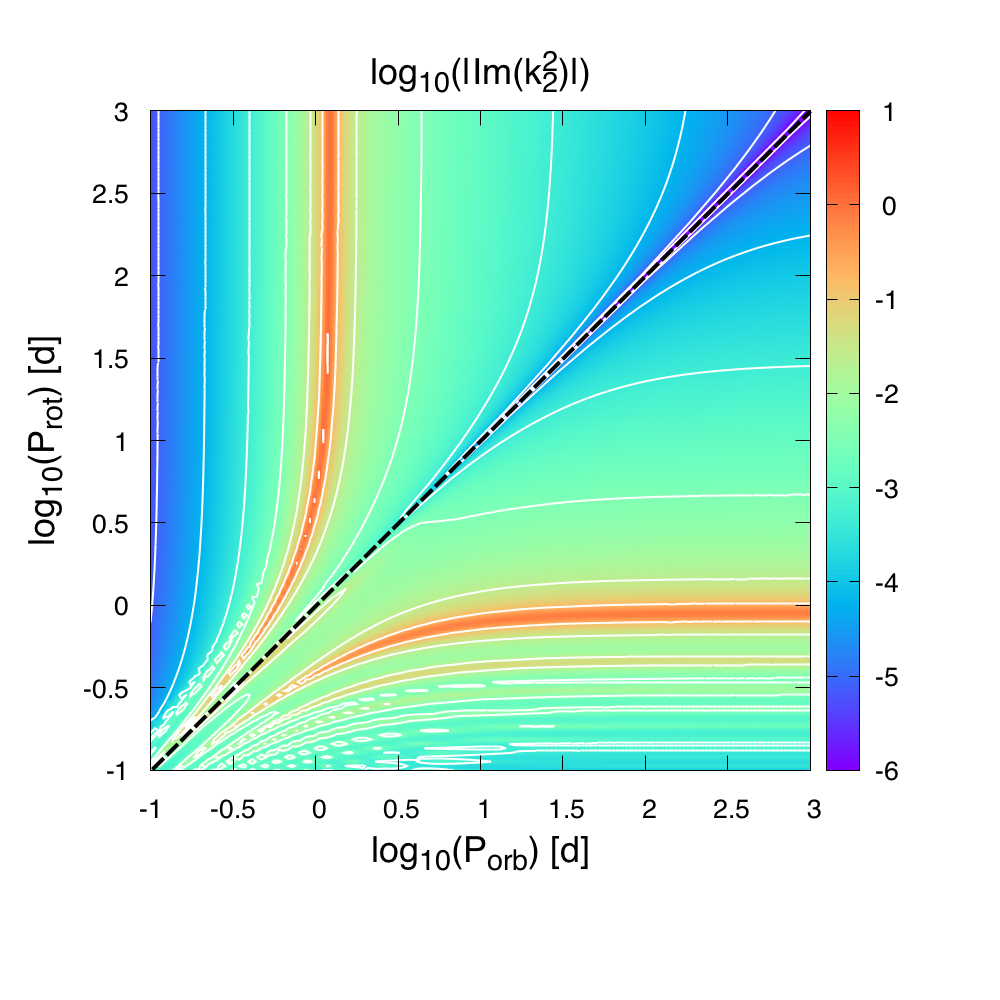} \hspace{0.1cm}
 \includegraphics[width=0.25\textwidth,trim = 2.5cm 6.3cm 3.0cm 3.5cm,clip]{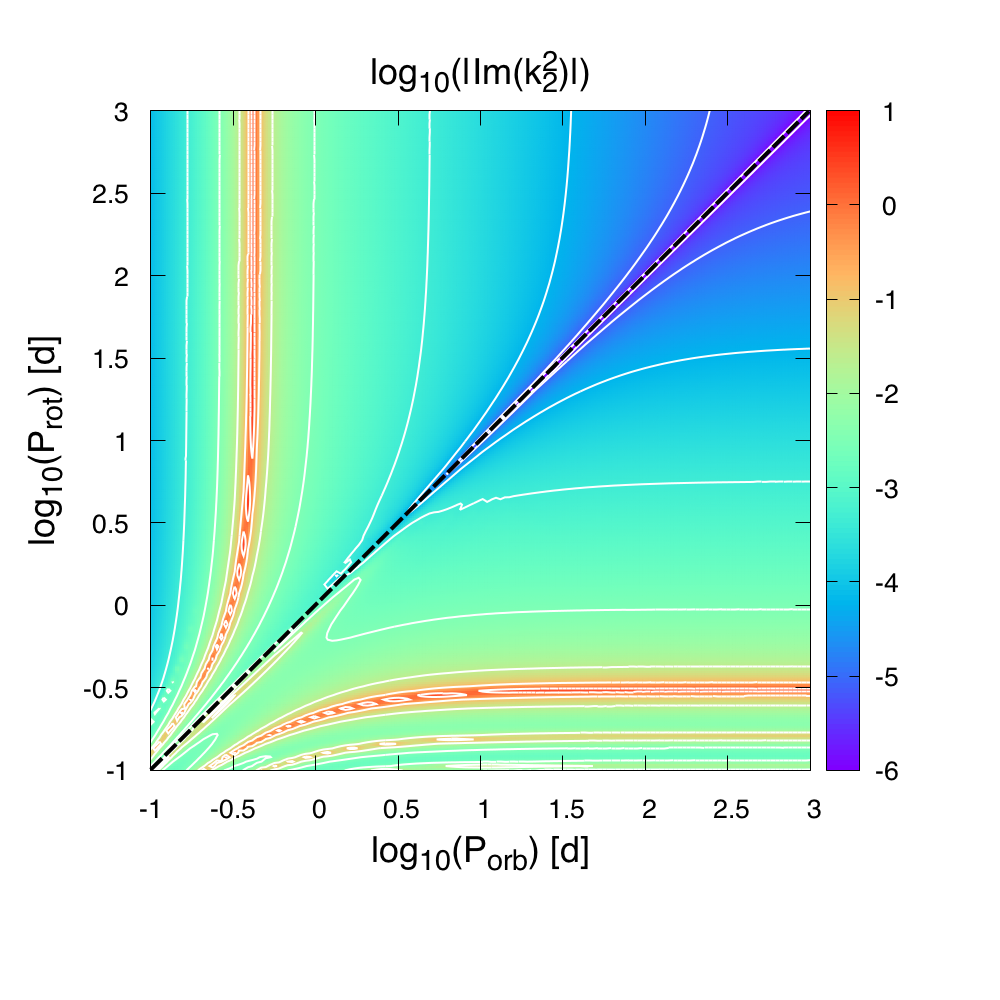} \hspace{0.1cm}
 \includegraphics[width=0.25\textwidth,trim = 2.5cm 6.3cm 3.0cm 3.5cm,clip]{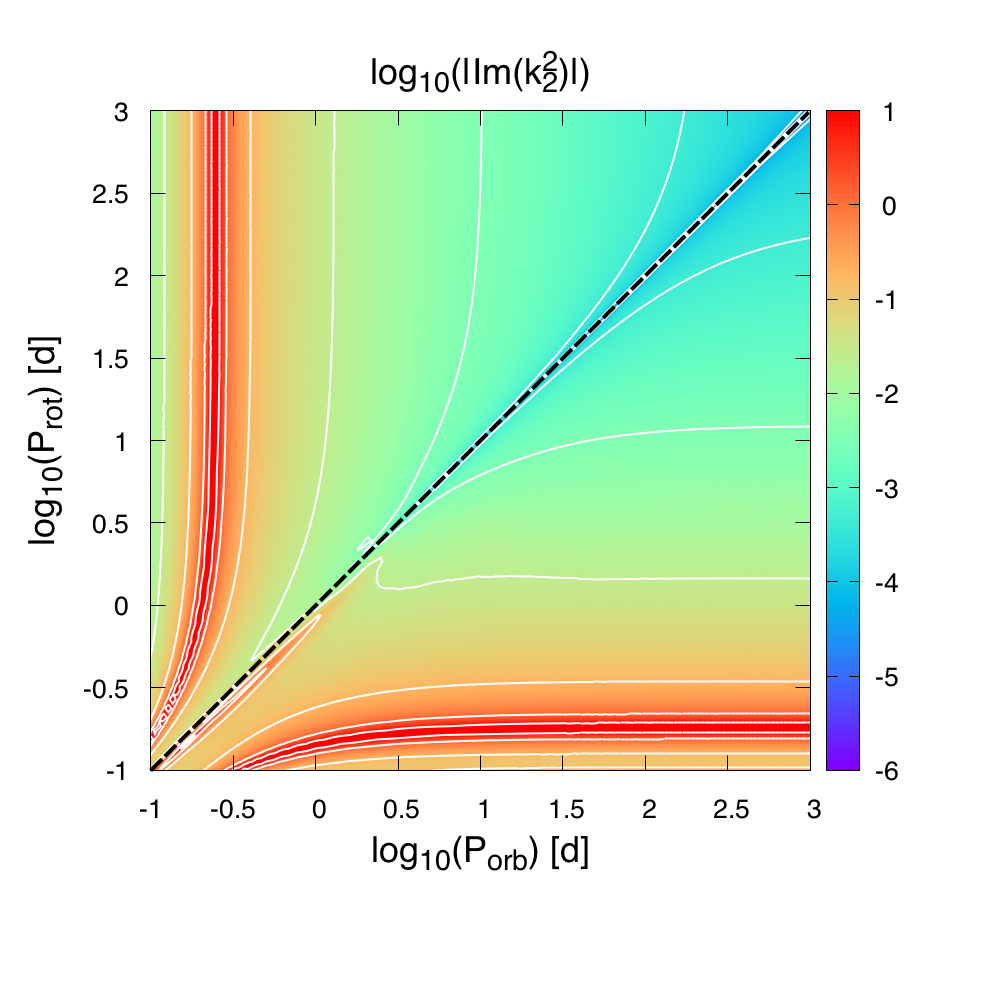} \\[0.1cm]
     \hspace{1.0cm}
   \includegraphics[height=0.3cm]{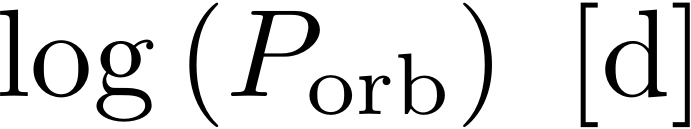} \hspace{3.0cm}
   \includegraphics[height=0.3cm]{auclair-desrotour_fig6o.pdf} \hspace{3.0cm}
   \includegraphics[height=0.3cm]{auclair-desrotour_fig6o.pdf} \hspace{0.0cm} \\[0.3cm]
 \includegraphics[height=0.45cm]{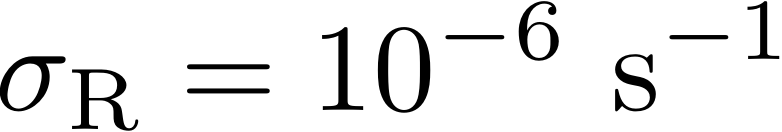} 
  \hspace{0.8cm}
   \includegraphics[height=0.3cm]{auclair-desrotour_fig6c.pdf} \hspace{2.5cm}
   \includegraphics[height=0.3cm]{auclair-desrotour_fig6d.pdf} \hspace{2.5cm}
   \includegraphics[height=0.3cm]{auclair-desrotour_fig6e.pdf} \hspace{2.0cm}~ \\[0.2cm]
    \raisebox{1.0cm}{\includegraphics[width=0.02\textwidth]{auclair-desrotour_fig6f.pdf}} \hspace{0.1cm}
   \raisebox{1.0\height}{\includegraphics[width=0.015\textwidth]{auclair-desrotour_fig6g.pdf}}
  \includegraphics[width=0.25\textwidth,trim = 2.5cm 6.3cm 3.0cm 3.5cm,clip]{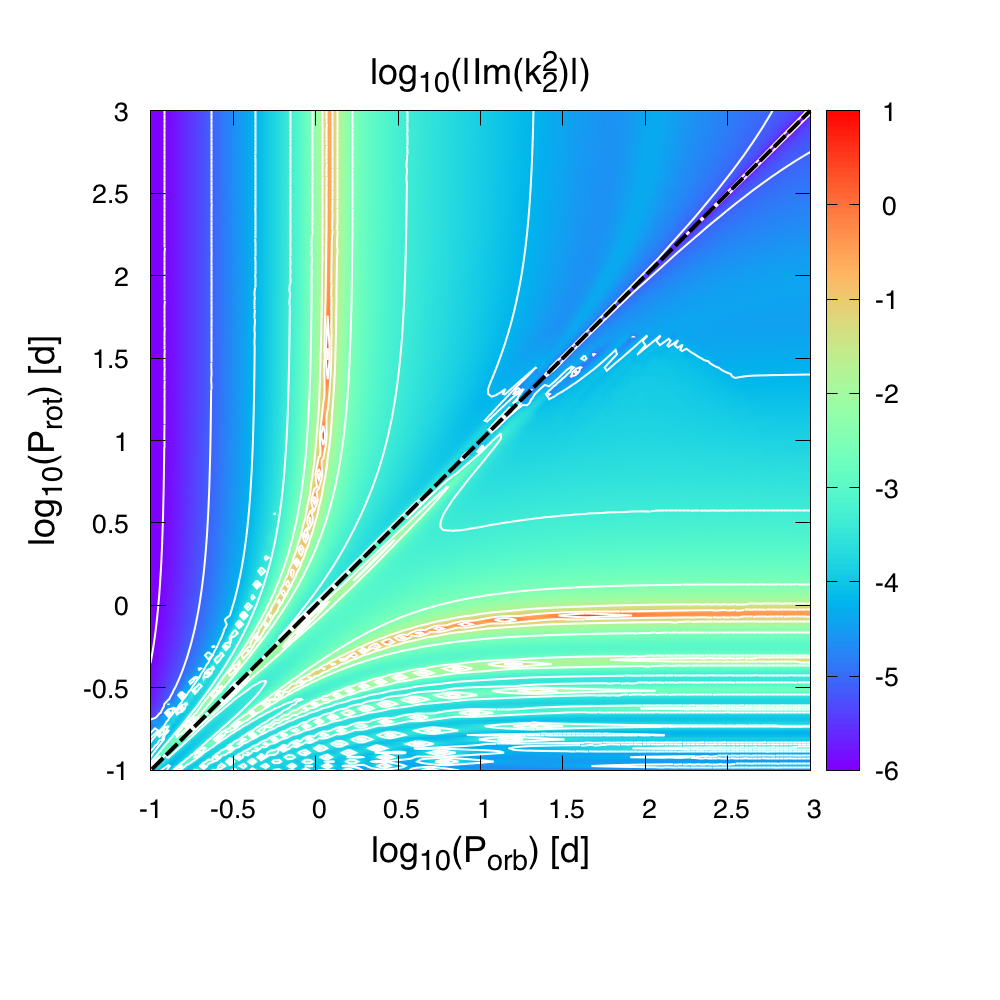} \hspace{0.1cm}
  \includegraphics[width=0.25\textwidth,trim = 2.5cm 6.3cm 3.0cm 3.5cm,clip]{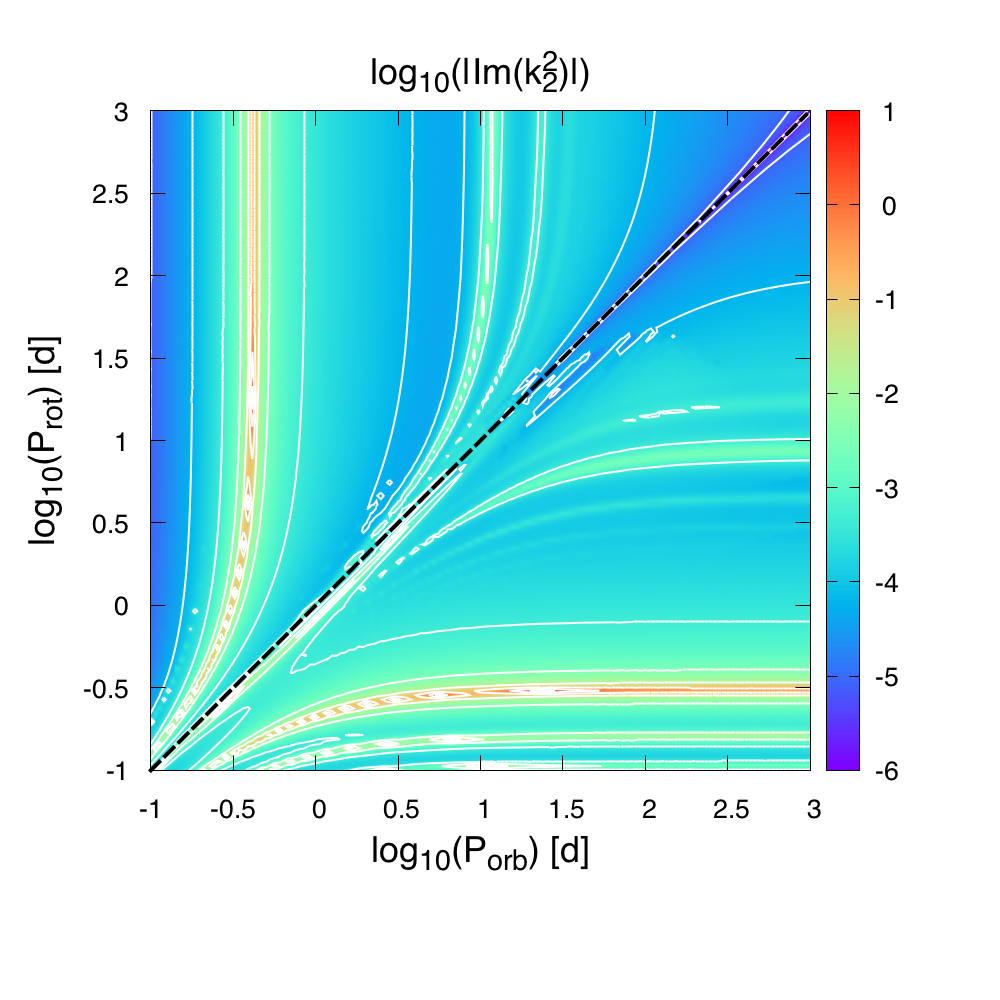} \hspace{0.1cm}
  \includegraphics[width=0.25\textwidth,trim = 2.5cm 6.3cm 3.0cm 3.5cm,clip]{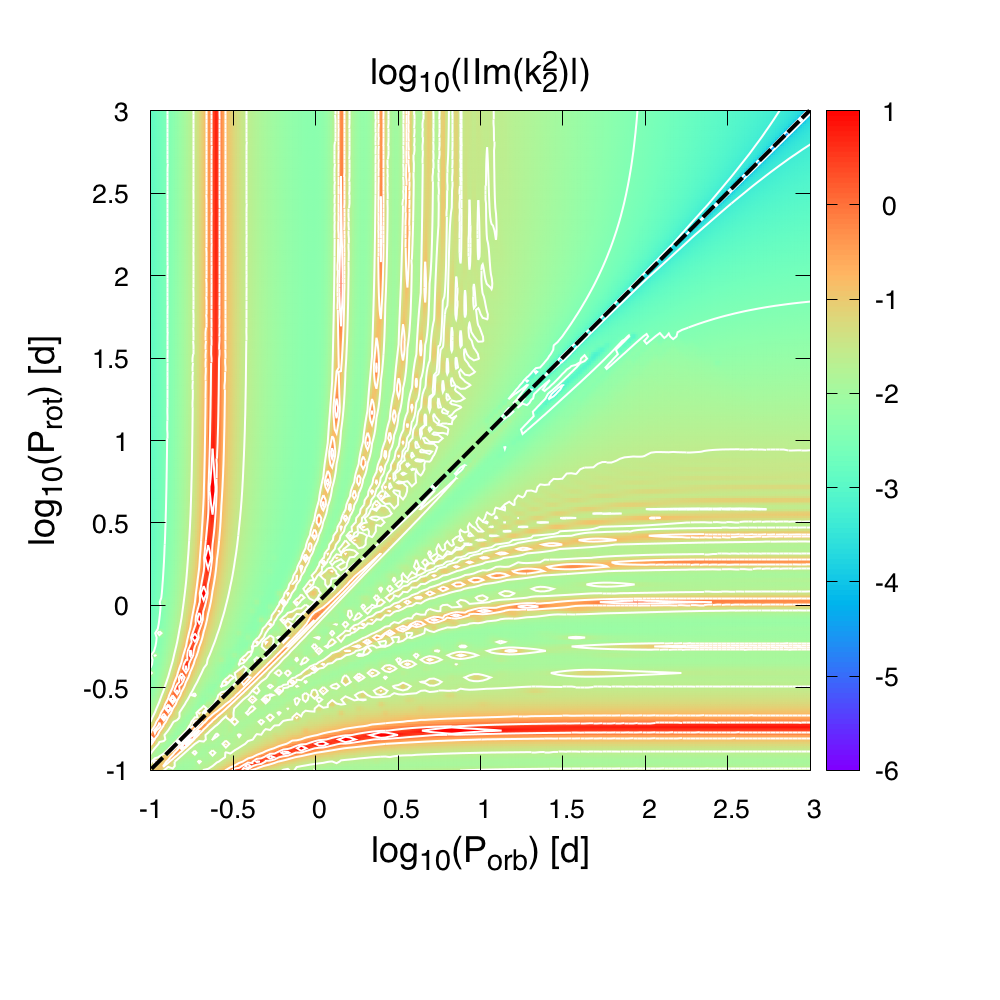} \\[0.1cm]
    \raisebox{1.0cm}{\includegraphics[width=0.02\textwidth]{auclair-desrotour_fig6k.pdf}} \hspace{0.1cm}
   \raisebox{1.0\height}{\includegraphics[width=0.015\textwidth]{auclair-desrotour_fig6g.pdf}}
 \includegraphics[width=0.25\textwidth,trim = 2.5cm 6.3cm 3.0cm 3.5cm,clip]{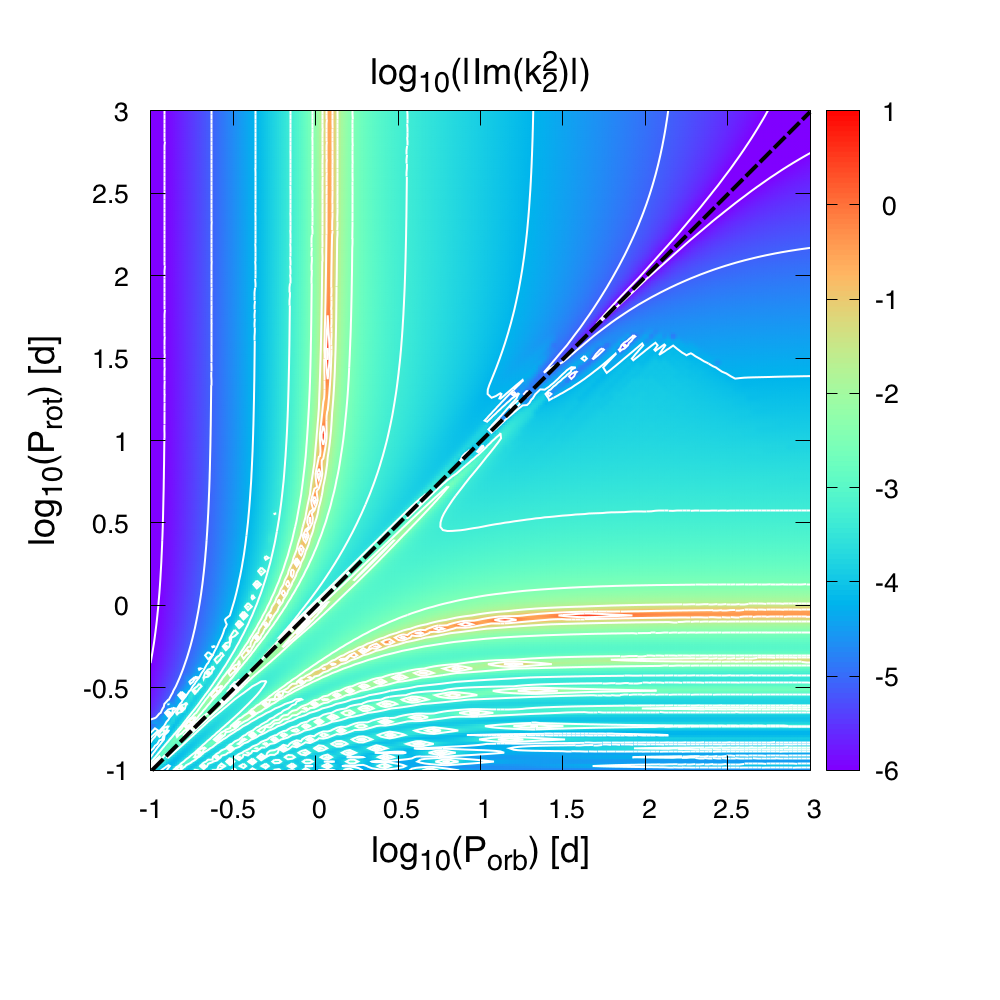} \hspace{0.1cm}
 \includegraphics[width=0.25\textwidth,trim = 2.5cm 6.3cm 3.0cm 3.5cm,clip]{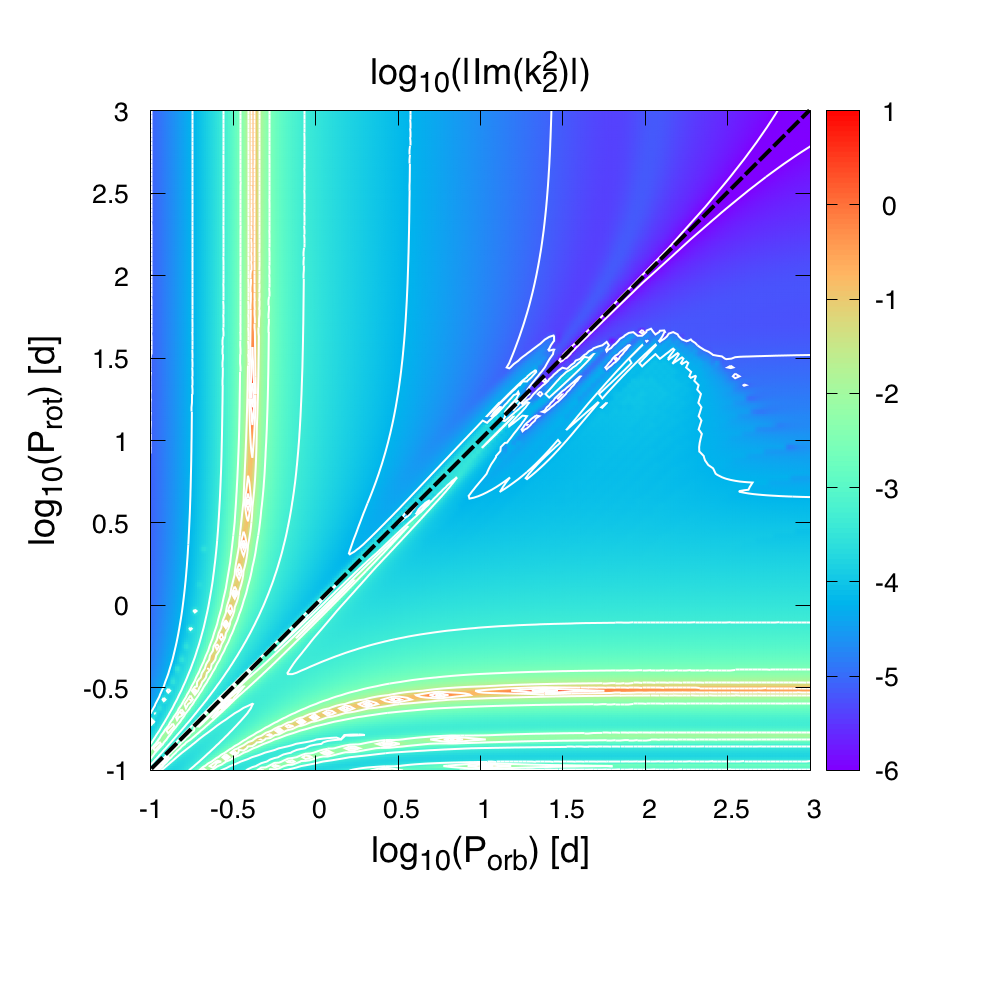} \hspace{0.1cm}
 \includegraphics[width=0.25\textwidth,trim = 2.5cm 6.3cm 3.0cm 3.5cm,clip]{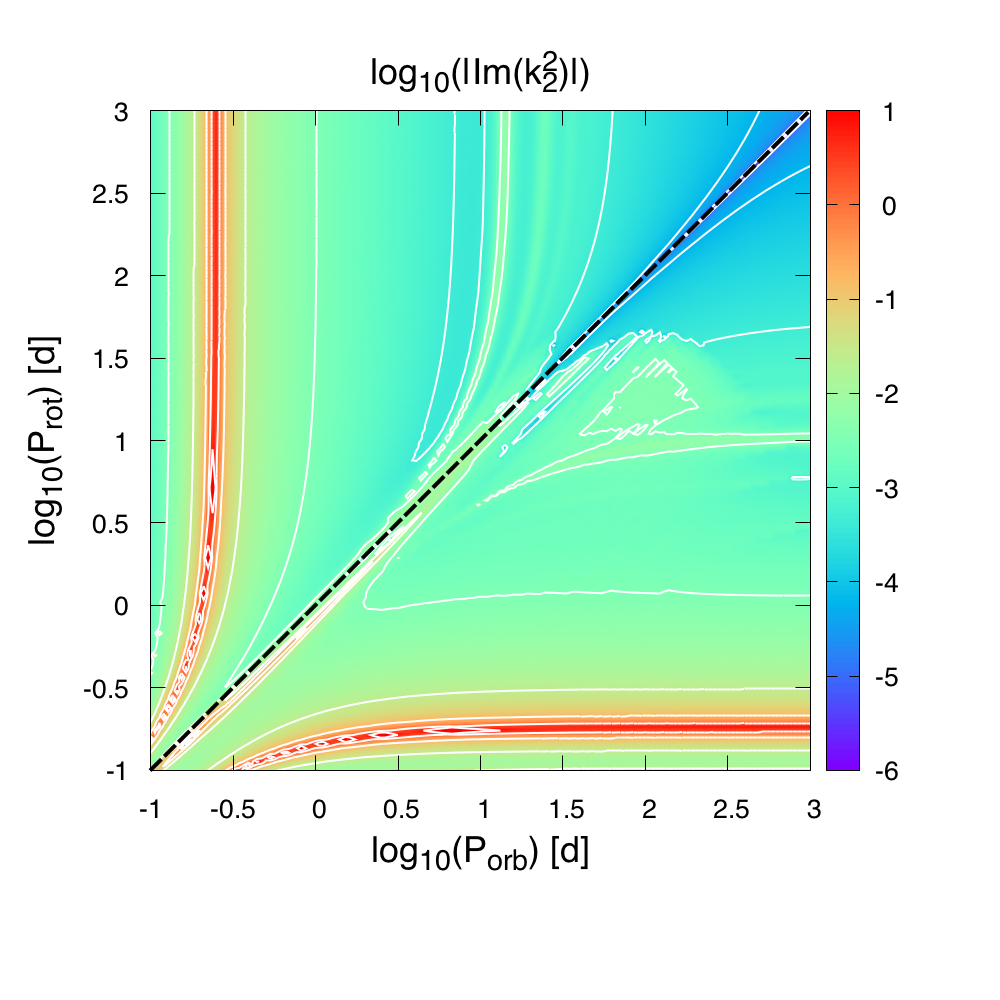} \\[0.1cm]
    \hspace{1.0cm}
   \includegraphics[height=0.3cm]{auclair-desrotour_fig6o.pdf} \hspace{3.0cm}
   \includegraphics[height=0.3cm]{auclair-desrotour_fig6o.pdf} \hspace{3.0cm}
   \includegraphics[height=0.3cm]{auclair-desrotour_fig6o.pdf} \hspace{0.0cm}
\caption{\label{fig:explo_Imk22} Imaginary part of the tidal Love number $ \Im \left\{ k_2^{2} \right\} $ associated with the quadrupolar oceanic tide. The logarithm of $ \Im \left\{ k_2^2 \right\} $ is plotted as a function of the orbital (horizontal axis) and rotation (vertical axis) periods in logarithmic scale by using Eq.~(\ref{k22_thin}) \padc{for $ \sigma_{\rm R} = 10^{-6} \ {\rm s^{-1}} $ (bottom) and $ \sigma_{\rm R} = 10^{-5} \ {\rm s^{-1}} $ (top),} and various values of $ H $ and $ N$. Horizontally, $ H = 10 $~km (left), $ H = 100 $~km (middle) and $ H = 1000 $~km (right). Vertically, $ N = 10^{-4} \ {\rm s^{-1}} $ (bottom) and $ N = 10^{-5} \ {\rm s^{-1}} $ (top). Colors correspond to logarithmic decades (color bars on the right). The diagonal \padc{black dashed line} corresponds to spin-orbit synchronization.}  
\end{figure*}

\begin{figure*}[htb]
 \centering
   \includegraphics[height=0.6cm]{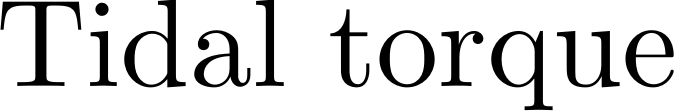} \\[0.3cm]
   \includegraphics[height=0.45cm]{auclair-desrotour_fig6b.pdf} 
   \hspace{0.8cm}
   \includegraphics[height=0.3cm]{auclair-desrotour_fig6c.pdf} \hspace{2.5cm}
   \includegraphics[height=0.3cm]{auclair-desrotour_fig6d.pdf} \hspace{2.5cm}
   \includegraphics[height=0.3cm]{auclair-desrotour_fig6e.pdf} \hspace{2.0cm}~ \\[0.2cm]
   \raisebox{1.0cm}{\includegraphics[width=0.02\textwidth]{auclair-desrotour_fig6f.pdf}} \hspace{0.1cm}
   \raisebox{1.0\height}{\includegraphics[width=0.015\textwidth]{auclair-desrotour_fig6g.pdf}}
  \includegraphics[width=0.25\textwidth,trim = 2.5cm 6.3cm 3.0cm 3.5cm,clip]{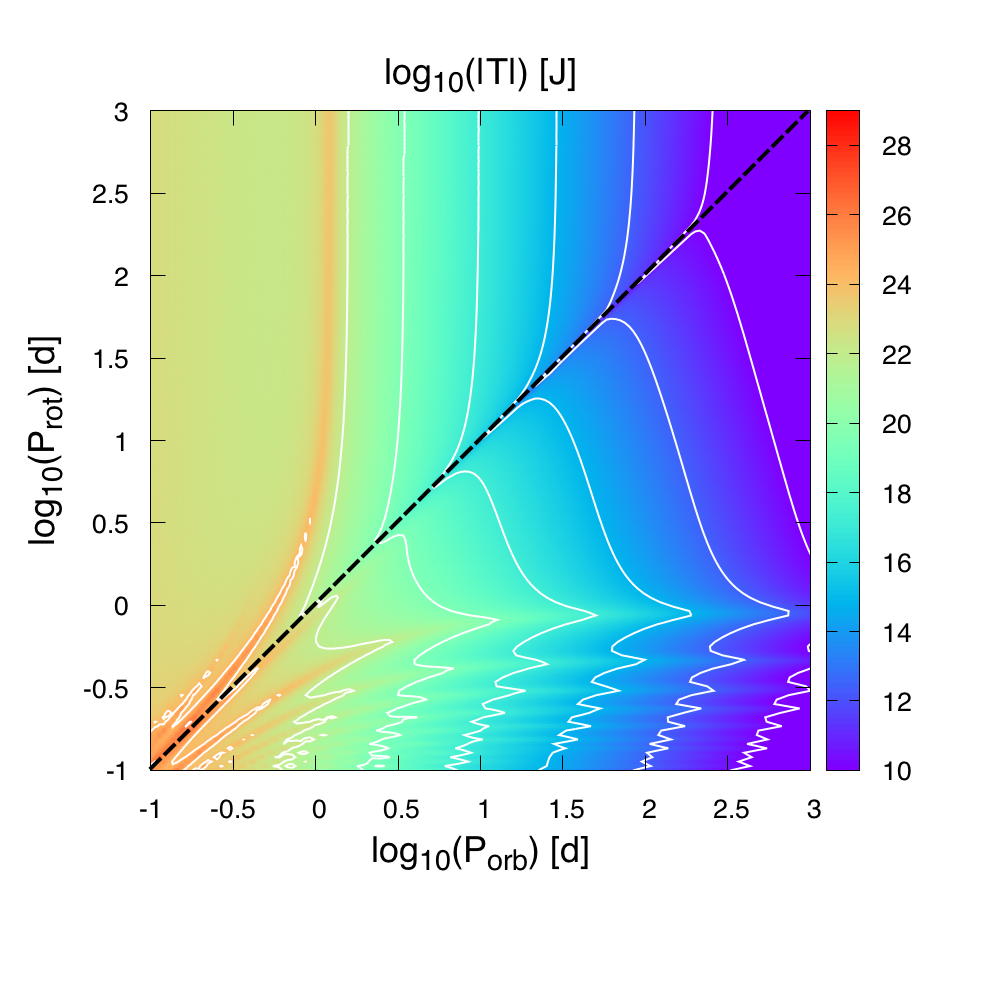} \hspace{0.1cm}
  \includegraphics[width=0.25\textwidth,trim = 2.5cm 6.3cm 3.0cm 3.5cm,clip]{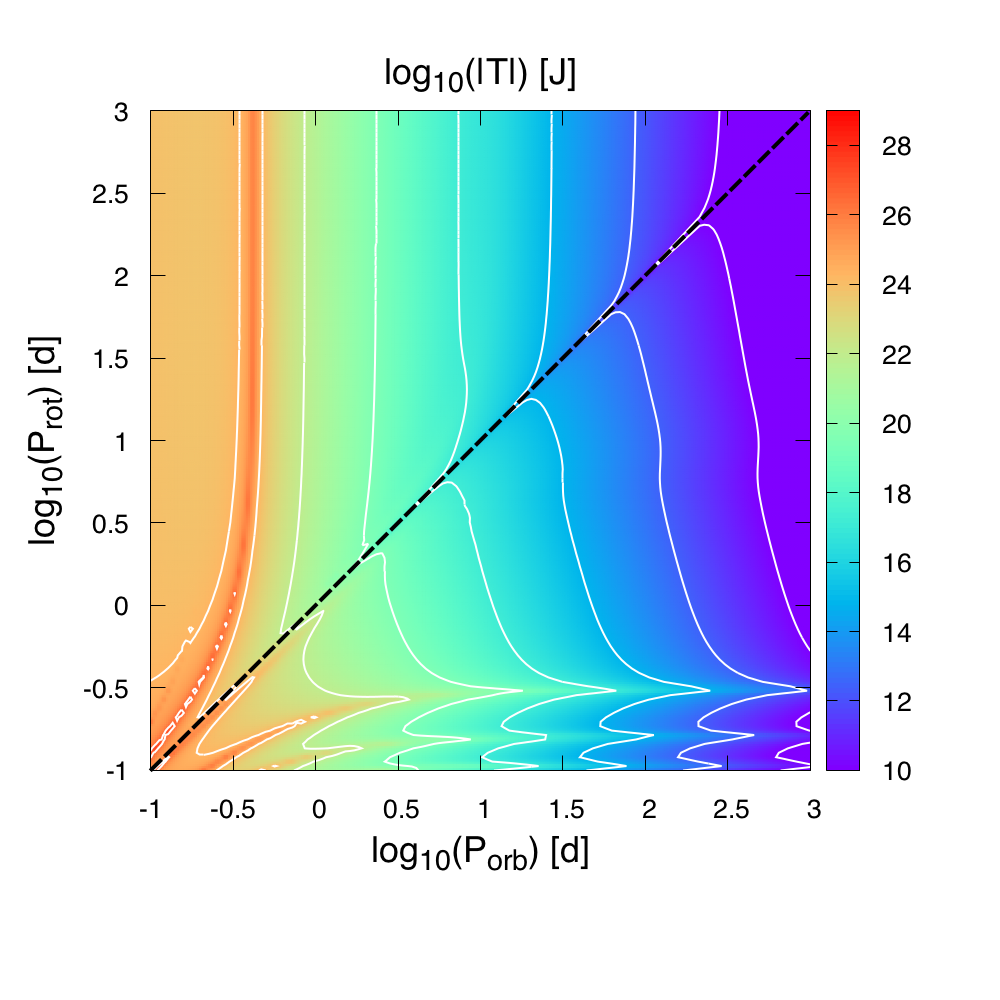} \hspace{0.1cm}
  \includegraphics[width=0.25\textwidth,trim = 2.5cm 6.3cm 3.0cm 3.5cm,clip]{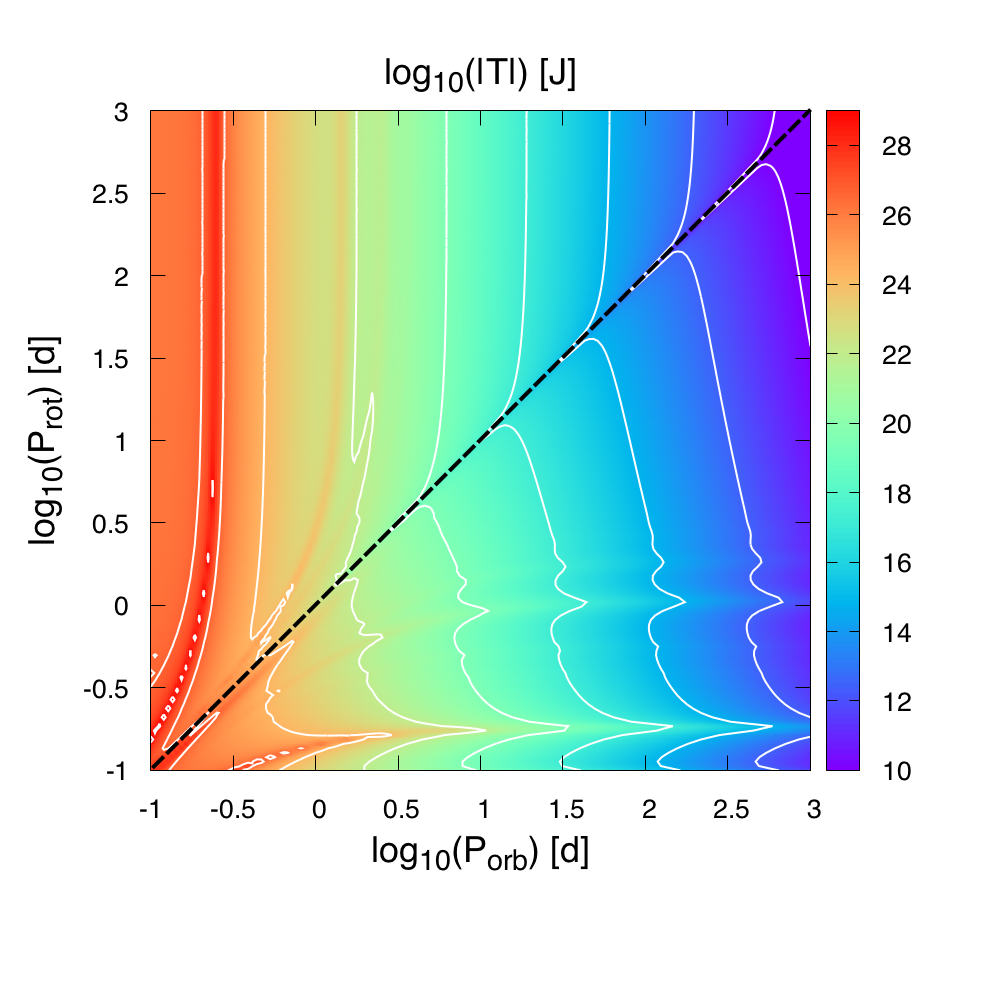} \\[0.1cm]
    \raisebox{1.0cm}{\includegraphics[width=0.02\textwidth]{auclair-desrotour_fig6k.pdf}} \hspace{0.1cm}
   \raisebox{1.0\height}{\includegraphics[width=0.015\textwidth]{auclair-desrotour_fig6g.pdf}}
 \includegraphics[width=0.25\textwidth,trim = 2.5cm 6.3cm 3.0cm 3.5cm,clip]{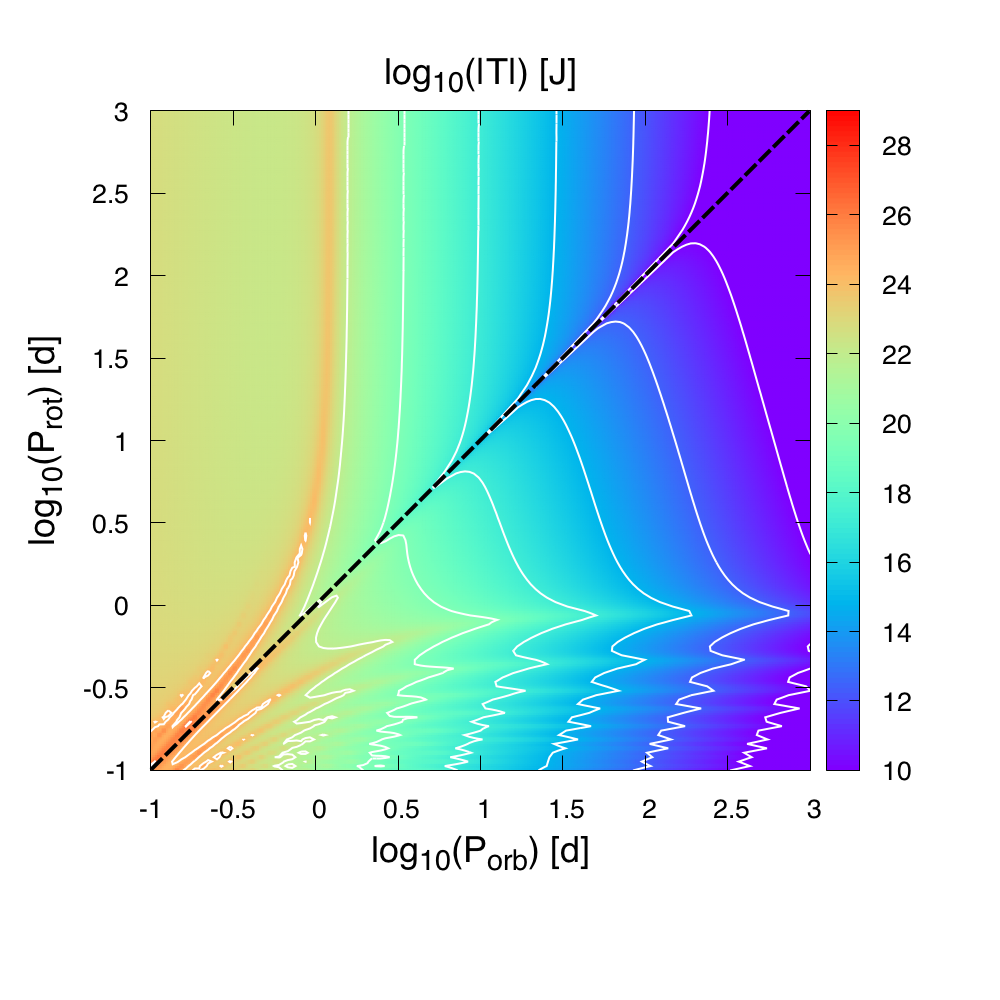} \hspace{0.1cm}
 \includegraphics[width=0.25\textwidth,trim = 2.5cm 6.3cm 3.0cm 3.5cm,clip]{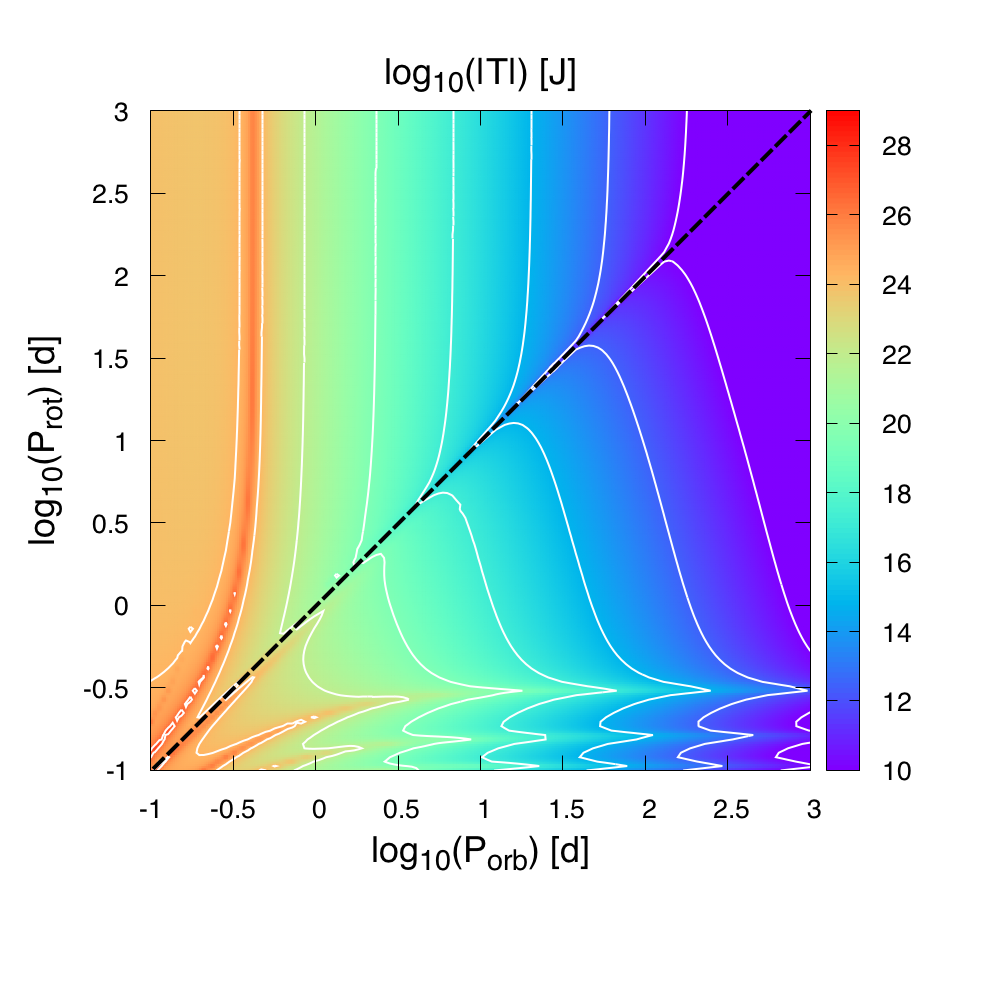} \hspace{0.1cm}
 \includegraphics[width=0.25\textwidth,trim = 2.5cm 6.3cm 3.0cm 3.5cm,clip]{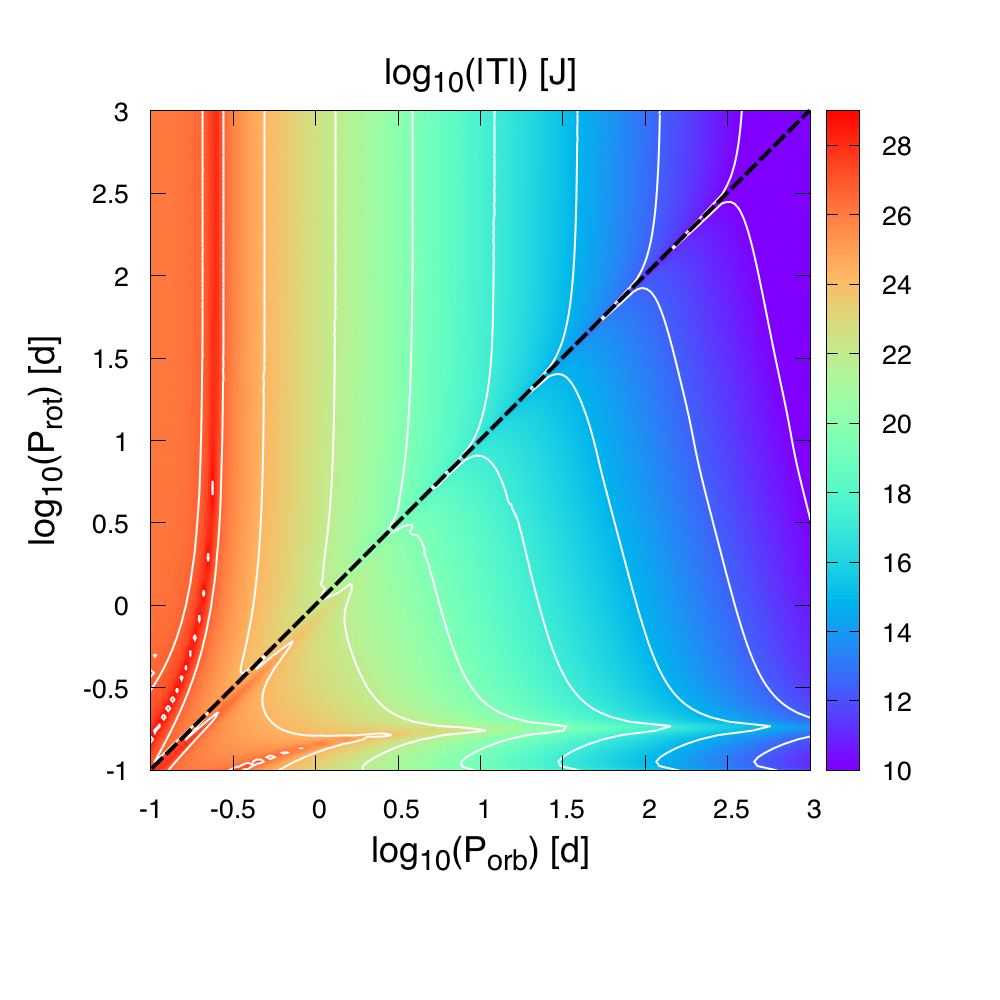} \\[0.1cm]
     \hspace{1.0cm}
   \includegraphics[height=0.3cm]{auclair-desrotour_fig6o.pdf} \hspace{3.0cm}
   \includegraphics[height=0.3cm]{auclair-desrotour_fig6o.pdf} \hspace{3.0cm}
   \includegraphics[height=0.3cm]{auclair-desrotour_fig6o.pdf} \hspace{0.0cm} \\[0.3cm]
 \includegraphics[height=0.45cm]{auclair-desrotour_fig6p.pdf} 
  \hspace{0.8cm}
   \includegraphics[height=0.3cm]{auclair-desrotour_fig6c.pdf} \hspace{2.5cm}
   \includegraphics[height=0.3cm]{auclair-desrotour_fig6d.pdf} \hspace{2.5cm}
   \includegraphics[height=0.3cm]{auclair-desrotour_fig6e.pdf} \hspace{2.0cm}~ \\[0.2cm]
    \raisebox{1.0cm}{\includegraphics[width=0.02\textwidth]{auclair-desrotour_fig6f.pdf}} \hspace{0.1cm}
   \raisebox{1.0\height}{\includegraphics[width=0.015\textwidth]{auclair-desrotour_fig6g.pdf}}
  \includegraphics[width=0.25\textwidth,trim = 2.5cm 6.3cm 3.0cm 3.5cm,clip]{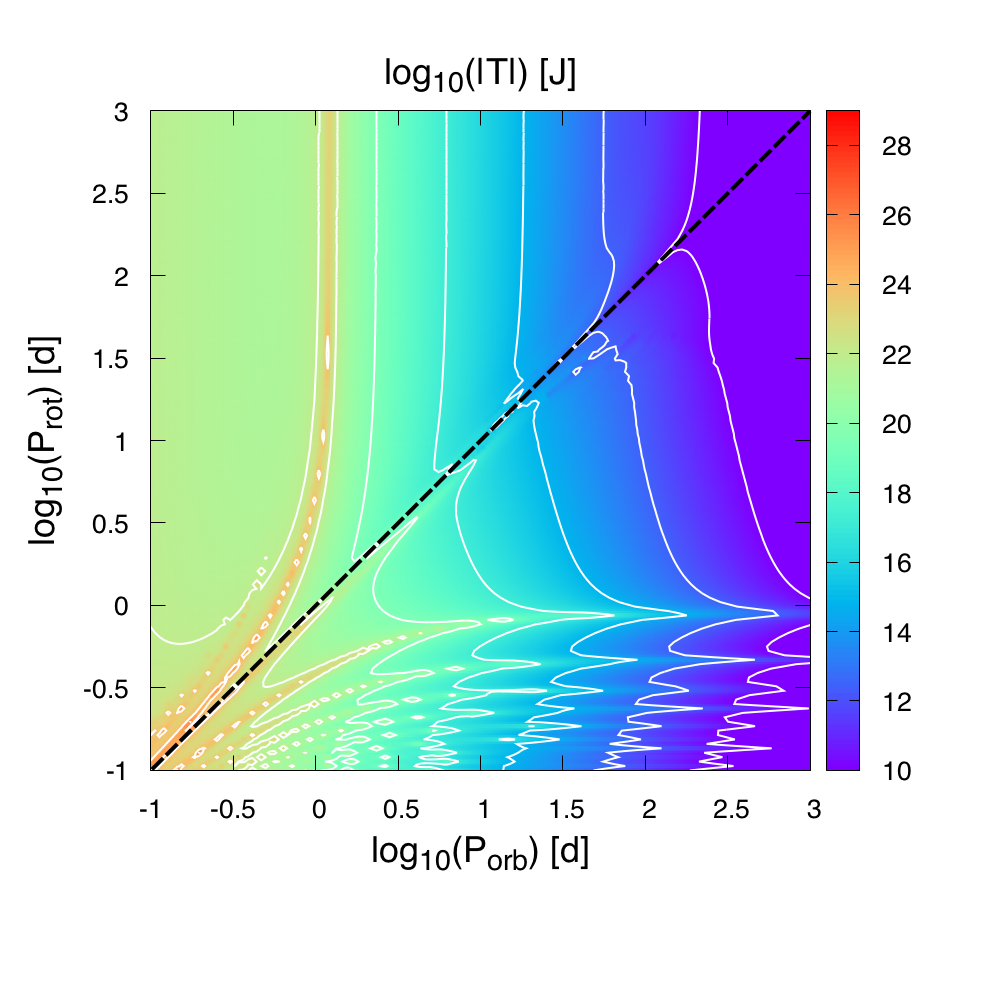} \hspace{0.1cm}
  \includegraphics[width=0.25\textwidth,trim = 2.5cm 6.3cm 3.0cm 3.5cm,clip]{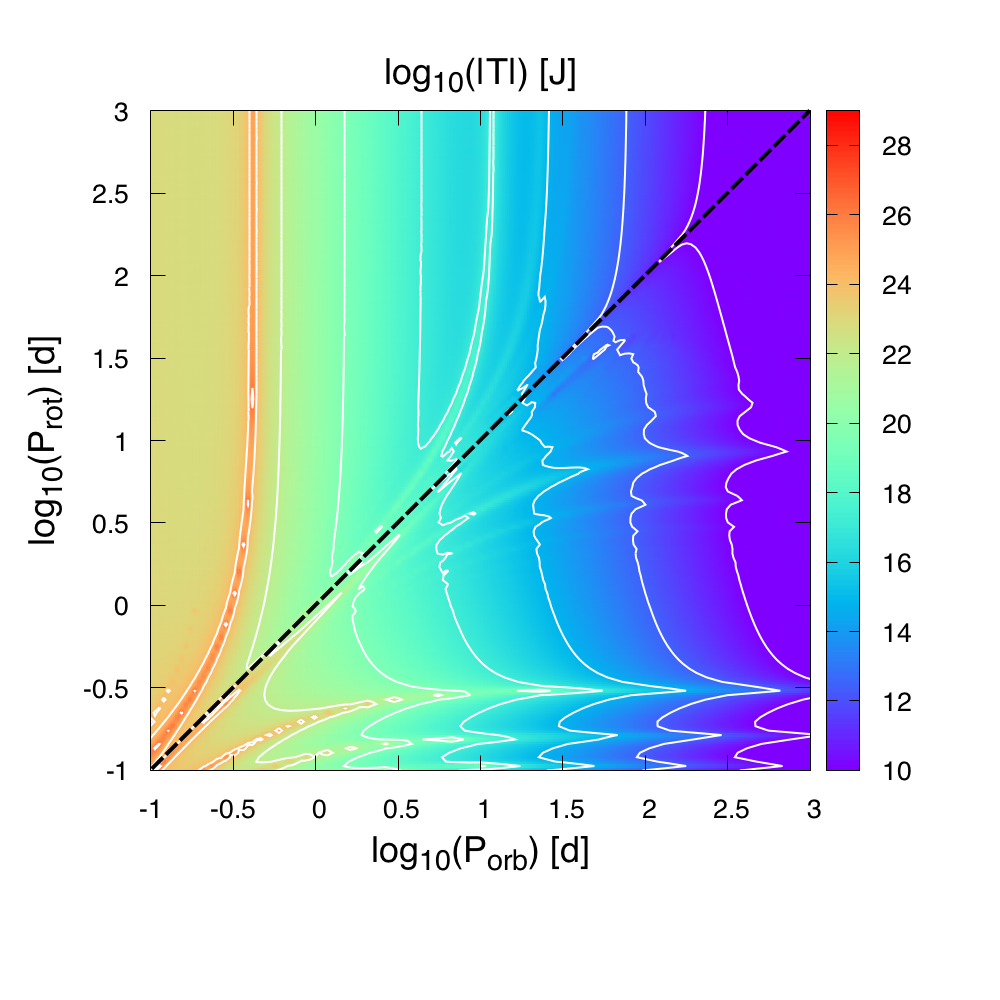} \hspace{0.1cm}
  \includegraphics[width=0.25\textwidth,trim = 2.5cm 6.3cm 3.0cm 3.5cm,clip]{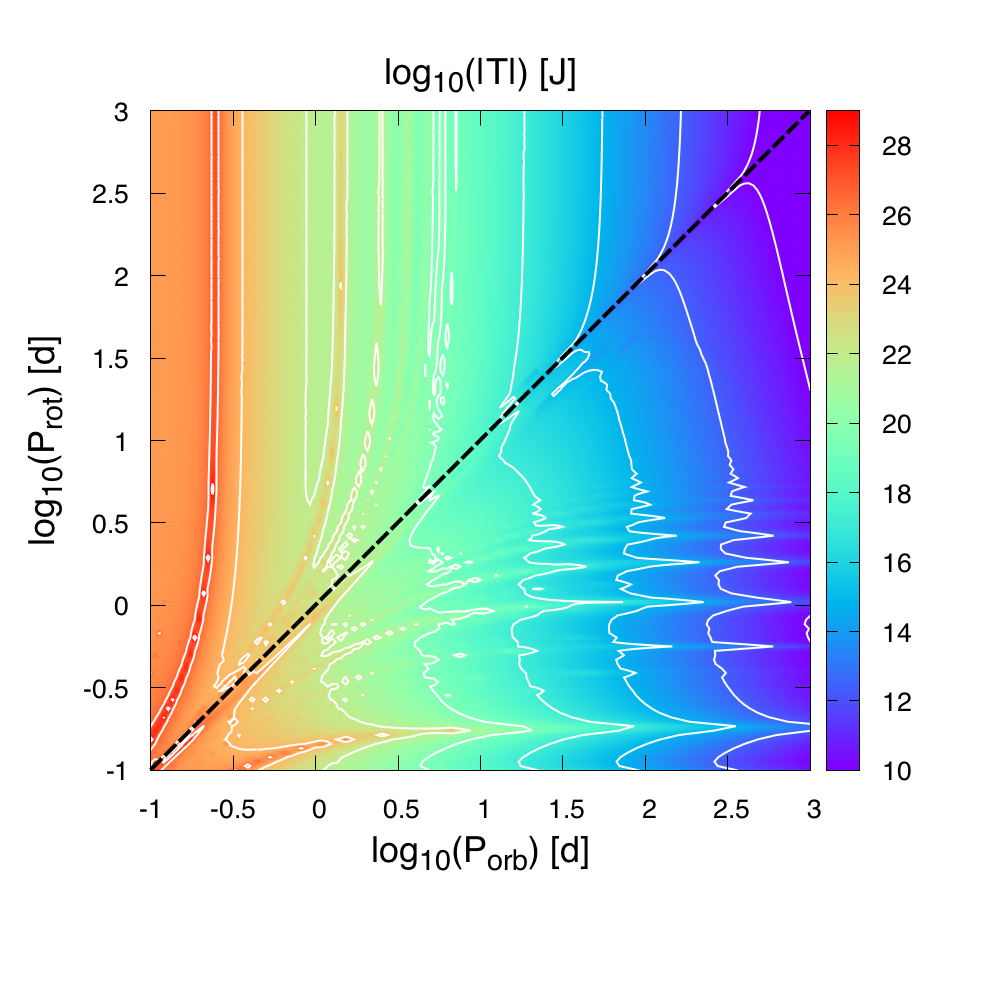} \\[0.1cm]
    \raisebox{1.0cm}{\includegraphics[width=0.02\textwidth]{auclair-desrotour_fig6k.pdf}} \hspace{0.1cm}
   \raisebox{1.0\height}{\includegraphics[width=0.015\textwidth]{auclair-desrotour_fig6g.pdf}}
 \includegraphics[width=0.25\textwidth,trim = 2.5cm 6.3cm 3.0cm 3.5cm,clip]{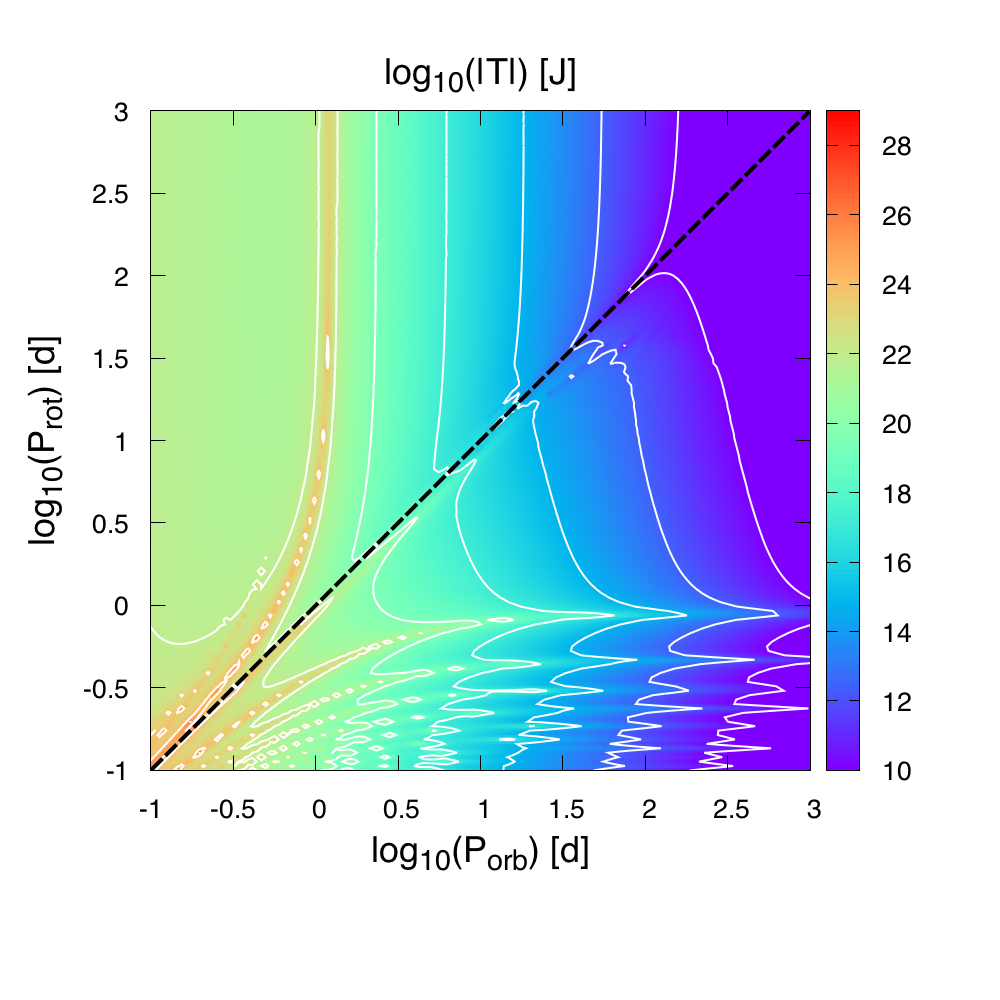} \hspace{0.1cm}
 \includegraphics[width=0.25\textwidth,trim = 2.5cm 6.3cm 3.0cm 3.5cm,clip]{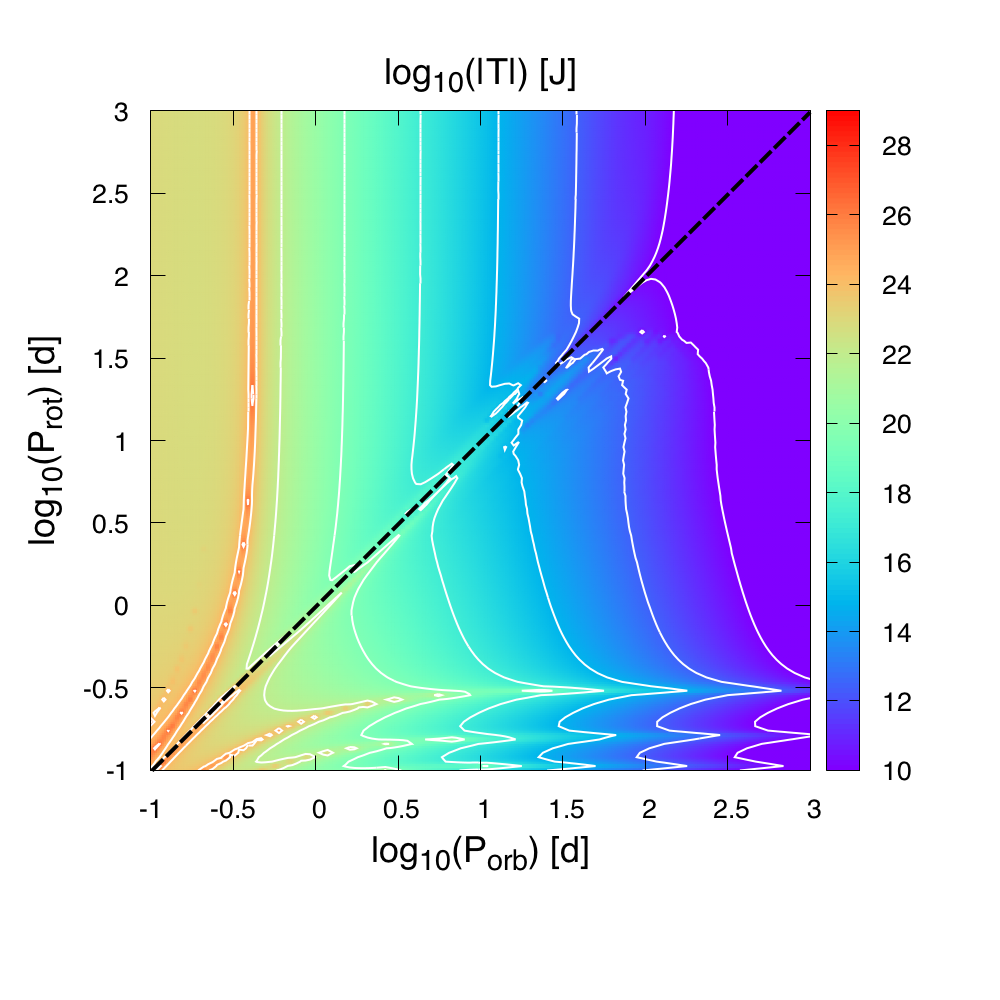} \hspace{0.1cm}
 \includegraphics[width=0.25\textwidth,trim = 2.5cm 6.3cm 3.0cm 3.5cm,clip]{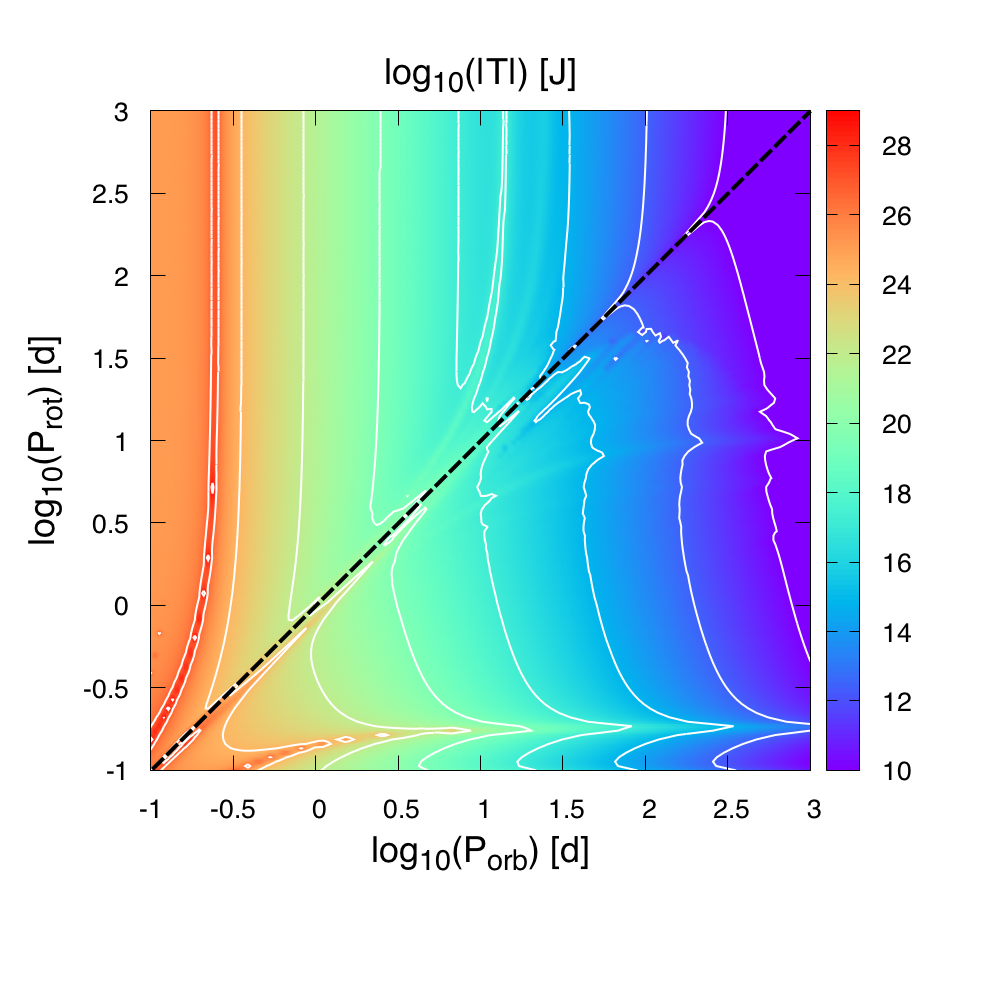} \\[0.1cm]
    \hspace{1.0cm}
   \includegraphics[height=0.3cm]{auclair-desrotour_fig6o.pdf} \hspace{3.0cm}
   \includegraphics[height=0.3cm]{auclair-desrotour_fig6o.pdf} \hspace{3.0cm}
   \includegraphics[height=0.3cm]{auclair-desrotour_fig6o.pdf} \hspace{0.0cm}
\caption{\label{fig:explo_torque} Tidal torque associated with the Lunar quadrupolar oceanic tide. The logarithm of $ \mathcal{T} $ is plotted as a function of the orbital (horizontal axis) and rotation (vertical axis) periods in logarithmic scale by using \smc{Eq.~(\ref{torque_thin})} \padc{for $ \sigma_{\rm R} = 10^{-6} \ {\rm s^{-1}} $ (bottom) and $ \sigma_{\rm R} = 10^{-5} \ {\rm s^{-1}} $ (top),} and various values of $ H $ and $ N$. Horizontally, $ H = 10 $~km (left), $ H = 100 $~km (middle) and $ H = 1000 $~km (right). Vertically, $ N = 10^{-4} \ {\rm s^{-1}} $ (bottom) and $ N = 10^{-3} \ {\rm s^{-1}} $ (top). Colors correspond to logarithmic decades (color bars on the right). The diagonal \padc{black dashed line} corresponds to spin-orbit synchronization.}  
\end{figure*}

\section{From shallow to deep oceans}
\label{sec:explo_para}

In the previous sections, we have identified the depth of the ocean and the stability of its stratification as the key structure parameters characterizing the oceanic tidal response. The depth determines the eigenvalues of resonant surface gravity modes, the Brunt-Väisälä frequency the frequency range of internal gravity modes. Besides, the contribution of these modes is weighted by the dimensionless number $ N^2 H / g $, which compares the Archimedean force to the gravity force. Therefore, we examine in this section the sensitivity of tidal dissipation to $ H $ and $ N $. We consider the case of idealized Earth sisters hosting a global ocean and submitted to the semidiurnal gravitational forcing of any perturber orbiting in their equatorial plane. These planets are characterized by various orders of magnitude of ocean depths ($ H = 10, 100 , 1000 $ km) and Brunt-Väisälä frequency ($ N =10^{-4}, 10^{-3} \ {\rm s^{-1}}$), all the other parameters being unchanged (see Table~\ref{para_cases}). \padc{Following the discussion of the previous section, the frequency associated with the Rayleigh drag is set to $ \sigma_{\rm R} = 10^{-5} \ {\rm s^{-1}} $ in a first case (friction time scale of the Earth's ocean). In a second case, it is set to $ \sigma_{\rm R} = 10^{-6} \ {\rm s^{-1}} $, which corresponds to a weaker friction, to broaden the explored parameter space.} Hence, for each planet, we plot \smc{in} Fig.~\ref{fig:explo_Imk22} the imaginary part of \smc{the} tidal Love number as a function of the orbital period of the perturber and the rotation period of the planet. Then, assuming that the perturber is a Moon-like satellite ($M_{\rm pert} = 7.346 \times 10^{22} $~kg), we plot the corresponding tidal torque exerted on planets \smc{in} Fig.~\ref{fig:explo_torque}. 


 \rec{As it may be noted,} the three maps plotted for $ H = 10 $~km (bottom panels of Figs.~\ref{fig:explo_Imk22} and \ref{fig:explo_torque}) have the same aspect whatever the value of $ N $. \rec{This can be interpreted as the fact that the tidal response generated by the gravitational forcing exerted on the ocean is mainly composed of surface gravity waves. In other words, the tidal response is dominated by its barotropic component, which induces that the role played by \smcc{the} stratification appears to be negligible in this case, as mentioned before.} We identify the resonances due to surface \rec{inertial-gravity} modes observed in the case of the Earth in the previous section (Fig.~\ref{fig:Earth_spectre}, left column). They form a typical \emph{wings pattern}, which is proper to these waves \citep[see e.g.][Fig.~7, in the case of white dwarves]{FL2014}. However, note that this pattern is not symmetrical with respect to the diagonal line designating synchronization. This is a consequence of \smc{the effects of the Coriolis acceleration}. In the super-synchronous regime (area located below the diagonal line), \smc{they} induce a distortion that couples the quadrupolar forcing to several Hough modes. These \rec{inertial-gravity} waves tend to be mixed with acoustic waves while $ P_{\rm rot} \rightarrow 0 $. In the sub-synchronous regime (area located above the diagonal line), the forcing frequency is greater than the inertia frequency (super-inertial regime) and the only mode coupled with the forcing is the gravity mode of degree $ n = 0 $. As a consequence, only one resonance appears in the super-inertial regime. As the tidal potential scales as $ U_2^{2,\sigma} \, \propto \, n_{\rm orb}^2 $ (see Eq.~\ref{U22}), the tidal torque scales as $ P_{\rm orb}^{-4} $ (see \smc{Eq.~\ref{torque_thin}}), which corresponds to the horizontal color gradient of the maps plotted on Fig.~\ref{fig:explo_torque}. 

By moving to deeper oceans, we observe that the resonances are translated towards the small-period range, following the scaling law $ \sigma_n^{\pm} \, \propto \, \sqrt{gH} $ identified above. For $ N = 10^{-3} \ {\rm s^{-1}} $, the stratification is sufficiently strong to allow internal gravity waves to propagate. Therefore, new resonances appear in the frequency range $ \left| \sigma \right| \lesssim N$ (see \padc{right panels} of Figs.~\ref{fig:explo_Imk22} and \ref{fig:explo_torque}). This effect is enhanced for $ H = 1000 $~km. As the eigenfrequencies of internal gravity waves do not depend on $ \Omega $ like those of surface gravity waves, their wings patterns are symmetrical with respect to the diagonal axis. Moreover, given that the contribution of internal waves is weighted by the dimensionless number $ N^2 H / g $, the non-resonant background is amplified in the frequency range $ \left| \sigma \right| \lesssim N$. \padc{The effects of the resonances associated with internal gravity waves are attenuated by the Rayleigh drag. When $ \sigma_{\rm R} \rightarrow 0 $, resonances tend to increase the dependence of the the tidal torque and Love number on the forcing frequency \citep[as expected from the general scaling laws derived in][]{ADMLP2015}.}

\section{Discussion}
\label{sec:discussion}

In this work, we have opted for a simplified linear analysis allowing us to explore widely the domain of parameters with a reasonable computational cost. This approach is convenient to examine the frequency-resonant behaviour of a global ocean and highlight the key parameters of the problem, which can be \smc{explained in more details} in a second phase. Particularly, we identify and quantify in a consistant way the contribution of internal gravity waves restored by the stable stratification. This contribution can be important in the case of deep oceans, where the \smc{combined} effects of tides and stratification induce important density fluctuations. However, we shall discuss here the main simplifications assumed in the model:

\begin{itemize}
\item[$\bullet$] \emph{Global ocean of uniform depth} -- The large scale topographical features of the oceanic floor and continents are not taken into account. Because of this spherical symmetry, the obtained oceanic tidal response is very regular and exhibits resonances corresponding to global spherical modes (see e.g. Fig.~\ref{fig:Earth_spectre}). The case of a hemispherical ocean centred at the equator was treated in early studies in the shallow water approximation \citep[][]{Proudman1936,Doodson1938,LH1970,Webb1980}. Meridian continental shelves prevent large scales gravity waves to propagate. They thus couple the global gravitational forcing to modes determined by the scale of the basin. This modifies significantly the aspect of the frequency spectrum of the tidal response \citep[see e.g.][]{Webb1980}.

\item[$\bullet$] \emph{Rayleigh friction} -- All of the dissipative mechanisms damping the tidal response are reduced to a single parameter, the effective Rayleigh coefficient $ \sigma_{\rm R} $. This approximation is \rec{common in litterature \citep[e.g.][]{Webb1980,ER2001,ER2003,Ogilvie2009,Tyler2011} and} assumed for convenience. In addition to the traditional approximation, discussed in the next paragraph, it allows us to separate the $ x $ and $ \theta $ coordinates in \smc{the} dynamics. Rayleigh friction can be reasonably used to describe the drag caused by small scales topographical features homogeneously distributed around the planet, as in the numerical model of \cite{ER2001}. It can also be considered as a \rec{simplified} approximation of viscous and turbulent frictions if a lengthscale $ L $ such that $ \Delta  \textbf{V} \sim V / L^2 $ is introduced. However, it does not \smc{model} at all of the local dissipation due the breaking of waves on continental shelves whereas this effect stands for the major part of the \smc{tidal} energy in the case of the Earth.

\item[$\bullet$] \emph{Traditional approximation} -- This simplification is commonly used in \smc{the} literature to solve the latitudinal and vertical structure of the tidal response separately \citep[e.g.][]{Unno1989}. It consists in neglecting the latitudinal component of the rotation vector in the Coriolis acceleration. This means that we ignore both the radial component of the Coriolis force and the horizontal component of the Coriolis force associated with radial motions. Consequently, this approximation is well appropriate to 2D models, where vertical motions are negligible relatively to horizontal ones. In the case of deeper oceans, its domain of applicability depends on the hierarchy of the inertia ($2\Omega$), Brunt-Väisälä ($N$) and forcing ($ \sigma $) frequencies. In the super-inertial asymptotic regime ($ \left| 2 \Omega \right| \ll \left| \sigma \right| $), \smc{the forcing time scale is small compared to the rotation period}, which makes the traditional approximation appropriate. However, when this condition is not satisfied, motions are fully tridimensional in the case of a neutrally-stratified ocean ($ N \approx 0 $). This requires to use 3D \smc{(2D if solutions $ \propto \, \ed^{i m \varphi} $ are assumed)} numerical \smc{or} semi-analytical methods \citep[see e.g.][]{OL2004}. The Archimedean force acts as a restoring force on fluid particles in the vertical direction. Therefore, if \jlc{this restoring force} is sufficiently strong, it prevents vertical motions to be comparable in amplitude to horizontal ones. This condition is expressed as $ \left| 2 \Omega \right| \ll N $ \citep[e.g.][]{ADLM2017a}. The validity of the traditional approximation has been examined in various fluid layers, from planetary oceans and atmospheres \citep[e.g.][]{GS2005,GZ2008} to stellar interiors \citep[e.g.][]{Mathis2009,Tort2014,Prat2017}.

\item[$\bullet$] \emph{Solid rotation} -- As we consider in our model that the oceanic layer rotates uniformly with the planet at the angular velocity $ \Omega $, we ignore the complex interplay existing between tidal waves and mean flows \smc{\citep[e.g.][]{Favier2014,Guenel2016}}. By studying the propagation of internal gravity waves in a shear flow, \cite{BB1967} showed that the amplitude of oscillations is attenuated by a factor depending on the local Richardson number of the fluid, \smc{$ R_{\rm i} = N^2  \left| d  \textbf{V}_0  / dr \right|^{-2} $} (where $ \textbf{V}_0 $ stands for the velocity vector of the shear flow), as waves pass through a critical level at which their horizontal phase speed is equal to the zonal velocity of the flow. By this way, tidal waves can transfer horizontal moment to mean flows, which modifies the tidal response.

\item[$\bullet$] \emph{Free-surface boundary condition} -- In this work, a free-surface boundary condition is used to derive analytic solutions. This condition corresponds well to the case of external oceanic layers, where the upper surface can move freely. However, it is not appropriate to \smc{subsurface} oceans such as those presumedly hosted by Europa-like icy satellites \smcc{\citep[e.g.][]{Khurana1998,Kivelson2002}}. \rec{In order to study these objects, results obtained in this work can easily be applied to the case of a rigid lid by replacing for instance the standard free-surface condition by a rigid wall condition}.

\item[$\bullet$] \padc{\emph{Coupling with the solid part} -- In order to simplify the analysis, the solid part of the planet has been considered as a sphere of infinite rigidity. In reality, the solid part behaves as a visco-elastic body \citep[e.g.][]{Efroimsky2012} forced both by the tidal potential and the loading induced by the tidal distortion of the ocean. This solid-fluid coupling affects the frequencies given by Eq.~(\ref{sigmanpm}) and generates an additional coupling between tidal modes \citep[e.g.][]{Matsuyama2014}.} \rec{We simply ignored this coupling so that we do not need to consider the parameters of the solid response.}

\end{itemize}

\section{Conclusion}
\label{sec:conclusion}

To better understand the tidal response of the deep global oceans potentially hosted by recently discovered extrasolar planets, we developed a linear model reducing the physics of tides to the essentials. Our approach leaves the traditional 2D modeling inherited from the study of the Earth's \smc{oceans} to include three-dimensional effects resulting from the properties of the oceanic vertical structure. Particularly, it introduces the contribution of internal gravity waves induced by the stable stratification and takes into account dissipative mechanisms with a \smc{Rayleigh} drag, following the early work by \cite{Webb1980}. Hence, we wrote the equations describing the tidal response of a deep ocean in the general case, as well as the associated tidal torque, Love numbers and quality factor. We then simplified these equations in the framework of the thin-layer approximation to compute an analytic solution. This solution was finally used to explore the domain of parameters. We first treated the cases of idealized Earth and TRAPPIST-1 f planets to constrain the Rayleigh coefficient ($ \sigma_{\rm R} $) and illustrate asymptotic regimes. In a second phase, we examined in a systematic way the dependence of the tidal Love number and torque associated with the semidiurnal tide on the ocean depth ($ H $) and Brunt-Väisälä frequency ($ N $). 

In the thin layer limit, we recover the behaviour identified by early studies \citep[e.g.][]{LH1970,Webb1980}. The tidal response is composed of resonant surface \rec{inertial-gravity} modes due to the conjugated effects of gravity and rotation. In this case, the role played by stratification is negligible. The fluid compressibility can affect the tidal response in the high-frequency range, where horizontally propagating Lamb modes can be excited. The \rec{importance of the role} played by the stratification grows with the ocean depth. The stable stratification allows internal gravity waves to propagate. It induces internal density fluctuations characterized by a frequency-resonant behaviour, like surface gravity waves. The contribution of this component is not negligible in the case of deep oceans and sensitively increases the evolution timescales of the planet-perturber system. 

Although the linear analysis developed in this work is \jlc{simplified} with respect to numerical  models, \jlc{it provides a} very convenient \jlc{way} to explore the \smc{broad} domain of \smc{planetary} parameters and unravel the complex dynamics of oceanic tides with a reasonable computational cost and few physical \smc{control} parameters. \rec{These analytical results can be used in the future as benchmark validations for fully 3D GCM tidal studies. It should be considered as a next step to introduce meridian boundaries in the model in order to study the effect of blocking the progression of the tidal response and thus better characterize the role played continental shelves in the oceanic tidal dissipation.  }

\begin{acknowledgements}
The authors acknowledge funding by the European Research Council through ERC grants SPIRE 647383 and WHIPLASH 679030, \smc{the Programme National de Planétologie (INSU/CNRS) and the CNRS PLATO grants at CEA/IRFU/DAp}. \rec{They wish to thank the referee, Robert Tyler, for his helpful suggestions and remarks.} 
\end{acknowledgements}

\bibliographystyle{aa}  
\bibliography{auclair-desrotour} 

\end{document}